\documentclass[a4paper,preprintnumbers,floatfix,onecolumn,aps,prb,unsortedaddress,superscriptaddress,showkeys]{revtex4-2}
\usepackage[utf8x]{inputenc}
\usepackage{verbatim}
\usepackage{graphicx,floatrow}
\usepackage{subfig}
\usepackage{color}
\usepackage{amsmath,amsfonts,amssymb}
\usepackage{miller}
\usepackage{lineno}
\usepackage{url}
\usepackage{verbatim}
\usepackage{multirow}

\renewcommand{\vec}[1]{\ensuremath{\mathbf{#1}}}
\renewcommand{\Hat}[1]{\ensuremath{\hat{\mathbf{#1}}}}
\newcommand{\mat}[1]{\ensuremath{ \underline{\underline{\mathbf{#1}}}} }

\begin{document}

\title{Atomistic Study of Irradiation-Induced Plastic and Lattice Strain in Tungsten}
\author{Jintong Wu}
\affiliation{Department of Physics, University of Helsinki, Post-office box 43, FIN-00014 University of Helsinki, Finland}
\author{Daniel R. Mason}
\affiliation{UK Atomic Energy Authority, Culham Science Centre, Oxfordshire OX14 3DB, UK}
\author{Fredric Granberg}
\email{fredric.granberg@helsinki.fi}
\affiliation{Department of Physics, University of Helsinki, Post-office box 43, FIN-00014 University of Helsinki, Finland}

\begin{abstract}

We demonstrate a practical way to perform decomposition of the elasto-plastic deformation directly from atomistic simulation snapshots. Through molecular dynamics simulations on a large single crystal, we elucidate the intricate process of converting plastic strain, atomic strain, and rigid rotation during irradiation. Our study highlights how prismatic dislocation loops act as initiators of plastic strain effects in heavily irradiated metals, resulting in experimentally measurable alterations in lattice strain. We show the onset of plastic strain starts to emerge at high dose, leading to the spontaneous emergence of dislocation creep and irradiation-induced lattice swelling. This phenomenon arises from the agglomeration of dislocation loops into a dislocation network. Furthermore, our numerical framework enables us to categorize the plastic transformation into two distinct types: pure slip events and slip events accompanied by lattice swelling. The latter type is particularly responsible for the observed divergence in interstitial and vacancy counts, and also impacts the behavior of dislocations, potentially activating non-conventional slip systems.
\end{abstract}


\maketitle

\section{Introduction}

\sloppy The consequences of irradiation in materials, mainly seen as microstructural evolution, originates from the interaction between the incident particles and the lattice atoms of the material. Among them, radiation damage produced during the collision cascade process leads to swelling, embrittlement, and creep. These mechanical properties determine the lifetime of nuclear plant parts. Tungsten (W) has been chosen to play a pivotal role in fusion test reactors under construction, such as the International Thermonuclear Experimental Reactor (ITER)~\cite{ITER1}. Here it will be used as the material facing the plasma inside the reactor, owing to its high atomic mass, leading to a low sputtering yield~\cite{Low_sputtering_yield}, high melting temperature and high thermal conductivity~\cite{high_temp_W1,high_temp_W2,high_conductivity_W}. 

Irradiation induces significant changes in both the microstructure and mechanical properties of materials~\cite{microstructuralChange_1, microstructuralChange_2, microstructuralChange_3, Mechanic_change}. The response of tungsten to irradiation has been extensively studied both experimentally and computationally. On the experimental side, Reza et al.~\cite{W_TGS_exp} used transient grating spectroscopy (TGS) and found that 20 MeV self-ion irradiation showed defect saturation at doses between 0.06 and 0.1 dpa. Another positron annihilation spectroscopy study by Hollingsworth et al.~\cite{W_PAS_exp_new} showed that vacancies appear as small clusters at low doses at room temperature. Recent research suggests that the accumulation of small vacancy clusters due to irradiation can eventually lead to the formation of macroscopic voids, resulting in observable swelling of the material~\cite{crystal_2023}. Another study elucidates that this swelling phenomenon primarily arises from the agglomeration of self-interstitial defects into dislocation loops and extended dislocation networks. These dislocations are produced during collision cascades initiated by, for example, neutron irradiation, leading to the accumulation of uncompensated relaxation volumes~\cite{Dudarev_NuclFus2018}. Radiation promotes the glide of dislocations at high doses, thereby further extending the dislocation network. This process involves the thermally activated motion of screw dislocations, particularly evident in BCC metals~\cite{Xiao_plastic_glide}. Understanding the underlying mechanism of plastic flow at high-dose irradiation is crucial to gaining insights into the macroscopic swelling mechanism. On the simulation side, high-dose simulations are computationally heavy and more efficient methods have been proposed to generate characteristic high-dose microstructures. Derlet et al.~\cite{derletCRA} proposed the creation relaxation algorithm (CRA) to quickly reach high dose levels without full MD cascades. Defect densities obtained by the CRA are found to be higher than experimental values due to the lack of thermal effects~\cite{Boleininger_SciRep2023}. To address this, combination of CRA and cascade simulation, known as cascade annealing (CA), have been used and showed good agreement with full MD simulations at similar doses~\cite{MasonPRM_VoidDetection,Granberg2022}. In the high-dose irradiation simulation scenario, particularly in the cell following CRA for sped-up dose accumulation, the atomic structure becomes highly disordered, and the complex defect network hinders our ability to accurately characterize the deformation process within the atomic system during the CA process.

Nevertheless, understanding the intricate and nonlinear dynamics of the damage microstructure at high irradiation doses is a multifaceted task, encompassing various length and timescales, while also being influenced by exposure and environmental conditions~\cite{kiener2011situ}. Consequently, creating a concise atomistic model that accurately encompasses all relevant factors at the appropriate scales remains a formidable challenge. The detectable fluctuations in strains and stresses within irradiated materials provide an avenue for directly confirming the accuracy of real-space simulations. Elasticity equations serve as the cornerstone for connecting atomic-scale defects to macroscopic strains. In classical MD simulations, elasto-plastic deformation is inherently coupled, and quantifying the magnitude of elasto-plastic deformation to gain a deeper understanding of the underlying defect evolution mechanism remains challenging. Earlier works~\cite{PreStrain1,PreStrain2,PreStrain3,PreStrain4,PreStrain5,PreStrain6,PreStrain7,PreStrain8,PreStrain9,PreStrain10} explored the plastic deformation of nanoscale metal crystals at low temperatures using MD simulations, but these qualitative studies lacked quantitative magnitude of plastic deformation. Vo et al.~\cite{VoQuantitative} proposed a decomposition algorithm primarily based on dislocation motion-induced plastic deformation. However, it is only applicable to one-dimensional configurations. In cases of complex dislocation motion, such as high defect density cases, the interaction between dislocations often renders the decomposition results inaccurate. Another previous study provided a decomposition method~\cite{stukowski2012elastic}, based on the virtual intermediate configuration concept. However, if the identification algorithm fails to accurately determine the local atomic structure, it cannot perform elasto-plastic decomposition. For instance, a cell exposed to high-dose irradiation can cause the algorithm to fail. 

In addition to advances in simulation methodology, there have been recent advances in microstructural characterization. The Wigner-Seitz (W-S) method~\cite{WS} has been used extensively to identify and characterize isolated defects embedded in a perfect reference crystal, but fails when the reference lattice itself can evolve, or when large local displacements comparable with the lattice spacing exist~\cite{GeV_AM_SHI}. Recently Mason et al.~\cite{MasonPRM_VoidDetection} developed a method based on detecting the Wigner-Seitz cell \emph{locally}, and using this to detect void isosurfaces. We refer to this method as Void Isosurface Detection (VID). Machine Learning analyses have also been developed to detect defect features related to vacancies and interstitials~\cite{Bhardwaj_MSMSE2021,Goryaeva_NatComm2020}.

In summary, achieving a clear, unambiguous, and consistent interpretation of data from ion-irradiated materials, particularly in the high-dose regime, remains a significant challenge. Existing experimental interpretation models rely on kinetic equations that involve numerous parameters but fail to account for the microscopic, fluctuating stresses and strains driving defect interactions at the nanoscale, as evidenced by prior studies~\cite{anderson2017theory,dudarev2010langevin,mason2014elastic}. Therefore, leveraging recent algorithmic advancements, we attempted to answer the following key research inquiries in this paper:
\begin{enumerate}
    \item How can we construct a quantitative explanatory model for irradiation effects at high doses? Can we effectively employ the microscopic fluctuating strain model as a bridge connecting atomic-scale defects to macroscopic strains~\cite{Dudarev_NuclFus2018}? Additionally, can various defect detection methods accurately signify the accumulation or abrupt release of internal loading resulting from high-dose irradiation at the atomic level?
    
    \item In simulations, lattice strain measurements can be derived from atomic positions via diffraction pattern peak positions. However, performing this calculation for overlapping cascade MD simulations involving tens of thousands of frames is prohibitively resource-intensive. Are there alternative approaches to acquire plastic and lattice strain information, directly from real-space simulations, even in the high dose regimes?
    
    \item Under different dosage conditions, how do dislocation loops behave, and do they induce plastic deformation? If so, does the slip only occurs on certain planes? Furthermore, can we further categorize the specific type of plastic transformation involved?
\end{enumerate}

In this article we provide a quantitative interpretation of the role played by prismatic dislocation loops as initiators of plastic strain effects in heavily irradiated metals. This phenomenon leads to experimentally measurable alterations in lattice strain. First, by employing different defect characterization techniques, we introduce the ``anomalies" observed in high-dose conditions (as detailed in section~\ref{sec:WSVID}). To find the mechanisms behind these phenomena at high doses and provide answers to pertinent questions, we show how atomic and plastic strain can be identified from snapshots of molecular dynamics simulations (section~\ref{sec:atomic_plastic_strain}). Importantly, we do this without tracking the relative positions of the atoms, as irradiation mixing makes this difficult. In section~\ref{sec:atomic_deformation} we show how the accumulation of lattice defects at low dose leads to an increasing homogeneous lattice strain, and how individual defects can reorient themselves spontaneously to reduce elastic energy. Finally in section~\ref{sec:plastic_deformation} we show how plastic strain starts to emerge at high dose, and so dislocation creep and irradiation-induced swelling both arise spontaneously through the agglomeration of dislocation loops into a dislocation network. We conclude that simulated irradiation shows both lattice and plastic strain responding to the evolution of defects on molecular dynamics timescales. This demonstrates unambiguously that a high density of irradiation induced defects self-organises almost instantaneously to reduce elastic energy density by dislocation-induced plastic slip.

\section{Simulation Methods}
\subsection{Irradiation simulations}

All simulations were conducted with LAMMPS~\cite{lammps}, implemented with an adaptive timestep~\cite{adaptive_timestep1,WS,adaptive_timestep3}. The dimensions of the cell were $120 \times 120 \times120$ conventional BCC unit cells, resulting in 3.456 million atoms. Periodic boundary conditions were applied in all dimensions. A friction force was applied to atoms according to the energy loss table provided by ``Stopping and Range of Ions in Matter" (SRIM)~\cite{srim} code to include effects of electronic stopping power, when the kinetic energy of an atom exceeds 10 eV~\cite{elstop}. We used the embedded atom method (EAM) potential by Ackland and Thetford~\cite{eam} with the short-range modified by Zhong et al.~\cite{zhong1998}, AT-ZN. This potential has previously shown a good agreement with experimental results~\cite{granberg2021JNM}.

For full MD cascade simulations, a 30 keV PKA was introduced at the centre of the system, in a random direction. Most of the cell was able to evolve without any constraints, except a two nm thick layer at the borders which were thermally controlled to 300 K. No pressure control was active during this simulation. Each cascade was followed for 30 ps, followed by a relaxation period of 10 ps with temperature control at 300 K on all atoms, and stress-strain boundary conditions $\sigma_{x} = \sigma_{y} = \sigma_{z} = \varepsilon_{xy} = \varepsilon_{yz} = \varepsilon_{zx} = 0$. Thereafter, the cell was shifted before the next cascade event, in order to obtain a homogeneous irradiation. This procedure was repeated 4000 times to achieve the dose of $\sim$0.11 dpa (with a threshold displacement energy of 90 eV) according to the classical Norgett-Robinson-Torrens displacements per atom (NRT-dpa) model~\cite{NRT_N,NRT_R,NRT_T}. Following Ref.~\cite{MasonPRM_VoidDetection}, we find the canonical DPA level, cdpa, defined as the expected atomic fraction of vacancies produced. This value was obtained from the first 30 overlapping cascades to be $\mathrm{cdpa}=4.2\times 10^{-6} N_{\mathrm{casc}}$.

The CRA method~\cite{derletCRA} directly inserts FPs to quickly construct a damaged system. For this work, we randomly inserted 1000 interstitials and deleted 1000 atoms per step, and relaxed the system with the conjugate gradient method (CG), similarly to previous studies~\cite{derletCRA,MasonPRL}. This was repeated enough times to achieve doses of 0.01, 0.0143, 0.03, 0.1 and 0.2 cdpa. As the CRA method does not include either thermal annealing or cascade annealing (CA), to produce a high dose microstructure comparable with full MD simulations we subsequently performed CA with 1600 30 keV cascades (0.0078 cdpa) on the CRA simulation boxes. This combination of CRA+MD has been shown to produce simulated irradiation microstructures almost indistinguishable from MD alone~\cite{MasonPRM_VoidDetection,Granberg2022}. A long thermal annealing was also investigated, however, this method did not produce comparable results to full MD, like the CA did. More details about the results and confirmation of convergence in the combination of CRA+MD used in this work can be found in the Supplementary Information (SI).

All quantitative simulation results presented are the average of three different simulation runs carried out.

\subsection{Atomic vs plastic strain in a single crystal}
\label{sec:atomic_plastic_strain}

In this section we derive the relationship between lattice strain and plastic strain in a single crystal simulation. The \emph{total} strain, \mat{\varepsilon}, is well-defined, being the change in dimensions of the supercell. But this strain can be decomposed into different contributing factors. The \emph{elastic} strain, $\mat{e}$, is related by Hooke's law to the stress on the boundary via the elastic compliance tensor, ie $\mat{e} = \mat{C}^{-1} \mat{\sigma}$. In Mura's formulation~\cite{mura2013micromechanics}, the total strain $\mat{\varepsilon} = \mat{e} + \mat{\varepsilon}^{\star}$, where the second term $\mat{\varepsilon}^{\star}$, known as the \emph{eigenstrain}, is the \emph{stress-free} strain in the body due to defects. Here we are treating the atoms explicitly, and so can make a further distinction. A part of the strain is visible to the atoms, visible in the location of the x-ray diffraction peaks, comes from the eigenstrains arising from the irradiation induced defects. This strain affects the evolution of small irradiation-induced defects, as they have some freedom to rotate to compensate the elastic energy density. But a second part of the strain comes from plastic events, such as dislocation slip or creep, which after having occurred leave the crystal lattice essentially unaffected. These $\emph{plastic}$ strains appear in the diffraction pattern in the detailed fine-structure of each individual diffraction peak, not in its location.

Consider a periodic supercell with repeat vectors $\vec{A}_1$, $\vec{A}_2$, $\vec{A}_3$, containing a single crystal with primitive lattice vectors $\vec{b}_1$, $\vec{b}_2$, $\vec{b}_3$. We can compactly write these vectors as the matrix $\mat{A}$, whose columns are $\vec{A}_i$, and the matrix $\mat{B}$ whose columns are $\vec{b}_i$. Then for the lattice to be continuous across the supercell periodic boundary requires~\cite{MasonPRM_VoidDetection}
    \begin{equation}
        \mat{A} = \mat{B} \, \mat{N},
    \end{equation}
where $\mat{N}$ is a $3\times 3$ matrix of integers representing the number of lattice repeats. The number of lattice sites in the periodic supercell is $N_{\mathrm{latt}} = \mathrm{det}[\mat{N}] \times N_{\mathrm{motif}}$, where $N_{\mathrm{motif}}$ is the number of atoms in the motif of the primitive lattice. 

The two matrices $\mat{B}$ and $\mat{N}$ define a \emph{reference lattice} for the simulation, compatible with the simulation box size and shape.

Now consider the box of atoms after some simulated irradiation. If we are running with a general set of stress- and strain-boundary conditions, the periodic supercell now has repeat vectors $\mat{A}'$. The change in the box can be written $\mat{A}' = \mat{F} \, \mat{A}$, where the deformation is given by the product of linear atomic and plastic deformations, denoted with subscripts $_a$ and $_p$, respectively, i.e. $\mat{F} = \mat{F}_a \, \mat{F}_p$. The primitive cell can be atomically strained, and can also be rotated~\footnote{To see that the primitive cell can be rotated, we use proof by \emph{reductio ad absurdum}: The most extreme irradiation disordering could be taking out all the atoms, replacing them in amorphous positions, and annealing back to a crystal lattice. For no rotation to be true it is necessary that the exact same lattice orientation is always recovered from amorphisation + annealing.}, so the lattice vectors can also change. We assume here that no second phase emerges, and a single grain is present after irradiation. If we can compute the best fit $\mat{B}'$ to the atomic positions after irradiation, then we can write  $\mat{B}' = \mat{F}_a \, \mat{R} \, \mat{B}$, where $\mat{R}$ is a rotation matrix. We can use the method of polar decomposition to find the atomic strain
\begin{eqnarray}
\mat{F}_a \, \mat{F}_a^T &=& \left( 1 + \mat{\varepsilon}_a \right) \left( 1 + \mat{\varepsilon}_a \right)^T \nonumber\\    
&=& \left( \mat{B}' \mat{B}^{-1} \mat{R}^T \right)  \left( \mat{B}' \mat{B}^{-1} \mat{R}^T \right)^T,      \nonumber\\
\mat{\varepsilon}_a  &\approx& \frac{1}{2}\left( \left( \mat{B}' \mat{B}^{-1} \right) \left( \mat{B}' \mat{B}^{-1}  \right)^T - \mat{1} \right).
\end{eqnarray}
where to get the last line we have taken a linear approximation.
From the arguments above, the matrix of lattice repeats after irradiation must solve $\mat{A}' = \mat{B}' \mat{N}'$. As $\mat{A}'$ is read from the size of the MD simulation box, we can rearrange to give the plastic strain
    \begin{equation}
        \mat{F}_p = \mat{R} \, \mat{B} \, \mat{B}'^{-1} \mat{A}' \,  \mat{A}^{-1},
    \end{equation}  
or as a function of the lattice repeats, 
    \begin{equation}
        \label{eqn:plasticStrainDeformation}
        \mat{F}_p = \mat{R} \, \mat{B} \, \mat{N}' \, \mat{N}^{-1} \mat{B}^{-1}  .
    \end{equation}
Using polar decomposition we can find the plastic strain. As that the matrix product $\mat{N}' \, \mat{N}^{-1}$ in Eq.~\ref{eqn:plasticStrainDeformation} is not symmetric, we write $\mat{F}_p = \mat{R}' \left( 1 + \mat{\varepsilon}_p \right)$, where $\mat{\varepsilon}_p$ is the (symmetric) plastic strain tensor. Then
\begin{eqnarray}
        \mat{F}_p^T \, \mat{F}_p &=& \left( 1 + \mat{\varepsilon}_p \right)^T \left( 1 + \mat{\varepsilon}_p \right) \nonumber \\   
        &=& \left( \mat{R} \, \mat{B} \, \mat{B}'^{-1} \mat{A}' \, \mat{A}^{-1} \right)^T \left( \mat{R} \, \mat{B} \, \mat{B}'^{-1} \mat{A}' \,  \mat{A}^{-1} \right),      \nonumber\\
        \mat{\varepsilon}_p  &\approx& \frac{1}{2}\left( \left( \mat{B} \, \mat{B}'^{-1} \mat{A}' \,  \mat{A}^{-1} \right)^T \left( \mat{B} \, \mat{B}'^{-1} \mat{A}' \, \mat{A}^{-1} \right) - \mat{1} \right), \nonumber \\
        \label{eqn:plasticStrainTensor}        
    \end{eqnarray}
where the last line is a linear approximation. Alternatively we can express Eq.~\ref{eqn:plasticStrainTensor} in terms of the number of lattice repeats. 
    \begin{equation}
        \label{eqn:plasticStrainTensor2}
        \mat{\varepsilon}_p  \approx \frac{1}{2}  \left( \left( \mat{B} \, \mat{N}' \, \mat{N}^{-1} \mat{B}^{-1} \right) + \left( \mat{B} \, \mat{N}' \, \mat{N}^{-1}  \mat{B}^{-1} \right)^T \right) - \mat{1}         
    \end{equation}
We conclude that the deformation corresponding to plastic strain is directly related to the change in the reference lattice for the simulation, and can be simply computed from the change in number of lattice repeats. This implies that both transformations which shear the reference lattice, but preserve the number of lattice sites $(\mathrm{det}[\mat{N}']=\mathrm{det}[\mat{N}])$ \emph{and} transformations which change the number of lattice sites $(\mathrm{det}[\mat{N}']\neq\mathrm{det}[\mat{N}])$ can be represented within the same plastic deformation $\mat{F}_p$.

Hence in this work, the formalism for the plastic deformation gradient $\mat{F}_p$ contains both radiation-induced slip and swelling.

Now consider a general plastic transformation defined by a Burgers vector change $\vec{b}$ each time we move along the line $\vec{n}$, with no associated homogeneous lattice strain. The magnitude of the Burgers vector is $b=|\vec{b}|$. We can define the degree of the transformation with a single number $\gamma = b/|\vec{n}|$. $\gamma$ is simply the angle of shear if $\vec{b}$ is in the plane with normal $\vec{n}$. Then the periodic repeat vector $\vec{A}_1$ changes to $\vec{A}_1 + (\vec{A}_1\cdot \Hat{n}/|\vec{n}| ) \vec{b}$, where the caret $\Hat{}$, denotes a normal vector, and so the supercell changes to 
    \begin{equation}
        \mat{A}' = \mat{A} + \frac{\gamma}{b} \left( \begin{array}{ccc}
                                    (\vec{A}_1 \cdot \Hat{n}) \, \vec{b} & (\vec{A}_2 \cdot \Hat{n}) \, \vec{b} & (\vec{A}_3 \cdot \Hat{n}) \, \vec{b}               \\
                                    |   &   |   &   |   \\
                                    |   &   |   &   |   
                            \end{array} \right).
    \end{equation}
It is convenient at this point to consider as an illustrative example a cubic simulation cell, of side $m a_0$, i.e. $\mat{A} = m \, a_0 \mat{1}$. Then the change in such a supercell due to a slip event is with Burgers vector $\vec{b}$ on plane $\Hat{n}$ is given by the outer product
    \begin{equation}
        \mat{A}' = \mat{A} + \frac{m \gamma}{b}  \vec{b} \otimes \Hat{n},
    \end{equation}
and the linearised plastic strain is recognised as the Schmidt tensor,
    \begin{equation}
        \mat{\varepsilon}_p =  \frac{\gamma}{2 b} \left(  \vec{b} \otimes \Hat{n} + \Hat{n} \otimes \vec{b} \right).
    \end{equation}
The new number of lattice repeats, $\mat{N}' = \mat{B}^{-1} \mat{A}' = \mat{N} + ( \gamma / b ) \, \mat{N} \, (  \vec{b} \otimes \Hat{n} )$.

For BCC metals, $N_{\mathrm{motif}}=1$, so
    \begin{eqnarray}
        \mat{B}^{\mathrm{(BCC)}} &=& \frac{a_0}{2} \left( \begin{array}{ccc}
                        -1  &   1   &   1       \\
                        1   &   -1   &   1       \\
                        1   &   1   &   -1       \\
                    \end{array} \right) , \nonumber\\
        \mat{N}^{\mathrm{(BCC)}} &=& \left( \begin{array}{ccc}
                        0   &   m   &   m       \\
                        m   &   0   &   m       \\
                        m   &   m   &   0       \\
                    \end{array} \right),
    \end{eqnarray}
and the new number of lattice repeats in the BCC crystal due to a plastic event is
\begin{equation} 
        {\mat{N}^{\mathrm{(BCC)}}}' = \mat{N}^{\mathrm{(BCC)}}  \\ + \frac{m\gamma}{b} \left( \begin{array}{ccc}
                                     (b_2+b_3) n_1      &  (b_2+b_3) n_2        &   (b_2+b_3) n_3       \\
                                     (b_3+b_1) n_1      &  (b_3+b_1) n_2        &   (b_3+b_1) n_3      \\
                                     (b_1+b_2) n_1      &  (b_1+b_2) n_2        &   (b_1+b_2) n_3
                            \end{array} \right),
    \end{equation}
while a similar process using an FCC primitive cell gives

        \begin{equation}
        {\mat{N}^{\mathrm{(FCC)}}}' = \mat{N}^{\mathrm{(FCC)}} + \frac{m\gamma}{b} \left(\begin{array}{ccc}
                                     (-b_1+b_2+b_3) n_1     &  (-b_1+b_2+b_3) n_2       &   (-b_1+b_2+b_3) n_3       \\
                                     (b_1-b_2+b_3) n_1      &  (b_1-b_2+b_3) n_2        &   (b_1-b_2+b_3) n_3      \\
                                     (b_1+b_2-b_3) n_1      &  (b_1+b_2-b_3) n_2        &   (b_1+b_2-b_3) n_3
                            \end{array} \right)
    \end{equation}

Note that, if we impose the condition that the reference lattice is a single crystal, then the number of lattice repeats must be integer after a plastic event, this constrains permissible values of $\gamma$ within the simulation box. In particular, we see the smallest $\gamma$ must scale as $\sim 1/m$, so small plastic deformations are not permitted in a small simulation box.

\begin{figure}[t]
\begin{center} 
\includegraphics[width=.55\columnwidth]{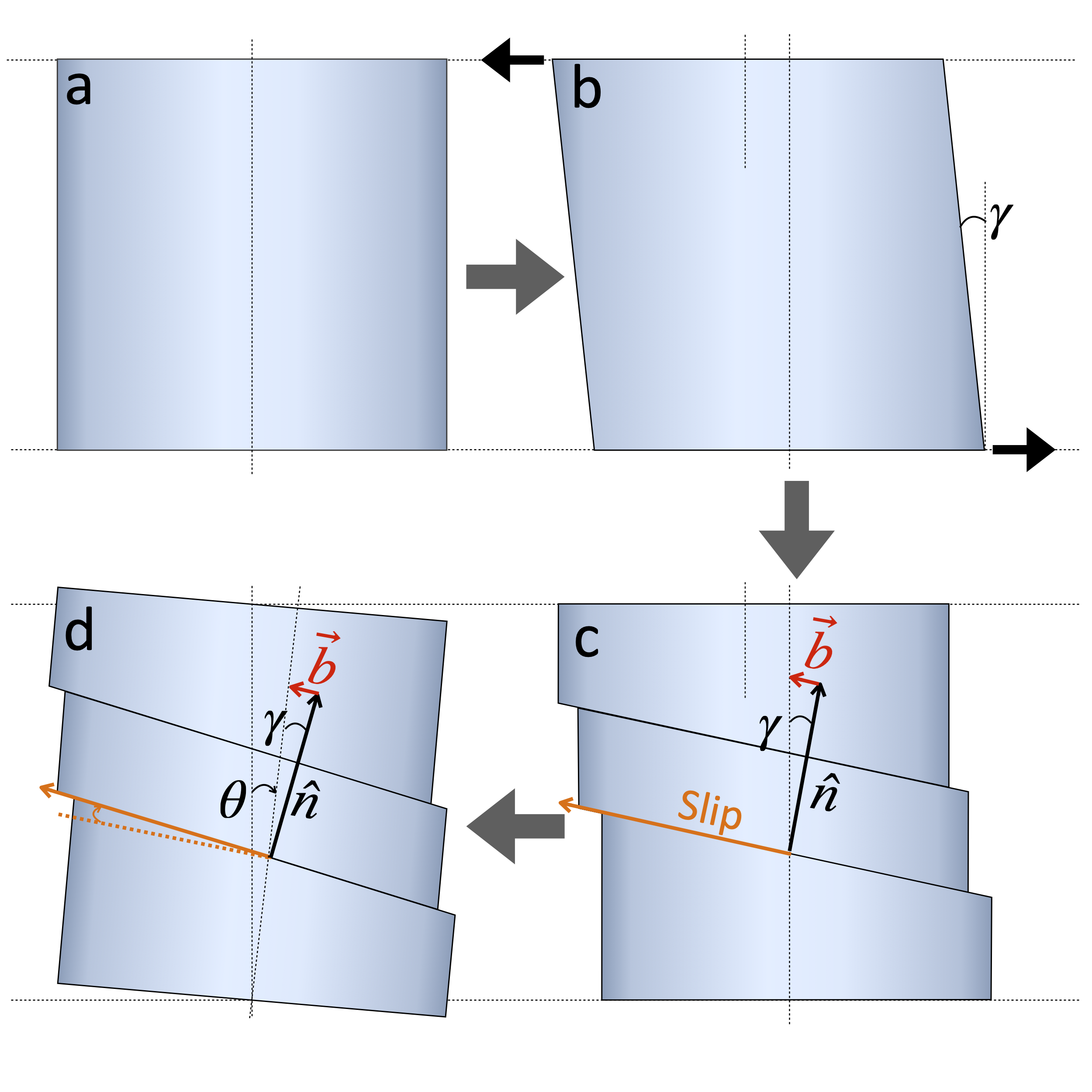}
\end{center}
\caption{Schematic representation depicting single crystal slip and corresponding rigid rotation. (a) Prior to irradiation, (b) during irradiation, as deformation initiates, (c) a pure slip event, and (d) spontaneous atomic system rotation, within the constraint of a simulation box disallowing shear.}
\label{SlipProcess}
\end{figure}

For the BCC/FCC cases, we find the change in lattice sites, $\Delta N_{\mathrm{latt}} = \mathrm{Det}[ \mat{N}' ] - \mathrm{Det}[ \mat{N} ]$, is given by 

    \begin{eqnarray}        
        \Delta N_{\mathrm{latt}}^{\mathrm{(BCC)}} &=& 2 \frac{m^3 \gamma}{b} \vec{b} \cdot \Hat{n},     \nonumber\\
        \Delta N_{\mathrm{latt}}^{\mathrm{(FCC)}} &=& 4 \frac{m^3 \gamma}{b} \vec{b} \cdot \Hat{n},
    \label{DeltaN}
    \end{eqnarray}
from which we confirm that pure slip events $(\vec{b} \cdot \Hat{n} = 0)$ are associated with no change in lattice sites, but other plastic events change the number of lattice sites and so cause swelling. In Fig.~\ref{SlipProcess}, we illustrate the relationship between the slip direction normal and the Burgers vector. Establishing a connection between atomistic observations and continuum slip events offers the potential to develop a comprehensive interpretative model for high-dose irradiation scenarios. In section~\ref{sec:plastic_deformation}, we outline both the similarities and distinctions between atomic-level slip phenomena and macroscopic models of plastic deformation.

\subsection{Computing the elasto-plastic strain decomposition from atomic positions}

The reference primitive lattice vectors making the matrix $\mat{B}$ can be found directly from the position of peaks in the diffraction pattern. But this is an expensive calculation, so below we describe an order-N approximation for $\mat{B}$, which we find works well in the high-dose irradiation cases described here.

Given a reference lattice $\mat{B}$, the reference atom positions are    
    \begin{equation}
        \vec{y} = \mat{B}[u,v,w]^T + \vec{m}_t,
    \end{equation}
where $[u,v,w]$ is a triplet of integers and $\vec{m}_t$ is the position of the $t^{th}$ motif point in the reference unit cell. Given the observed atom positions, $\{ \vec{x}_j \}$ , $j \in \{1,\ldots N\}$, the Wigner-Seitz occupation, $o_i$, is the number of atoms closest to of reference atom position $i$, i.e. the number of atoms for which $\left|\vec{x}_j - \vec{y}_i\right| < \left|\vec{x}_j - \vec{y}_k\right| \, \forall \, k \neq i$. The index $i \in \{ 1,\ldots N_{\mathrm{latt}}\}$, as defined above. Note that the sum of the occupation, $\sum_i o_i = N_{\mathrm{atoms}}$, but that $N_{\mathrm{atoms}}$ is not necessarily equal to $N_{\mathrm{latt}}$. The number of point defects is defined by $N_{\mathrm{PD}} = 1/2\, \sum_i |o_i - 1|$.

We can make the \emph{ansatz} that there exists a set of reference atom positions $\{ \vec{y}_i \}$ for which $N_{\mathrm{PD}}$ is minimised. For a single crystal, we can in principle find this reference by minimising $N_{\mathrm{PD}}$ with respect to a vector offset common to the motif points $\{ \vec{m}_t \}$ and the elements of $\mat{N}$.

We can find a new reference lattice $(\mat{B}',\mat{N}')$ from atom positions as follows. We compute the fit function on atom $i$:
    \begin{equation}
        S_i = \mat{T}_i \sum_j \left( \vec{x}_j - \vec{x}_i - \delta \vec{y}_k \right) + \vec{\delta}_i,
    \end{equation}
where $\delta \vec{y} = \mat{B}[u',v',w']^T + \vec{m}_t - \vec{m}_1$ is an expected separation between reference atom sites. This we minimise with respect to the nine matrix elements of $\mat{T}_i$, displacement vector $\vec{\delta}_i$, and matching of neighbours to $\{\delta \vec{y}_k\}$. We take the sum over neighbours with cutoff between 3rd and 4th nearest neighbours. This method is slower than polyhedral template matching~\cite{Larsen_MSMSE2016}, but robust when point defects are present. The matrix $\mat{T}_i = \mat{R}_i ( 1 + \mat{\varepsilon}_i )$ is the deformation gradient, a product of atomic strain and rotation at atom $i$. The rotation matrix and strain can be separated using polar decomposition. To compute an approximate homogeneous lattice strain and orientation, we need an appropriate average of the spatially varying deformation gradient.  To do this, we compute an arithmetic mean atomic deformation gradient $\langle \mat{T} \rangle = 1/N \sum_i \mat{T}_i$, and mean strain $\langle \mat{\varepsilon} \rangle = 1/N \sum_i \mat{\varepsilon}_i$ separately. Then we compute an appropriate rotation matrix, $\mat{R}$, from the arithmetic mean using polar decomposition. Finally we compute the homogeneous atomic deformation gradient compatible with the unitary rotation matrix and mean strain, as $\mat{T} = \mat{R} ( 1 + \langle \mat{\varepsilon} \rangle )$.

The method above finds new reference lattice vectors $\mat{B}' = \mat{T}\,\mat{B}$ in $\mathcal{O}(N)$ time, and from this we find $\mat{N}' = \mat{A}' \mat{B}'^{-1}$ to complete the reference lattice. We then search for an optimal vector offset in $\{ \vec{m}_t \}$. The Wigner-Seitz occupations $\{o_i\}$ reported here therefore minimise $N_{\mathrm{PD}}$ subject to a $\mathcal{O}(N)$ estimation of the atomic deformation gradient.

\subsection{Further analysis methods}

The ``Open Visualization Tool" (OVITO)~\cite{ovito} was used for visualization and the dislocation extraction algorithm (DXA)~\cite{DXA} implemented was used for dislocation identification. For the results obtained with our W-S analysis, a cluster size distribution analysis was performed. In cases where a cluster contains both vacancies and interstitials, we determined the net content of defects to accurately assess the defect size. It is important to highlight that such mixed clusters were infrequent in our study because of our process of finding the reference lattice described above, and the disparity in the cluster's overall content before and after calculating the net defects was approximately 1\%, even at the higher doses. The cutoff value for vacancies and interstitials are intermediate value between the nearest neighbor (1NN) and 2NN, and between the 3NN and 4NN, respectively~\cite{Vac_cutoff,Inter_cutoff}. The shorter cutoff for vacancies compared to previous studies was chosen to give an estimate for possible deuterium retention, as for vacancies in clusters to collectively add to the empty volume they need to be in the nearest position to another vacancy.

\section{Results and Discussion}
\subsection{Defect evolution during irradiation}
\label{sec:WSVID}

\begin{figure}[t]
\begin{center} 
\includegraphics[width=.8\columnwidth]{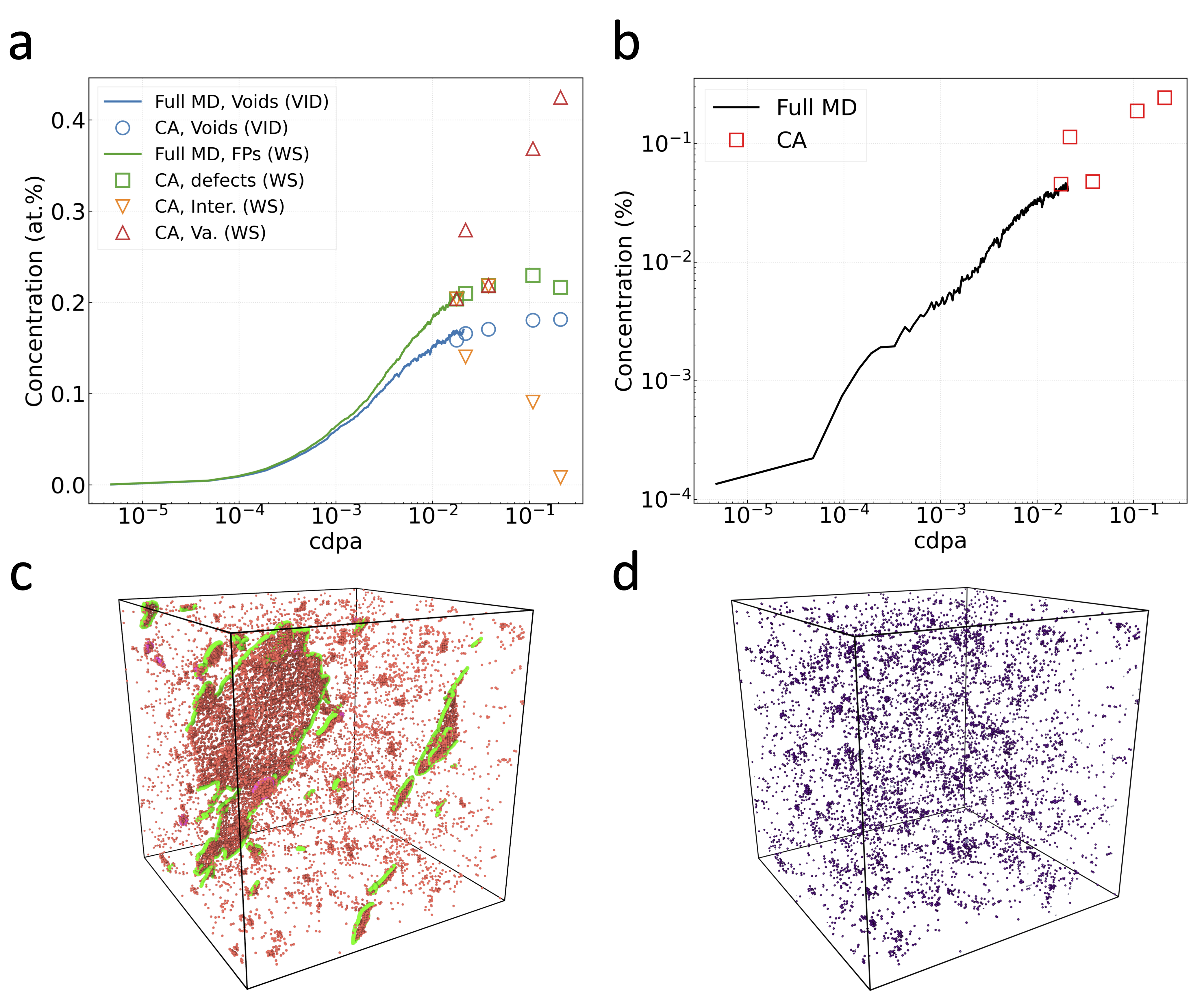}
\end{center}
\caption{(a) Point defect evolution as a function of dose characterized by different methods. (b) The concentration of vacancies in loops calculated as the difference between total unoccupied sites by the W-S method and vacancies using the VID method. Solid lines: full MD simulations. Red squares: after 1600 CA steps. (c) Vacancy distribution obtained by W-S and (d) the void distribution obtained by VID after CA, cdpa = 0.2047.}
\label{pointDef}
\end{figure}

In this section, we highlight the core question of this paper by contrasting the similarities and differences in the accumulation of radiation-induced damage through various characterization methods. Fig.~\ref{pointDef}(a) shows the evolution of point defects as a function of dose. The atomic fraction of vacancies, computed from void isosurfaces using VID, is similar to the atomic fraction of unoccupied lattice sites ($o_i = 0$) computed with Wigner-Seitz below $10^{-3}$ cdpa, but the levels diverge above this dose. Fig.~\ref{pointDef}(b) shows the difference between these two methods. Unoccupied lattice sites which do not correspond to an open volume are associated most commonly with vacancy loops. This can be seen by comparing Fig.~\ref{pointDef}(c) and Fig.~\ref{pointDef}(d). The former, computed with W-S, clearly shows vacancy loops bounded by dislocation lines in addition to monovacancies and vacancy clusters, whereas the latter, computed with VID, shows only monovacancies and vacancy clusters. Multiply occupied Wigner-Seitz lattice sites ($o_i > 1$) are associated with interstitials, either indivdually as crowdions or in interstitial clusters and dislocation loops. We see in Fig.~\ref{pointDef}(a) that the count of Wigner-Seitz unoccupied and multiply occupied lattice sites diverges above $10^{-2}$ cdpa. Previous studies have shown that similar apparent artefacts by W-S might appear if the perfect lattice structure of a whole layer is identified as interstitials~\cite{Wu_JMCC}. In Fig.~\ref{pointDef}(c), the alignment of vacancies by W-S and dislocations by DXA confirms that the significant increase in vacancy concentration observed in our work is not an artefact. Hence, we pose the following question: What factors contribute to the emergence of these vacancy loops and the divergence in the evolution of vacancies and interstitials? This question will be answered in detail in section~\ref{sec:plastic_deformation}. Moreover, a full description of defect evolution and clustering under our simulation conditions and a convergence study for the CA can be found in the SI (see Table 1 and Figs.~S1--S8).

\subsection{Microscopic deformation processes contributing to atomic and plastic strain}

In this section, we will illustrate swelling and dislocation creep phenomena and their associated atomic and plastic strain responses that take place during the irradiation process. 

\begin{figure}[t]
\begin{center} 
\includegraphics[width=.6\columnwidth]{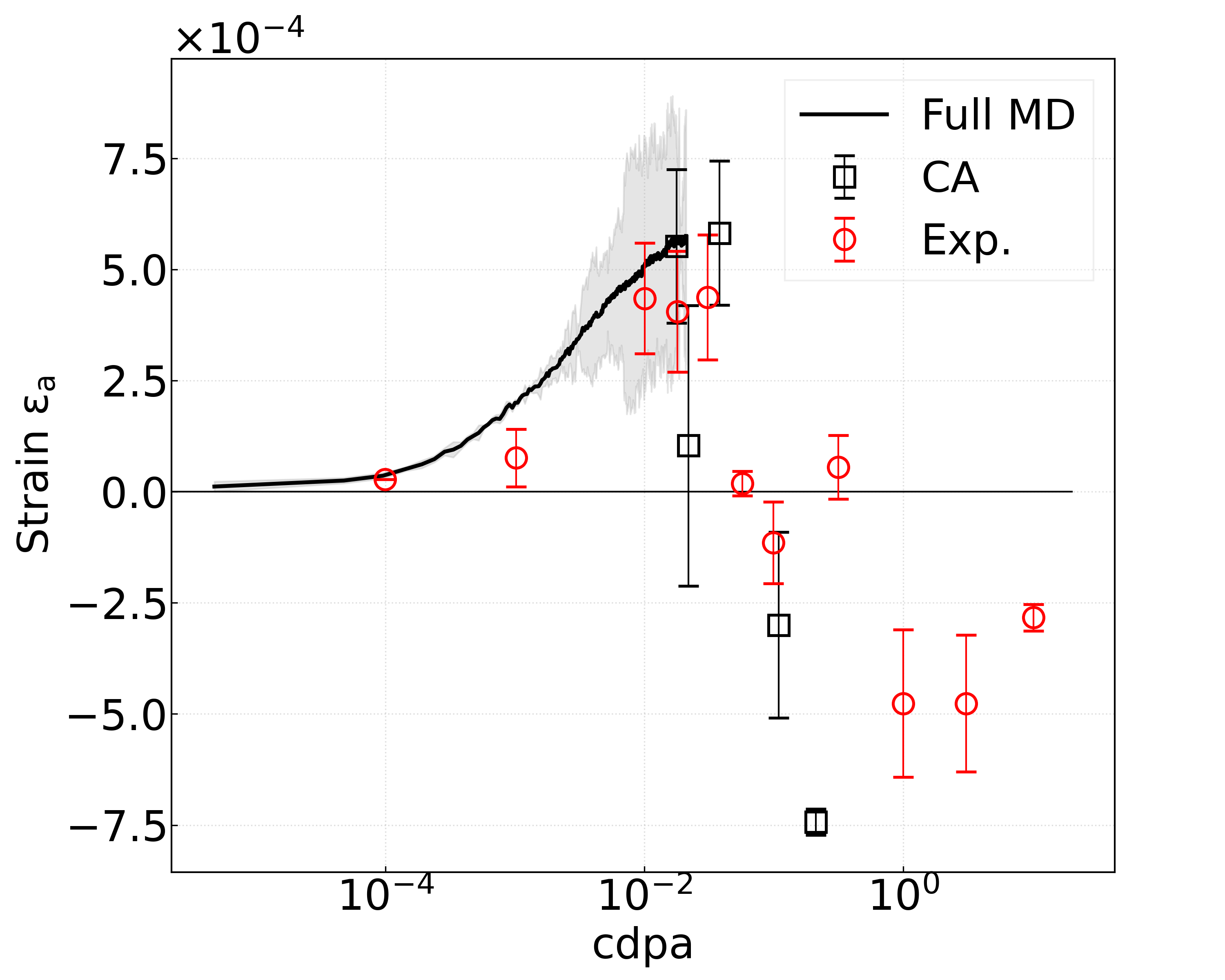}
\end{center}
\caption{Lattice strain derived from simulations. The gray shaded area represents the error fluctuation range of the full MD section, showing the range of three simulations. The error bars for the CRA+CA points is also associated with the three different runs. The experimental strain data is from Ref.~\cite{MasonPRL}.}
\label{Overallatomic}
\end{figure}

First we consider the average atomic strain, computed as $1/3 \, \mathrm{Tr}[ \varepsilon_a ]$. The evolution as a function of dose is shown in Fig.~\ref{Overallatomic}. These results are qualitatively similar to the experimental and simulation results reported in Ref.~\cite{MasonPRL}. The atomic lattice strain is tensile at low dose, during the accumulation of interstitial clusters and vacancies. Interstitials have a large positive relaxation volume, while the vacancies have a small negative relaxation volume, so if the numbers of each balance, the strain is positive. At high dose, the interstitials have coalesced into a network, leaving behind a vacancy-dominated microstructure with compressive atomic strain. Here we show an improved quantitative match to the experimental strain compared to Ref.~\cite{MasonPRL}. This improvement is largely because we use MD cascade annealing here, whereas the previous work used CRA only. We should, however, note that the experimental boundary conditions were different - the experimental study considered self-ion irradiation into a slab, and so expansion was only possible in the $z$-direction. The micro-Laue x-ray measurement detected $\varepsilon_a$ in the $z$-direction. As noted above, in this work the simulation cell was allowed to expand in the $x$-, $y$-, and $z$-directions. The change of sign in the strain at a dose of 0.01 cdpa corresponds to the point in Fig.~\ref{pointDef}(a) where the number of lattice sites is changing, and the count of unoccupied and multiply-occupied W-S sites diverges. In Ref.~\cite{MasonPRL} this was attributed to interstitial plane formation and here we consider the change in atomic and plastic deformation more generally.

The full evolution for both atomic and plastic deformation, rotation and the other measurables are found in the SI, for all our investigated samples. In the following subsections we are going through the main mechanisms and their implication for selected examples.

\subsubsection{Atomic deformation}
\label{sec:atomic_deformation}

\begin{figure}[t]
\begin{center} 
\includegraphics[width=.6\columnwidth]{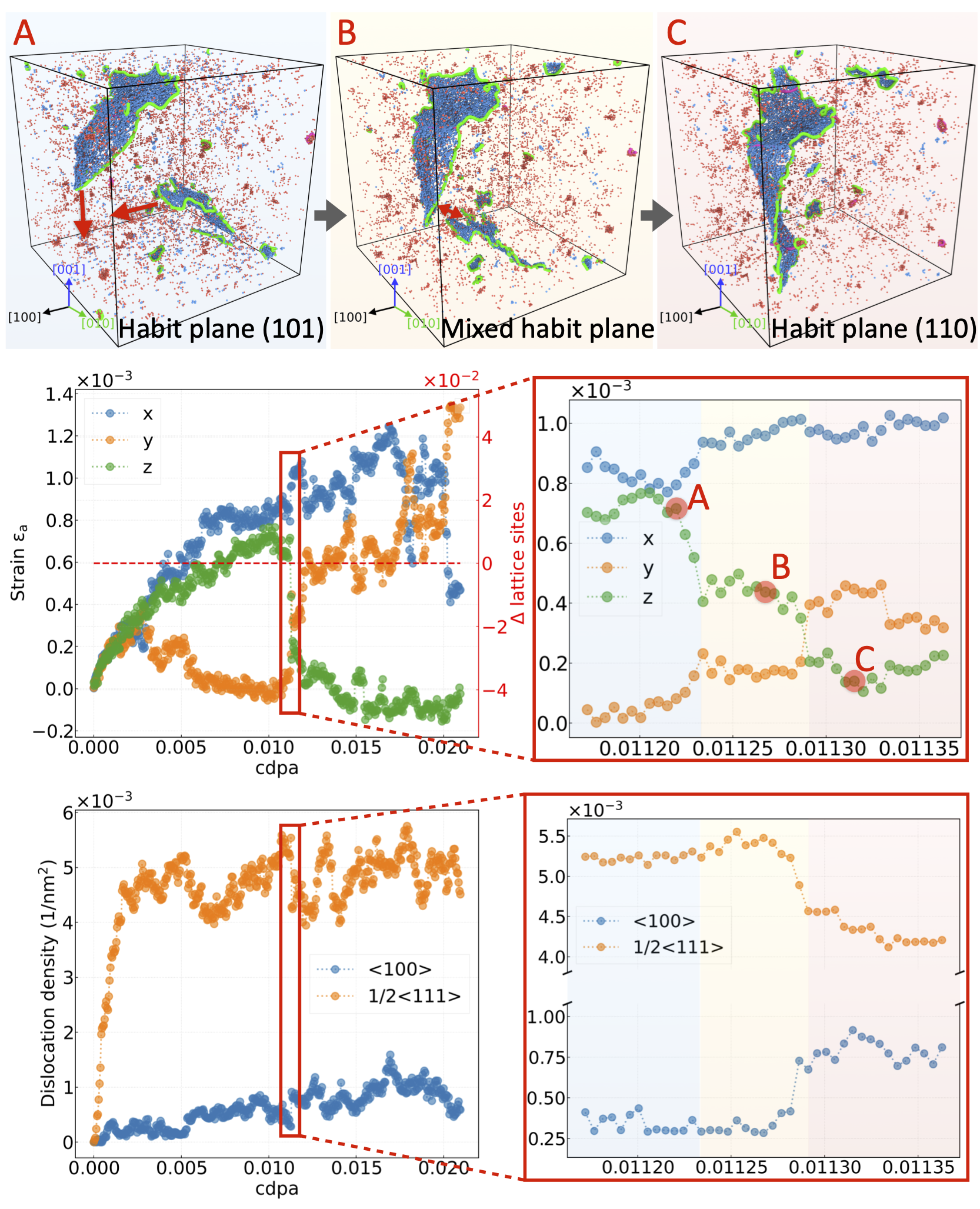}
\end{center}
\caption{Bottom row: the overall dislocation density evolution, with a zoomed-in view of the period during which the strain drop occurs. Middle row displays the evolution of atomic strain throughout the full MD simulation, with a closer look at the strain drop during 20 rounds of irradiation. The lattice sites have no change during the whole process, which indicates no plastic transformation happens. To distinguish between three different states of the evolution, a semi-transparent blue, yellow, and red background is used in the zoomed-in figures. Upper row: to illustrate the strain transition process, three representative frames are selected (\textbf{A}, \textbf{B} and \textbf{C}), namely before the release of $\epsilon_a(z)$, during the release, and after the release. The figures show the migration of interstitial loops at these three different stages. Red arrows indicate the direction of loop movements, while red and blue atoms represent vacancies and interstitials, respectively. Dislocation lines with 1/2\hkl<111> and \hkl<100> Burgers vectors are shown in green and pink lines, respectively.}
\label{fullMD_epse}
\end{figure}

We first consider the atomic strain in one of the simulations at a dose around 0.01 cdpa. At this point, the interstitial loops have grown large, but a network has not formed. Fig.~\ref{fullMD_epse} shows the atomic strain $\varepsilon_a$ in the $x$-, $y$-, and $z$-directions as a function of dose. In a small simulation cell, it is quite expected that the components of the strain are different, as they are determined by relatively few dislocation loops. In Fig.~\ref{fullMD_epse} we see the $x$- and $z$-components are large below 0.01 cdpa, with the $y$-component smaller. This is associated with the growth of a 1/2\hkl<111> interstitial loop with habit plane \hkl(101), which can be seen in the top panel of Fig.~\ref{fullMD_epse}, i.e. snapshot \textbf{A}. Also visible in the snapshot is a smaller loop with habit plane \hkl(-101). Over the course of 20 cascades (0.0001 cdpa), these two loops join together to produce a mixed-habit-plane object seen in snapshot \textbf{B}, and then coalesce to form a single large loop with habit plane \hkl(110) in snapshot \textbf{C}. The dislocation line length is shown in the bottom panel of Fig.~\ref{fullMD_epse}. There is no significant change in the fraction of W-S or VID defects during this event (as can be seen from Figs.~S2 and~S3 in SI).

This habit plane rotation from \hkl(101) to \hkl(110) has a dramatic effect on the \emph{atomic} strain components. In Fig.~\ref{fullMD_epse} middle panel we see the $y$-component rapidly increase while the $z$-component decreases. However, this event is not linked to a change in \emph{plastic} strain, as per the definition provided in this paper. This lack of association stems from the unaltered matrix of lattice repeats $\mat{N}$, as evidenced by the horizontal dot line, coupled with the fact that all plastic components remain at a constant value of 0. In the SI we provide several other examples of loop coalescence and habit plane rotation leading to significant and rapid changes in elastic stress, while not changing the plastic strain at all (see Figs.~S9--S15 and S17--S19).

These rapid fluctuations in local stress lead to rapid local changes in the microstructure. It is unclear from the current simulations whether such rapid local changes in the atomic stress will be so pronounced at greater length scales comparable to grain sizes. These results therefore show the ease with which the elastic dipole tensor of individual defects can be changed in response to stress~\cite{Dudarev_Acta2017}. At low dose, the habit planes of defects can shift, leading to a reorientation of their contribution to swelling~\cite{Dudarev_NuclFus2018}.

\begin{figure}[t]
\begin{center} 
\includegraphics[width=.8\columnwidth]{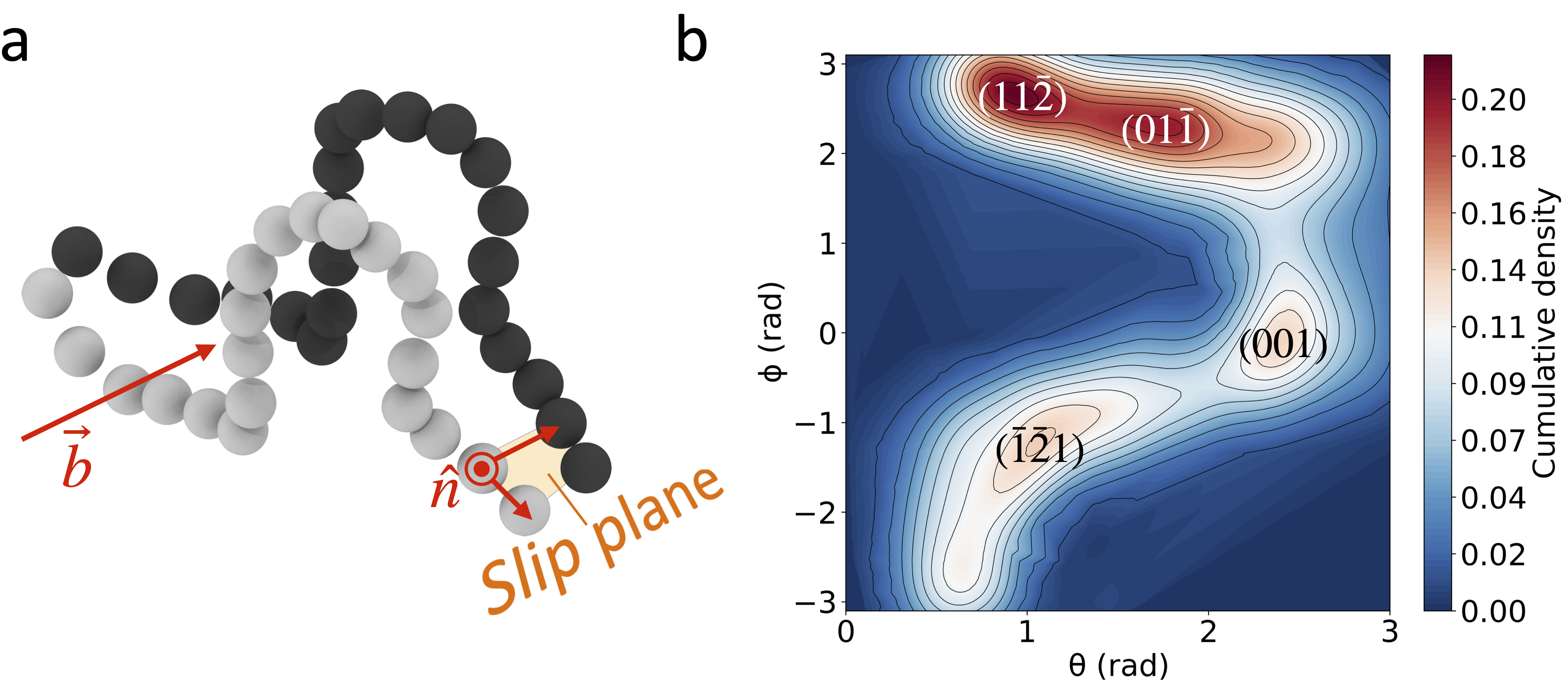}
\end{center}
\caption{Slip behavior of dislocations during the full MD case. (a) Schematic illustration of slip tracing extraction. The gray and black spheres symbolize dislocation segments before and after a plastic slip event (with the interval denoting a single cascade) extracted by the DXA method. The unit slip normal is shown as $\Hat{n}$. (b) The cumulative distribution of $\Hat{n}$ extracted over the course of 20 cascades from state \textbf{A} to \textbf{C} in the full MD case.}
\label{fullMD_slipnorm}
\end{figure}

The key role of Eq.~\eqref{DeltaN} lies in correlating atomistic observations with continuum slip events, simplifying the comprehension that rotation of the habit plane or a localized slip event does not impact the global cell, maintaining a zero plastic response. Nonetheless, local slips occur in the simulations, for instance, during the glide of dislocation lines/loops, as depicted in Fig.~\ref{fullMD_epse}. We formulated a straightforward slip tracing model to gain deeper insights into the dislocation movement, emphasizing that this method is an estimation, without meticulously converting each dislocation line network into a continuous field that accurately signifies the network's geometry and Burgers vector alterations~\cite{bertin2022sweep}. Notably, the minor slip amplitude from cascade events allows for easy tracking of dislocation line movements without employing the Sweep-tracing algorithm (STA), detailed in another study~\cite{bertin2022sweep}. Fig.~\ref{fullMD_slipnorm}(a) schematically illustrates the slip tracing model. The core concept of our model involves identifying the relevant slip facet by extracting dislocation lines exhibiting glide. This is achieved by pairing segments one-to-one before and after the slip event. The cross product of the slip vector and the unit tangent vector of the dislocation line symbolizes the $\Hat{n}_i$ for each slip facet. Fig.~\ref{fullMD_slipnorm}(b) displays the integral statistics of slip facet normals calculated on each pair of dislocation loops extracted from 20 full MD cascades, when the habit plane rotates from state \textbf{A} to \textbf{C} in Fig.~\ref{fullMD_epse}. The distribution of slip normals are plotted in projection as a function of the azimutal ($\theta$) and elevation ($\phi$) angle, by color mapping the density of the slip facet distributions. The high density slip normals are identified as \hkl(11-2) and \hkl(01-1). In BCC metals, the planes with the largest interplanar spacing are the \hkl{110} planes followed by the \hkl{112} and the \hkl{123} planes~\cite{BCCslip}. These slip planes often translates to lower energy barriers for slip compared to other planes, especially under irradiation which can enhance point defect mobility and dislocation line interaction. Other peaks are also observed (\hkl(001)), which can be the contribution from dislocation loops growing and rotating.

\subsubsection{Elasto-plastic deformation}
\label{sec:plastic_deformation}

To quantify the plastic strain, we proceed to examine the high-dose cases through CRA+CA simulations, which result in highly deformed microstructures, featuring abundant vacancies, offering the potential for plastic transformation.

\begin{figure*}[t]
\begin{center}
\includegraphics[width=0.8\columnwidth]{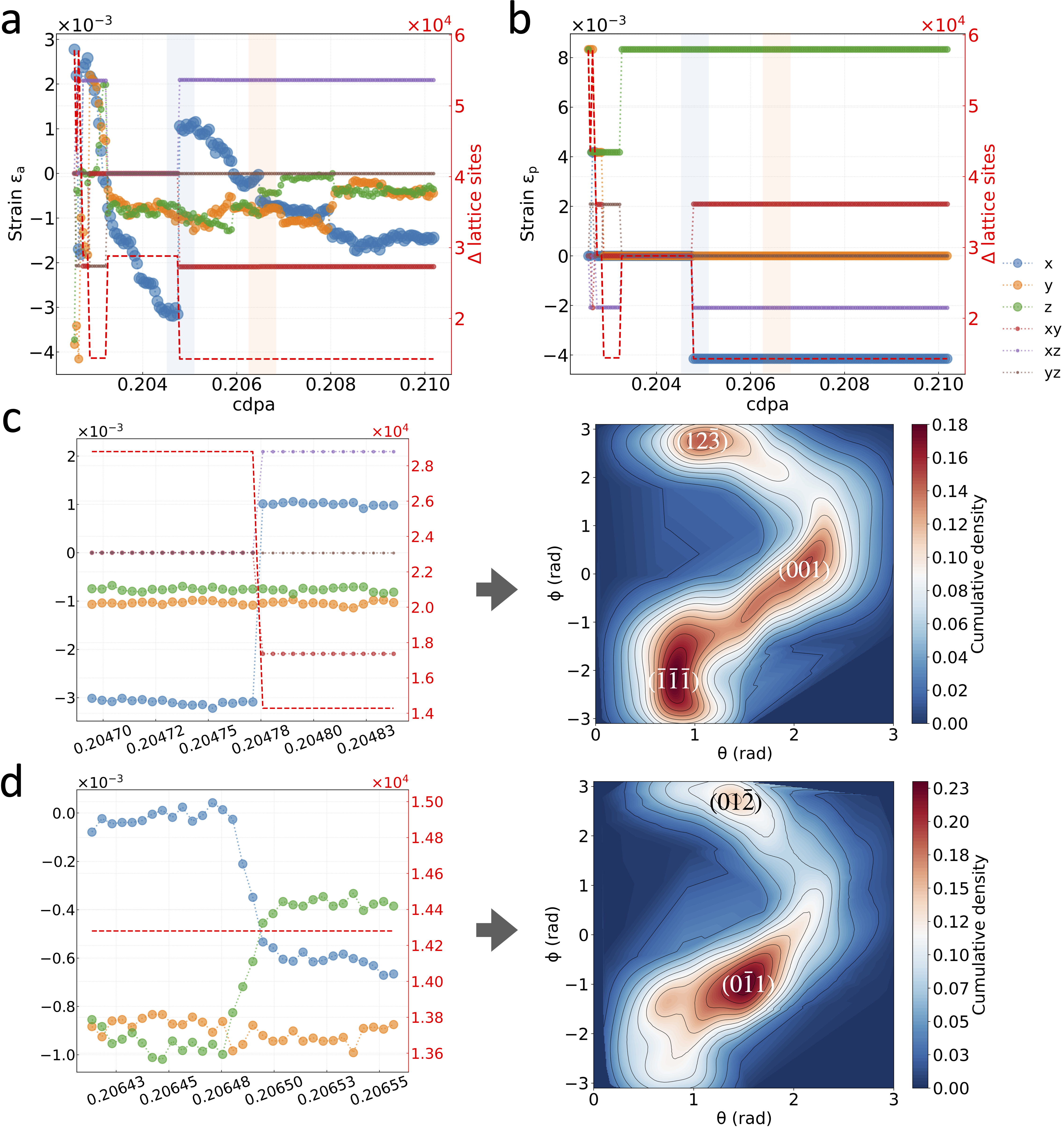}
\end{center}
\caption{(a) The elastic and (b) plastic responses as a function of dose at the highest starting dose of 0.2 cdpa during the 1st CA. The amount of change in lattice sites (according to the right axis) at different doses is indicated by the red dashed line. The two representative change of atomic strain, i.e. accompanying with global plastic deformation or not, are highlighted by the semi-transparent blue and orange backgrounds, respectively. Closer looks at local atomic strain and associated cumulative slip behavior during (c) global plastic deformation (highlighted in blue) and (d) pure elastic response (highlighted in orange).}
\label{CA700_1ep}
\end{figure*}

To begin with, the strain response at high doses is investigated. As an example, we consider the 1st CA case with the highest starting dose of 0.2 cdpa. As shown in Fig.~\ref{CA700_1ep}(a), the evolution of atomic strain $\varepsilon_a$ with both normal and shear components is observed. Initially, due to unstable isolated defects introduced by the CRA, the cell exhibits an ill-defined structure, leading to intense fluctuations in different indicators when the dose is below 0.203 cdpa. Subsequently, the sudden increases or decreases in excess lattice sites are seen, accompanied by abrupt changes in atomic strain in specific directions. These corresponding changes in shear components signify the occurrence of plastic transformation events, as shown in Fig.~\ref{CA700_1ep}(b). The lattice strain changes instantly when the excess lattice sites form or disappear, indicating that the plastic transformation happens immediately after a single cascade event. We select two representative regions for a detailed look, denoted as the blue and orange highlighted regions. These particular sections are chosen due to abrupt observable alterations in atomic strain. However, they exhibit different behavior: the region highlighted in blue undergoes a more significant change, coinciding with a global plastic slip event (Fig.~\ref{CA700_1ep}(b)), which is not observed in the region highlighted in orange.

Fig.\ref{CA700_1ep}(c) and (d) show the atomic responses and cumulative slip normal distributions for the two distinct cases, respectively. In Fig.\ref{CA700_1ep}(c), the instant change in atomic strain components is apparent, immediately following a single cascade event, accompanying the onset of global plastic deformation. An anomalous slip occurs in this instance (in addition to the conventional \hkl{123}), as depicted by the cumulative distribution of slip normals in Fig.\ref{CA700_1ep}(c). Throughout this transformation, the non-conventional slip systems are activated: the \hkl{111} slip normal prevails upon the occurrence of global plastic deformation. The growth of dislocation networks provide additional slip channels or influence the mobility and interaction of dislocations with \hkl{001} and \hkl{111} planes. Contrarily, Fig.~\ref{CA700_1ep}(d) shows a \hkl{110}-dominated slip duration when there is no global plastic response but a change of atomic strain. Subsequent examinations unveiled an additional instance of habit plane rotation. However, the difference here compared to the low-dose case shown in Fig.~\ref{fullMD_epse} is that the rotated dislocation loop is vacancy-type. Consequently, owing to the relaxation volume of a vacancy loop is negative~\cite{neg_vac_vol_3}, vacancies distributed in a material produce negative lattice strain, which can be readily observed using x-ray diffraction~\cite{neg_vac_vol_1,neg_vac_vol_2,neg_vac_vol_3,MasonPRL} in experiments. Therefore, an evident opposite change occurs in the \emph{atomic} strain components compared to the interstitial loop case shown in Fig.~\ref{fullMD_epse}.

Analyzing the local slip event, it can be found that, even though the habit plane rotation provides some additional slip channels, the slip planes should still be dominated by the ones with largest interplanar spacings. When global plastic deformation happens, this constraint no longer holds true. For the region highlighted in blue, this process is associated with a change in the number of lattice points, causing a plastic transformation characterized by swelling events. Consequently, the count of interstitials and vacancies diverges. This observed behavior is related to the Burgers vector $\vec{b}$ occurring across plane $n$, and it satisfies the relationship $\vec{b} \cdot \Hat{n} \neq 0$, as indicated by Eq.~\eqref{DeltaN}. It is important to highlight that this type of plastic event predominates in our high-dose simulations, which might be a consequence of the orthorhombic constraint imposed during the supercell relaxation. This observation suggests that swelling tends to occur spontaneously when a plastic transformation is triggered. The prevalence of this phenomenon underscores the strong correlation between plasticity and the occurrence of swelling in our simulations at high doses. The other cases experiencing plastic deformation are shown in SI (see Figs.~S16 and S20--S26).

\begin{figure}[t]
\begin{center}
\includegraphics[width=.8\columnwidth]{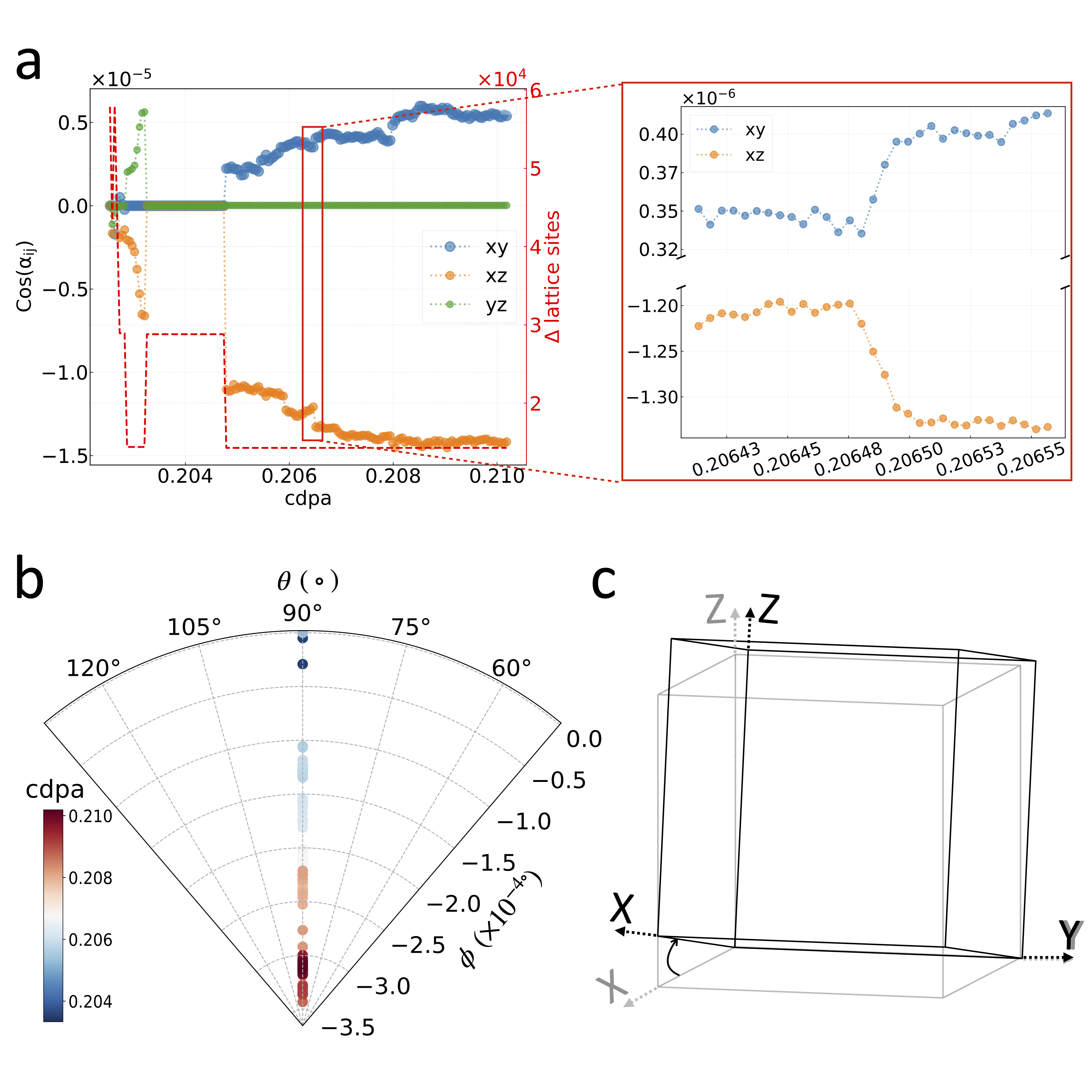}
\end{center}
\caption{(a) Rigid rotation and amount of change in lattice sites (right axis legend, red dashed line) at different doses for the highest starting dose (0.2 cdpa) during the 1st CA, $cos(\alpha_{ij})$ indicates the angle between the shear component and the normal axes, where $i,j={x,y,z}$. (b) Corresponding trajectories of rotation for \hkl[100] orientation, the doses are represented in a cold to warm color scale. (c) Schematic illustration of the crystal rotation (from the light grey to the black frame) of the atomic system with the dose above 0.205 cdpa.}
\label{CA700_1r}
\end{figure}

Furthermore, our observations reveal the occurrence of rigid rotation during plastic transformation. Notably, this rotation remains at zero during instances of elastic deformation, such as in the full MD cases. However, here, it is important to highlight that once the microstructure stabilizes, typically at a dose value above 0.205 cdpa, rigid rotation persists even in the absence of additional plastic slip events. Fig.~\ref{CA700_1r} illustrates this phenomenon, particularly in the zoomed-in detail as shown in Fig.~\ref{CA700_1r}(a), where vacancy loop habit plane rotation induces symmetric evolution trends in the $xy$- and $xz$-shear components. Clearly, the $xz$ components exhibit a more pronounced shear angle when compared to the $xy$ components. As a result, the entire cell undergoes a slight rotation about the $y$-axis as the axis of rotation, as depicted in Fig.~\ref{CA700_1r}(b) and (c). This phenomenon arises because shear strain is not permitted within the LAMMPS simulation box in this study. Consequently, when the orientation of the straining axis aligns favorably with one of the crystal's slip system, initiating slip along this primary slip vector would distort the cell's shape. However, the cell is constrained to remain aligned with the simulation box, necessitating a rotational adjustment to accommodate for this behavior. By monitoring the direction of rotation, we can establish a connection with the coherence between continuum and atomic-level slip events. Schmidt's prediction suggests that, a crystal should naturally rotate to align its predominant slip direction with the straining axis, based on purely geometrical considerations~\cite{meyers2008mechanical,zepeda2021atomistic}. When comparing the direction of rotation to that shown in Fig.~\ref{CA700_1ep}(b), we observe a consistent change in the direction of various shear components.

\begin{figure}[t]
\begin{center} 
\includegraphics[width=.9\columnwidth]{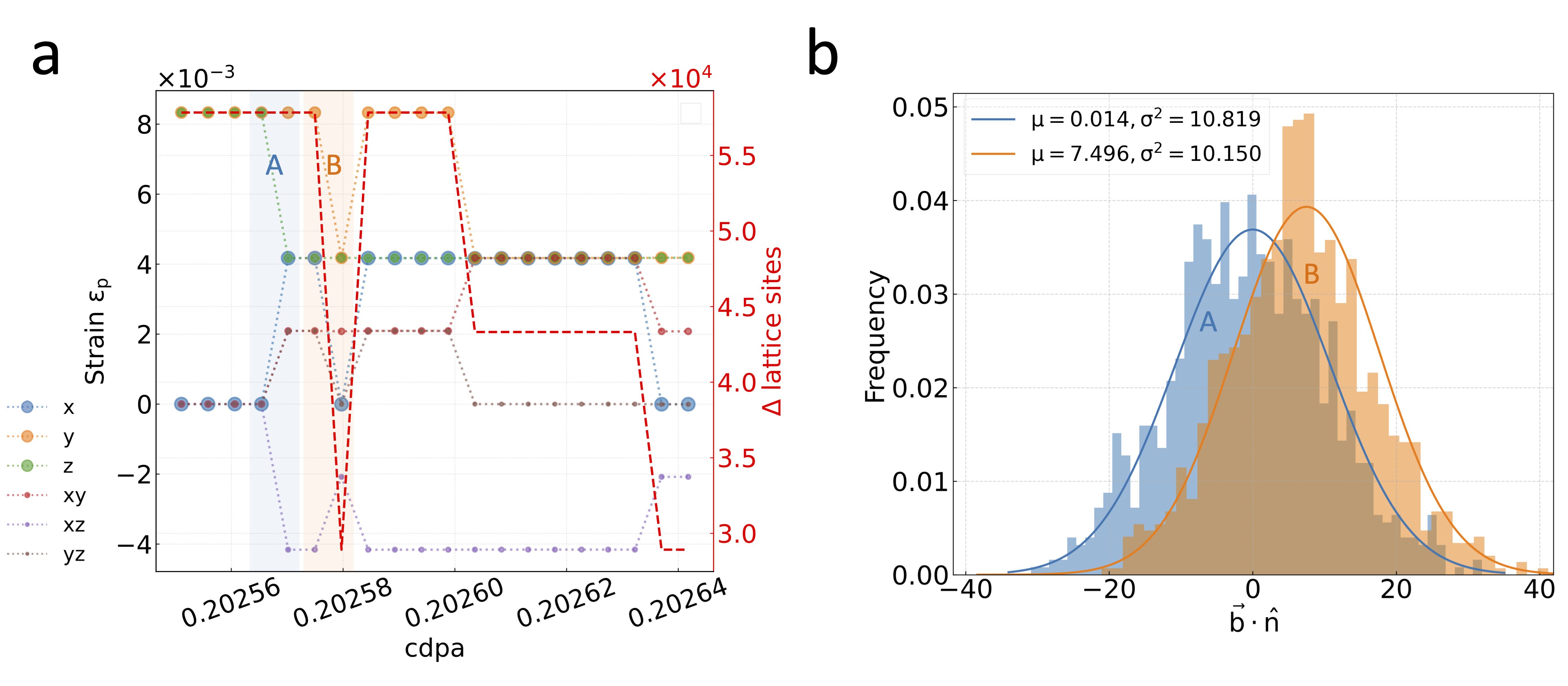}
\end{center}
\caption{Direct observations of two distinct slip events. (a) Plastic strain response (top panel) and corresponding defect concentration (bottom panel) as a function of dose at the highest starting dose of 0.2 cdpa during the 3rd CA case. The pure slip event and the slip event accompanied by lattice swelling are highlighted by the semi-transparent blue and orange backgrounds, respectively. The amount of change in lattice site (right axis legend) at different doses is indicated by the red dashed line. (b) Schematic illustration of dislocation slip. The gray and black spheres symbolize dislocation segments before and after a plastic slip event (with the interval denoting a single cascade) extracted by the DXA method. (c) Histogram of the distribution of $\vec{b} \cdot \Hat{n}$ after the Bootstrap method for events denoted as \textbf{A} and \textbf{B}. Normal distribution fitting curves are shown in colors matching their respective data sets.}
\label{CA700_3}
\end{figure}

In the preceding discussion, we explored the plastic transformation in cases of spontaneous swelling at high doses. Now, from Eq.~\eqref{DeltaN}, where we identify another distinct microstructural event: when $\vec{b} \cdot \Hat{n} = 0$, indicating a pure slip event. Upon careful examination, we discovered that such events are exceptionally rare. For a pure slip phenomenon to occur, there must be a spontaneous change in periodic lattice repeats, which only happens when the elements in the periodic lattice repeat matrix undergo symmetrical position exchanges. During our investigation, we encountered just one instance of this, at the very beginning in the 3rd CA case, with the highest starting dose of 0.2 cdpa. As depicted in Fig.~\ref{CA700_3}(a), the light-blue-highlighted region in the plastic strain $\varepsilon_p$ panel illustrates this pure slip event. Before and after this single cascade event, both the normal and shear components exhibit responses, but the number of excess lattice sites remains constant. A closer examination reveals that the matrix of primitive cell repeats changes from

\begin{equation}
        \mat{N}_{\mathrm{before}} = \left( \begin{array}{ccc}
                        0     &   121   &   121       \\
                        120   &   0     &   121       \\
                        120   &   121   &   0         \\
                    \end{array} \right) \nonumber
\hspace{1.5cm} \mathrm{to} \hspace{1.5cm}
       \mat{N}_{\mathrm{after}} = \left( \begin{array}{ccc}
                        0     &   121   &   121       \\
                        120   &   0     &   120       \\
                        121   &   121   &   0         \\
                    \end{array} \right). \nonumber
\end{equation}

To facilitate a direct comparison, we have annotated two cascade events, namely labeled as regions \textbf{A} and \textbf{B}. The transition in region \textbf{A} corresponds to a pure slip event, whereas the transition in region \textbf{B} represents a slip event accompanied by lattice swelling. In order to gain a visual understanding of the relationship between the angle defined by the Burgers vector and the slip normals, dot product with the spatial Burgers vectors is performed to check the perpendicularity. Fig.~\ref{CA700_3}(b) shows the histogram of $\vec{b} \cdot \Hat{n}$ after applying the Bootstrap method~\cite{efron1992bootstrap}. Employing normal distribution fitting, we obtained estimated result intervals within the 95\% confidence range for events shown in the region \textbf{A} and \textbf{B} respectively. The central values for these two cases are approximately 0 and 7. Consequently, it is interesting to note that by discerning alterations in the unit cell repeat number matrix, we can directly ``observe" the occurrence and type of slip events. Additionally, the infrequent pure slip events observed in our simulations could be attributed to the boundary conditions employed. It is conceivable that altering these conditions to include applied shear stress during simulated irradiation might yield different results. In our case, the practical result is that plastic events lead to material swelling.

\section{Conclusions}

With the combination of full defect build-up and accelerated simulated annealing MD simulations, we have investigated the mechanisms responsible for atomic and plastic strain during damage accumulation and evolution. We devised a strain decomposition method which works by resolving the homogeneous atomic strain field from the atom positions, and using this to infer the plastic strain required for a compatible single crystal. With our approach to dissecting elasto-plastic deformation at the atomic level, we bring clarity to the mechanism by which the gradual accumulation of lattice defects, as the dose increases, leads to the development of a uniform lattice strain.

The main conclusions of this study are:

\begin{enumerate}

    \item At low dose, plastic strain changes are absent, but atomic strain changes are brought about both by the accumulation of irradiation defects, and by the response of the defects to the strain. This was illustrated by the mechanism of habit plane rotation. In this work the defects were able to reduce elastic strain energy in the simulation within molecular dynamics timescales, indicating very low thermal activation barriers. In a real material, impurities such as carbon may slow the movement of sections of the dislocation loop~\cite{Castin_JNM2019}. It is unclear how the competition between the build-up of local stress and the retarding effects of impurities are balanced in technologically relevant conditions, and this merits further study.
        
    \item At high doses, both atomic and plastic strain can undergo alterations. The plastic transformation can be categorized into two types: pure slip events and slip accompanied by lattice swelling. The occurrence of pure slip events is infrequent in our simulations. In cases where the number of lattice sites remains unchanged, interstitials and vacancies show identical evolution trends. However, when the plastic transformation related to swelling occurs, a substantial change in excess lattice sites becomes apparent, which explains why the interstitial and vacancy count diverge. In this case, anomalous slip can occur. Lattice-swelling-related deformation could potentially activate slip on these planes despite them not being the primary slip planes in BCC tungsten, such as the \hkl{111} slip plane. The limited probability of pure slip events in the studied dose range accounts for the consistent material swelling observed during high-dose irradiation.
    
\end{enumerate}

This study presents a straightforward and effective method for decoupling the lattice strain field, which influences the evolution of defects resulting from irradiation, from the plastic strain field, which leads to material swelling. The methodology presented here can be used as an automatic analysis for future simulations, which can identify the elasto-plastic deformation. The results of this method can additionally be used to locate the interesting phenomena, which enables analysis by other methods, which cannot be used globally or on the the whole simulation series. This work is therefore a step towards describing radiation defect accumulation within the framework of finite element elasticity models. The findings underscore the detrimental impact of high-dose irradiation, leading to significant creep phenomena that pose challenges for the performance of advanced fission and fusion reactor components.

\section*{Acknowledgements}

The authors would like to thank Max Boleininger and Luca Reali for helpful discussions. This work has been carried out under the DEVHIS project, funded by the Academy of Finland (Grant number 340538). This work has been partly carried out within the framework of the EUROfusion Consortium, funded by the European Union via the Euratom Research and Training Programme (Grant Agreement No 101052200 — EUROfusion). Views and opinions expressed are however those of the author(s) only and do not necessarily reflect those of the European Union or the European Commission. Neither the European Union nor the European Commission can be held responsible for them. Computer time granted by the IT Center for Science -- CSC -- Finland and the Finnish Grid and Cloud Infrastructure (persistent identifier urn:nbn:fi:research-infras-2016072533) is gratefully acknowledged.

\bibliographystyle{apsrev4-2}
\bibliography{article}
\end{document}


\title{Supplementary Information for ``Atomistic Study of Irradiation-Induced Plastic and Lattice Strain in Tungsten"}
\author{Jintong Wu}
\affiliation{Department of Physics, University of Helsinki, Post-office box 43, FIN-00014 University of Helsinki, Finland}
\author{Daniel R. Mason}
\affiliation{UK Atomic Energy Authority, Culham Science Centre, Oxfordshire OX14 3DB, UK}
\author{Fredric Granberg}
\email{fredric.granberg@helsinki.fi}
\affiliation{Department of Physics, University of Helsinki, Post-office box 43, FIN-00014 University of Helsinki, Finland}

\maketitle

\section{Convergence tests}
\begin{figure*}[]
\begin{center} 
\subfloat[Cascade annealing]{\includegraphics[width=.45\columnwidth]{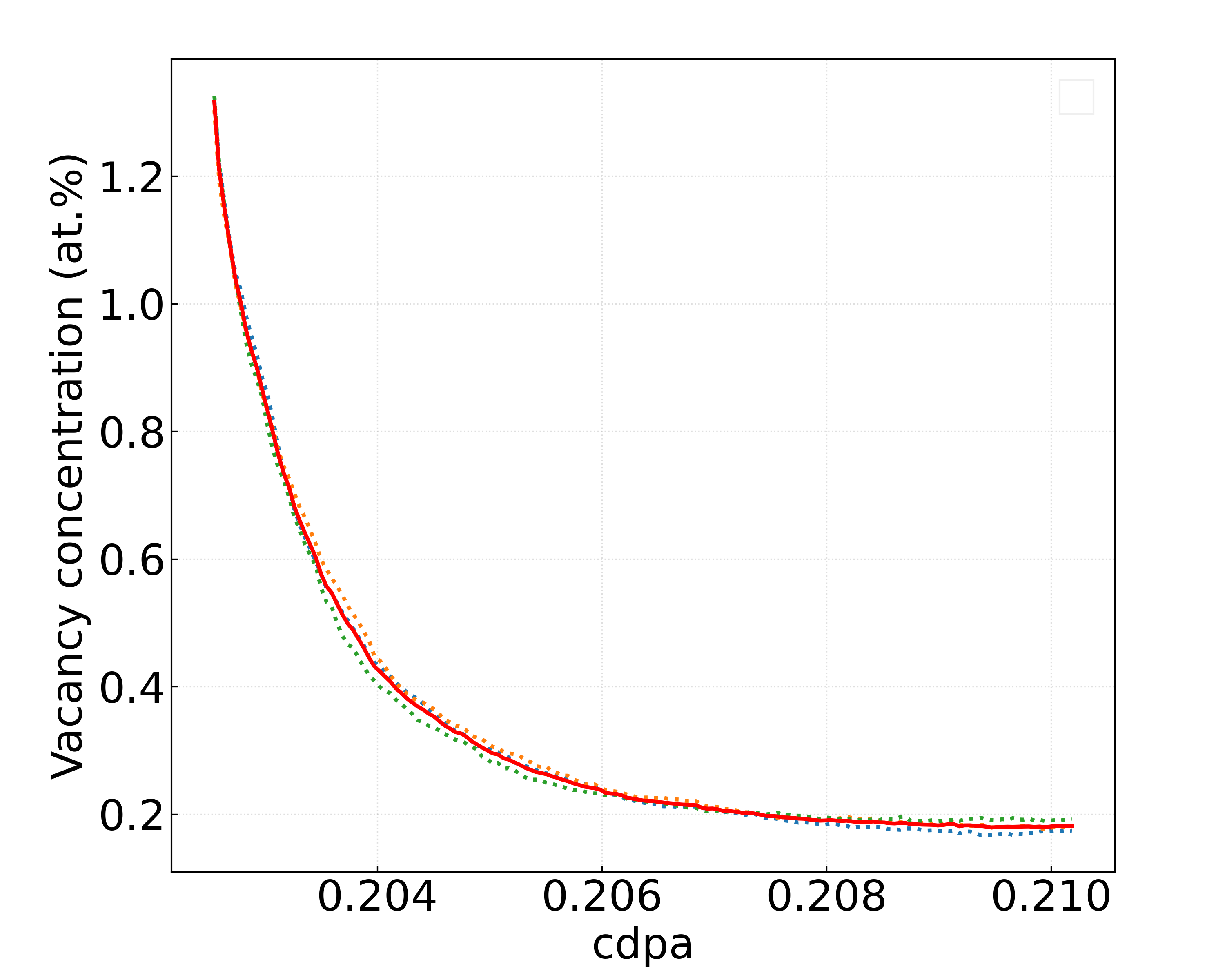}} 
\subfloat[Thermal annealing]{\includegraphics[width=.45\columnwidth]{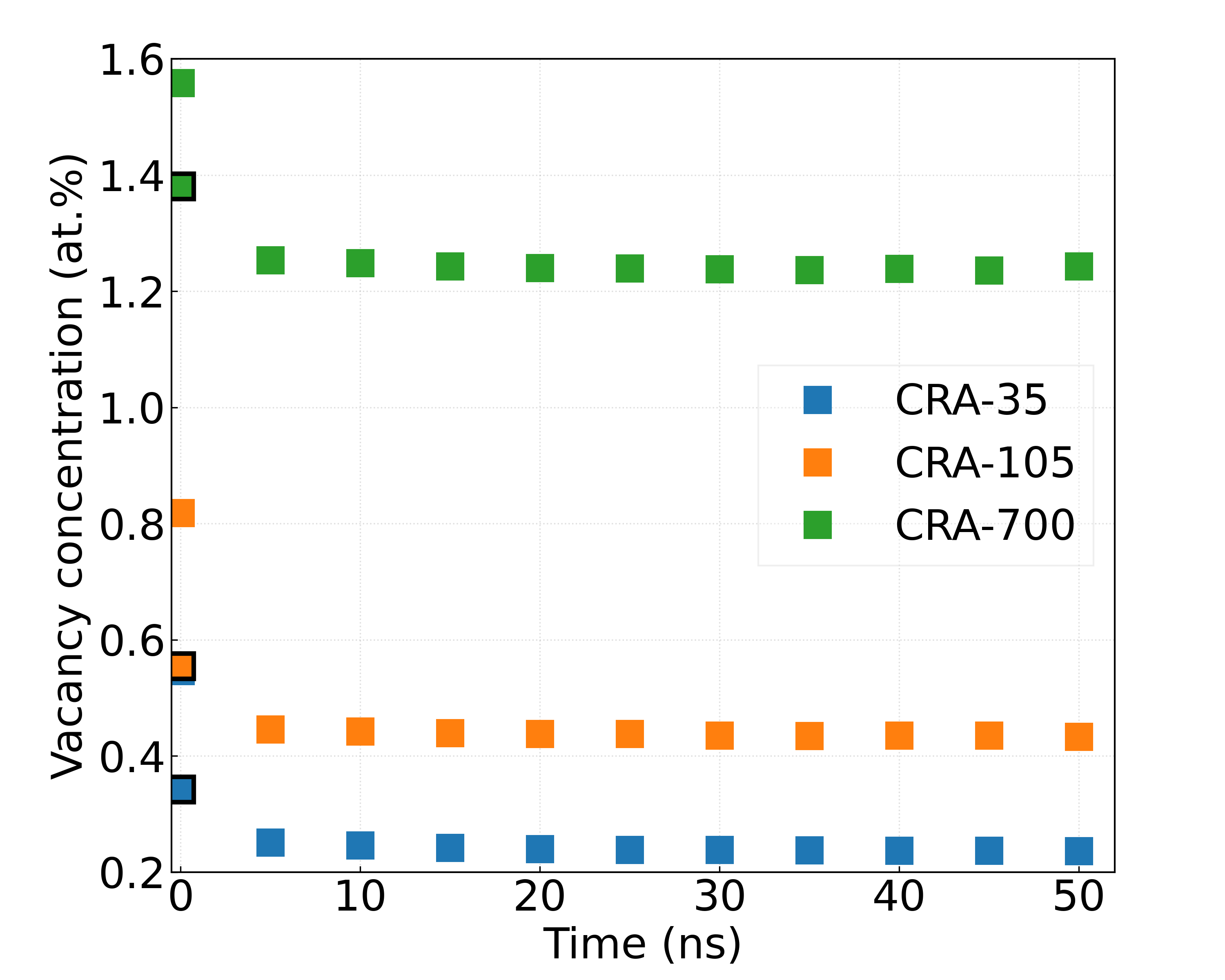}}
\end{center}
\caption{(a) Vacancy concentration evolution as a function of dose (CA steps) by the VID method for the highest dose case. Saturation is seen during the 1600 times of CA. Dots: three individual cases. Red line: average value. (b) Vacancy concentration as a function of time during TA, using the same color to indicate different initial doses. Symbols with black edges: after 30 ps relaxation (the initial cells for subsequent CA simulations). Symbols without edges: the evolution of vacancy concentration during the 50 ns thermal relaxation.}
\label{saturation}
\end{figure*}

To investigate how much of the annealing in cascade annealing (CA) is happening due to the thermal part and how much due to the cascades, a few CRA samples were subjected to only thermal annealing (TA). We look at both how much effect the initial heat up from 0 K to 300 K over 30 ps have on the evolution as well as a longer 50 ns relaxation. The 50 ns relaxation is the same duration as the full 1600 CA steps. Three independent simulations at 3 different doses at 0.01, 0.03 and 0.2 cdpa were selected for the longer thermal annealing. In Fig.~\ref{saturation}(a) and (b) the convergence during CA and TA for one of the cases is shown, respectively. In Fig.~\ref{saturation}(b), the solid symbols show the initial cells after 30 ps ``warm up" at room temperature (before subsequent CA simulations).

\section{Energy deposition}
After careful inspection we found that about 1400$\pm$100 times of CA can ensure that all defect levels reach saturation state at the different initial doses. An interesting idea arises based on this, if we combine the published results in Ref~\cite{Granberg2022}, for the 10 keV case, about 600$\pm$100 times of CA can reach the saturation level. In Ref~\cite{MasonPRM_VoidDetection}, this value is about 1600$\pm$100 times, for a different cell size and interatomic potential. Based on the total energy deposited in the system (the irradiation energies minus the electronic stopping powers, which are roughly 8000 eV and 2200 eV for 30 keV and 10 keV, respectively), a simple estimate on how much CA is needed in form of energy deposited per atom can be found, as shown in Table~\ref{tab:Edeposit}. We can see that it is between 7 -- 11 eV/atom needed in order to fully converge the results. This value can be used in further studies as an initially rough estimate on how many CA steps are needed for a certain energy and simulation cell size.

\section{Void evolution by the VID}
\begin{figure}[]
\begin{center} 
\includegraphics[width=.49\columnwidth]{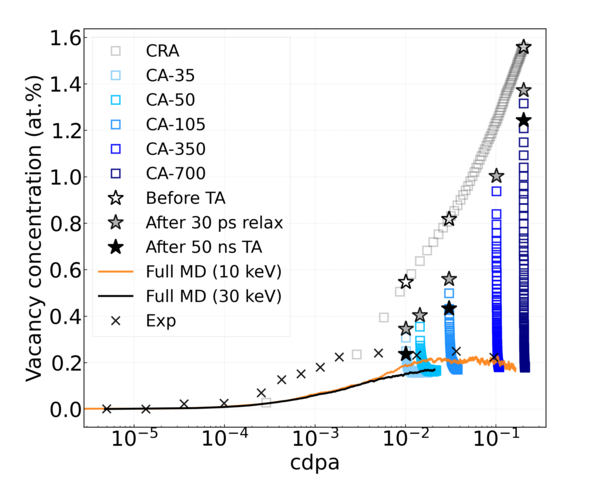}
\end{center}
\caption{Vacancy concentration (at.\%) evolution as a function of dose by the VID method. Solid lines: full MD simulations. Different shades of blue squares: the vacancy evolution during CA. Open star markers: the initial vacancy concentration after CRA (before TA). Gray filled star markers: after the first 30 ps relaxation at room temperature before starting the CA. Black filled star markers: after 50 ns TA. The ``$\times$'' markers (Exp) are experimental values from Ref.~\citenum{W_TGS_exp} and the orange line data from Ref.~\cite{Granberg2022}.}
\label{VoidVacEvolution}
\end{figure}

Looking at the vacancy concentration evolution at different doses analyzed by the Void Isosurface Detection (VID), Fig.~\ref{VoidVacEvolution}, we observe differences between different simulation methods. In order to facilitate the distinction, we adopted fixed number of inserted FPs/time + repetitions of this process to name the CA cases. E.g. ``CA-35" means repeating the process of inserting 1000 FPs 35 times to reach an initial dose of 0.01 cdpa. First examining previous full MD results (10 keV PKAs in a 0.5 million atoms cell~\cite{Granberg2022}), shown as an orange solid line, compared to our full MD results (30 keV PKAs in a $\sim$3.5 million atoms cell). Although naturally, increasing the PKA energy would introduce more defects in the cell, while if we convert the data to defect concentration as shown here, it could be found that the trend of vacancy concentration evolution is quite similar in the two cases. First an increasing trend is seen, and then a slow climb up until it reaches the saturation level. At the same doses as the previous results, we see a slightly lower defect concentration at the higher energy PKAs, however, the previous high-dose results are on a very similar level our results. We observe a good agreement with the saturation levels reported in the experiments from Ref.~\citenum{W_TGS_exp} (obtained by 20 MeV self-ion irradiation). Additionally, calculating the deuterium retention in our cells by assuming 5D per vacancy, we get a value of approximately 1.5\% deuterium retention, which is close to experimental values~\cite{MasonPRM_VoidDetection}. The CRA simulations show a dramatically increase in defect concentration during the whole process. During the non-thermodynamic CRA simulations make the randomly inserted FPs isolated and stable in the system. In addition, combined with previous reports on cascade overlap effects of W~\cite{Jesper_MoltenRegion} and the full MD results shown here, it can be seen that the linear increase in damage at the beginning (where no or little overlap is happening) will progressively become a trend of zero damage increase with increasing dose. This process, however, plays a negligible part during the CRA simulations, thus leading to a dramatic increase in the degree of lattice disorder. For the TA cases, depicted as black filled stars in Fig.~\ref{VoidVacEvolution} (also appearing in the other figures), it can be seen that the concentration of vacancies is further reduced by $\sim$0.05 at.$\%$ in 50 ns compared to the 30 ps case. Thermal annealing will indeed reduce the number of defects, but it does not anneal out everything, as there is a discrepancy between the methods, even though both of them are on the same timescales. The not so large difference between the short thermalization and the long TA, indicates that mainly the low barrier events and close-by defects are combined and annealed out. 

\section{Defect evolution by the W-S method}
\begin{figure}[]
\begin{center} 
\subfloat[Vacancy evolution]{\includegraphics[width=.49\columnwidth]{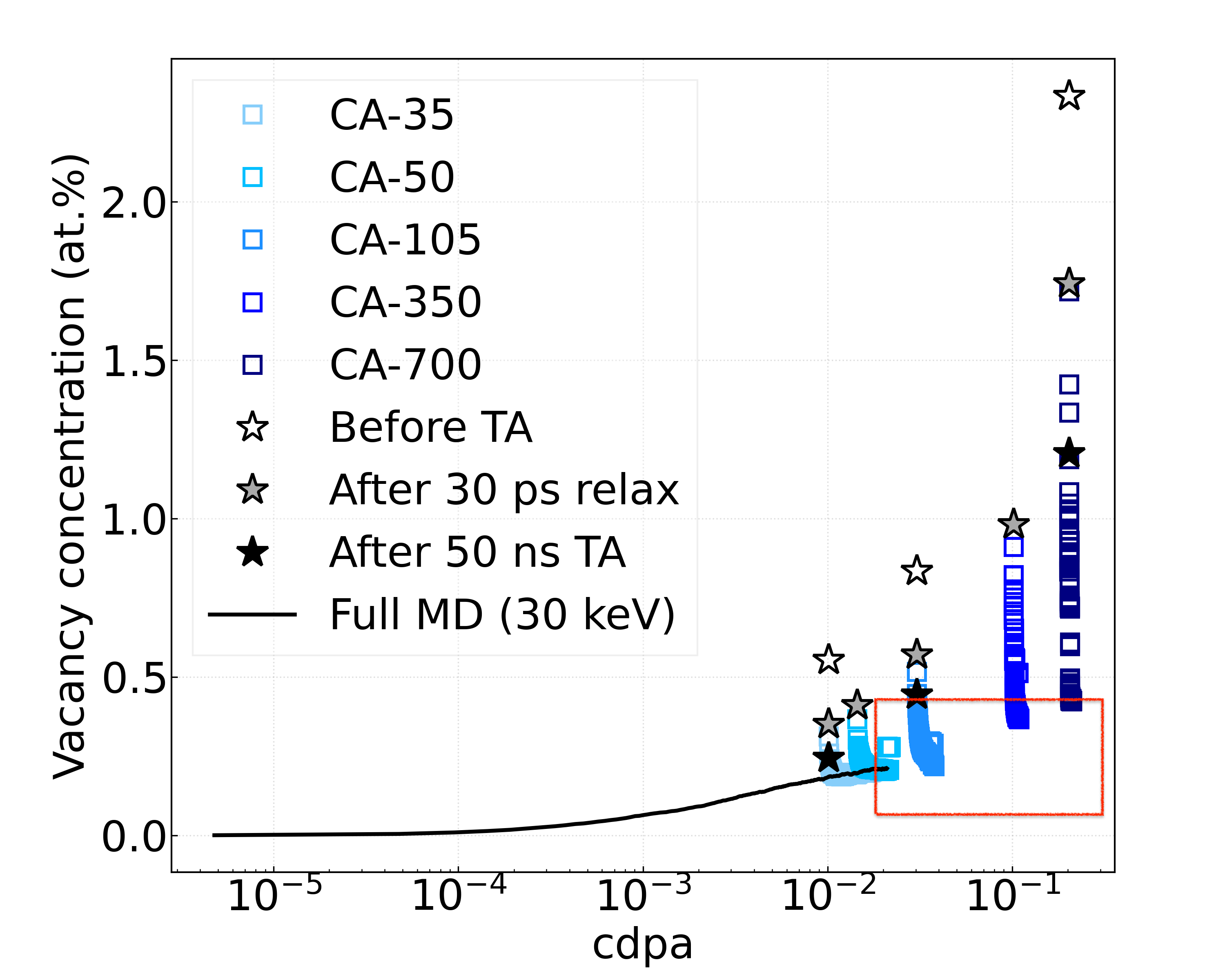}}
\subfloat[Interstitial evolution]{\includegraphics[width=.49\columnwidth]{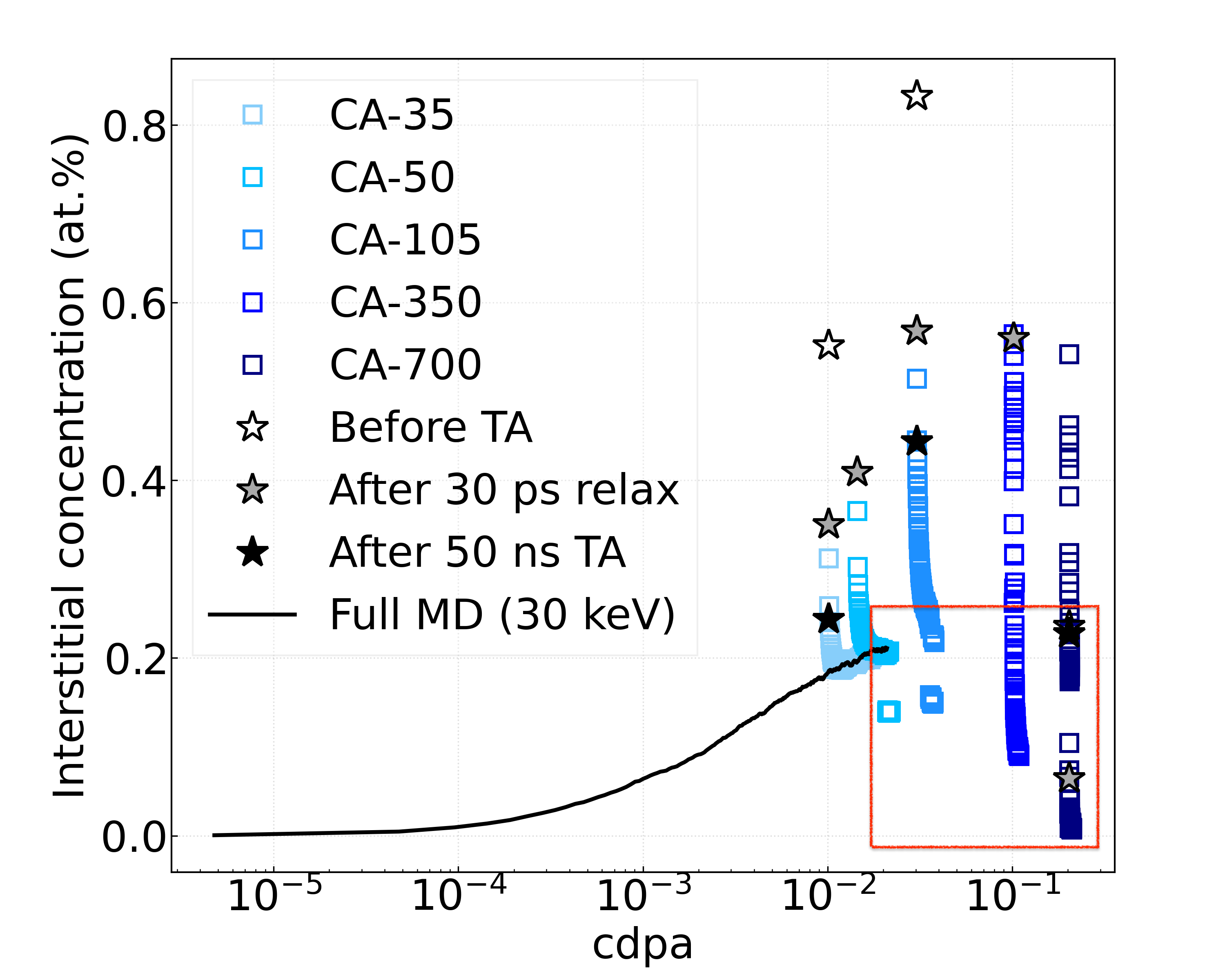}} \\
\subfloat[Defect evolution]{\includegraphics[width=.49\linewidth]{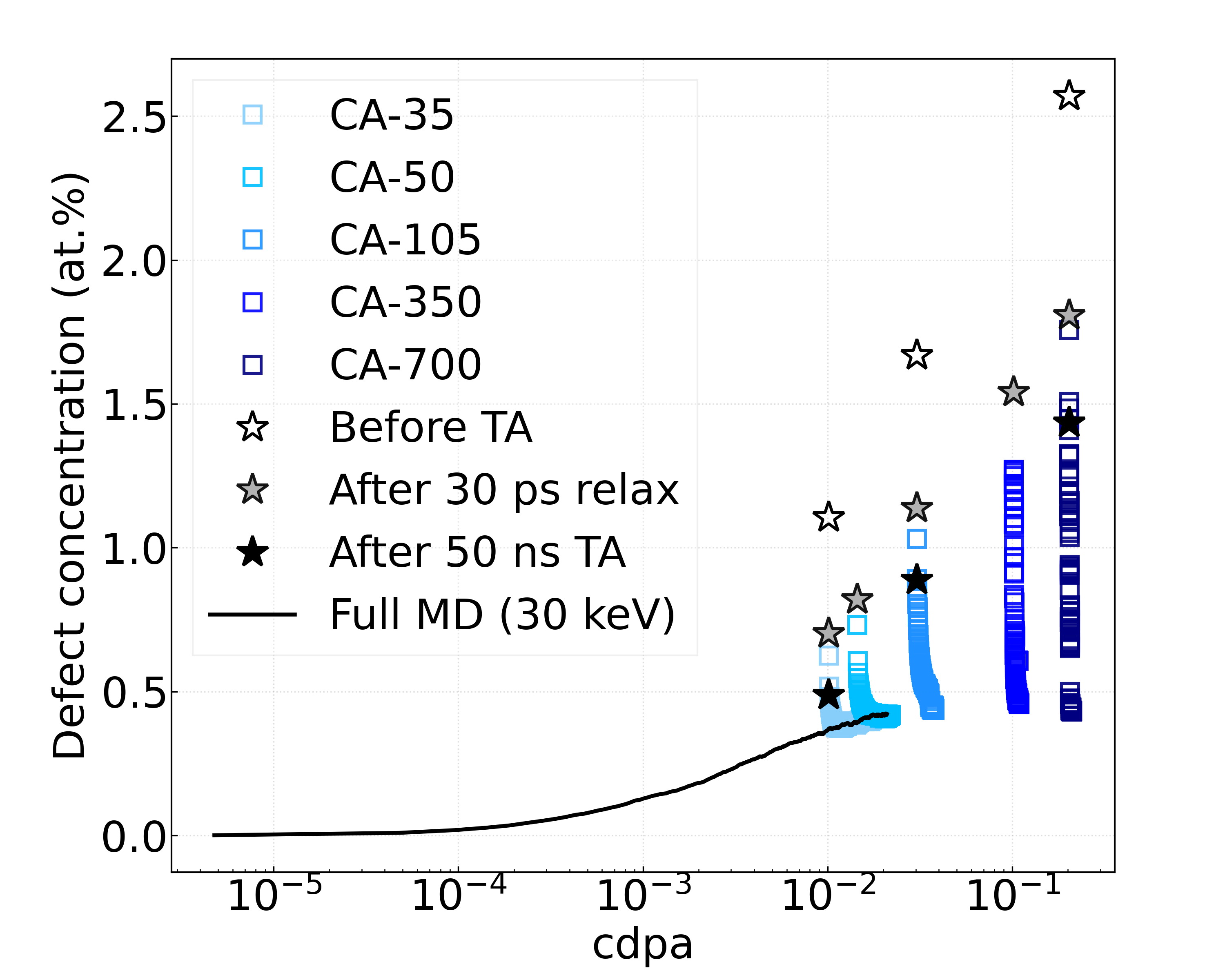}}
\end{center}
\caption{The (a) vacancy , (b) interstitial and (c) defect (sum of interstitials and vacancies) concentration as a function of dose determined by the W-S method. Solid lines: full MD simulations. Different shades of blue squares: the vacancy evolution during CA. Open star markers: the initial vacancy concentration after CRA (before TA). Gray filled star markers: after the first 30 ps relaxation at room temperature before starting the CA. Black filled star markers: after 50 ns TA. The red frame highlights the abrupt climbs and drops in the vacancy and interstitial concentrations after the CRA+CA.}
\label{WS_Vac_I_Evolution}
\end{figure}

Fig.~\ref{WS_Vac_I_Evolution} shows the at.\% defect concentration for all methodologies by the W-S method. 

\section{Volumetric swelling and potential increase}
\begin{figure}[]
\begin{center} 
\subfloat[Volumetric swelling]{\includegraphics[width=.45\columnwidth]{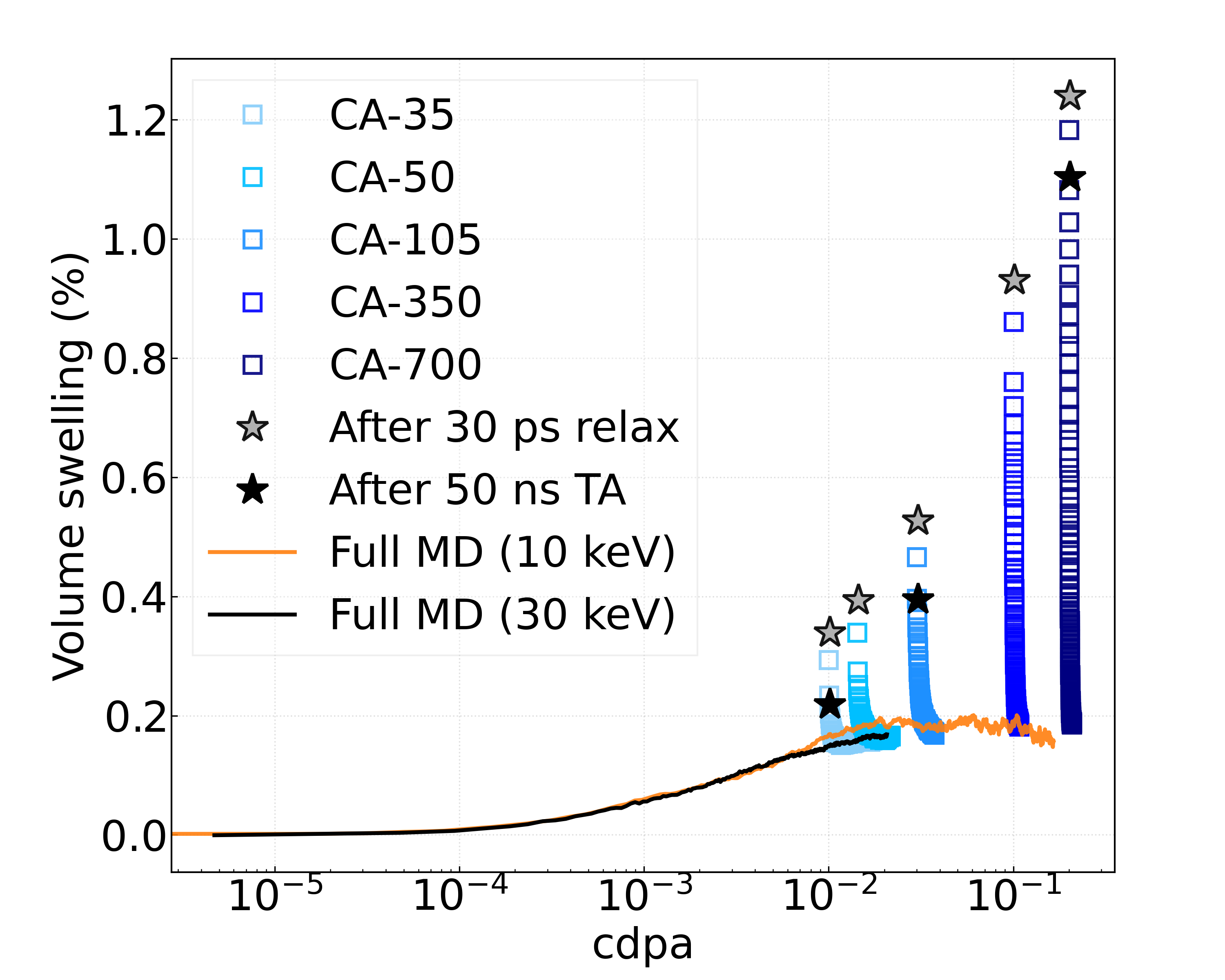}} 
\subfloat[Potential energy increase]{\includegraphics[width=.45\columnwidth]{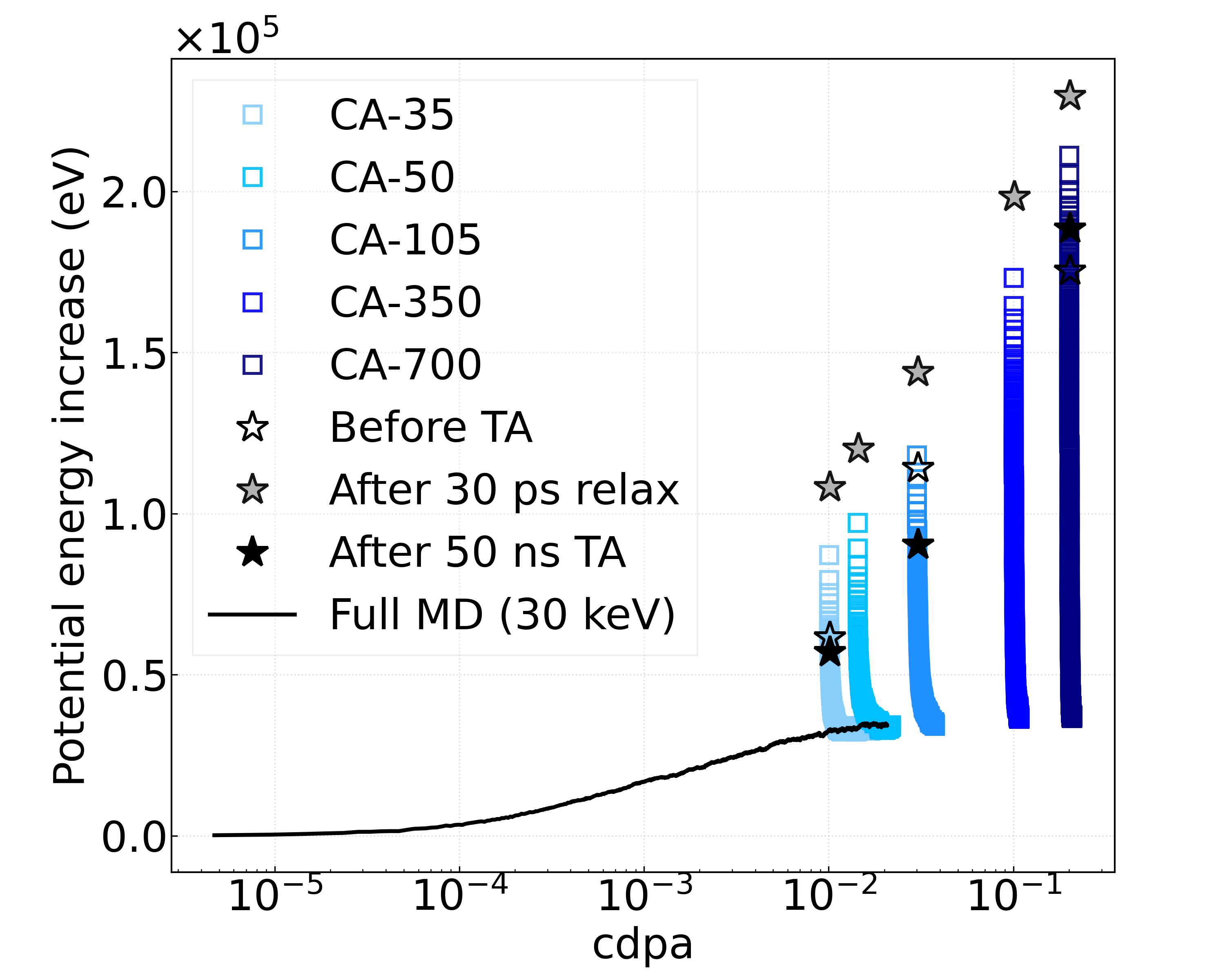}}  \\
\subfloat[Volumetric swelling during CA]{\includegraphics[width=.45\columnwidth]{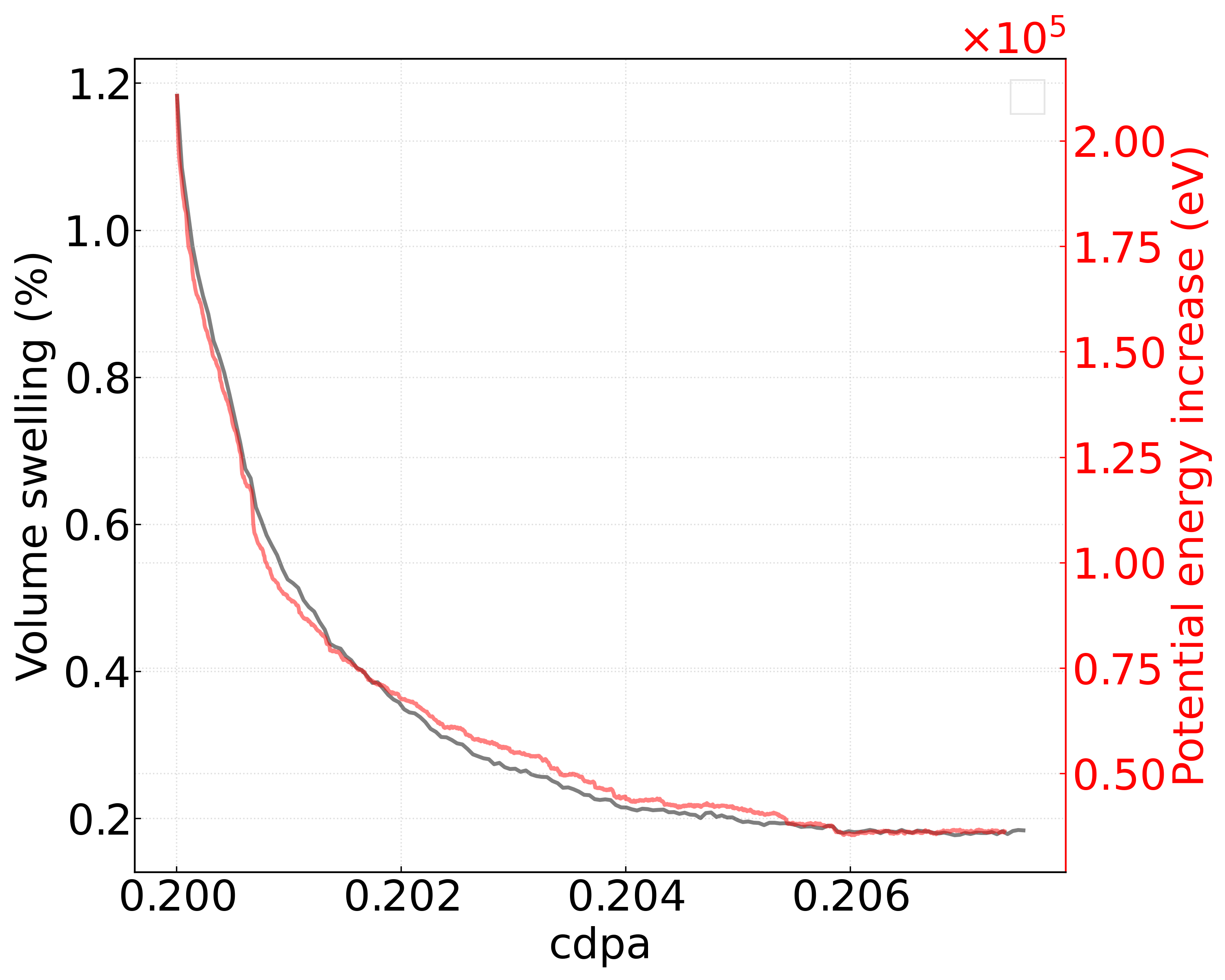}}
\end{center} 
\caption{(a) Volumetric swelling as a function of dose. Solid lines: full MD simulations. Black filled star markers represent the pure TA for three sets from CRA results at different doses for 50 ns, to compare to the 30 ps relaxation results (shown as gray filled star symbols) initially performed before the CA. The orange line data from Ref.~\cite{Granberg2022}. (b) Potential energy increase as a function of dose. (c) A closer examination to show the continuous decrease of volumetric swelling and potential energy during the CA with the highest starting dose.}
\label{continuity}
\end{figure}

In Fig.~\ref{continuity}, the evolution of the volumetric swelling and the potential energy increase as a function of dose are shown. A similar conclusion for a different setup is also reflected in the recent work of Boleininger et al.~\cite{MaxDXA}, where their simulations showed that the formation of extra planes only led to a sudden drop in the dislocation volume while the box volume was not affected. Note that the volumetric swelling remains unaffected by plastic deformation-induced slipping events associated with lattice sites swelling, which shown as the core part in the main manuscript. In addition, it can be found that the swelling caused by different energy irradiation is almost the same at low doses $\lessapprox$ 0.01 cdpa, with a value about 0.2\%, for both full MD and CRA+CA. A slight separation is observed at higher doses, both full MD and CRA+CA results for 30 keV PKAs are slightly lower than that of 10 keV PKAs. One of the reasons related to the PKA energy is the self-annealing effect of lower irradiation energy is relatively lower, which can explain why the damage level for 10 keV PKAs is higher than that of 30 keV PKAs at the same dose of $\approx$0.01 cdpa (Fig.\ref{VoidVacEvolution}). Despite minor differences, the values found in the experimental studies for different doses different temperatures vary in the range of 0.2–0.4$\%$~\cite{SwellExp_1,SwellExp_2}, close to both simulation cases.

\section{Cluster statistics}
\begin{figure*}
\begin{center} 
\subfloat[size = 1]{\includegraphics[width=.35\linewidth]{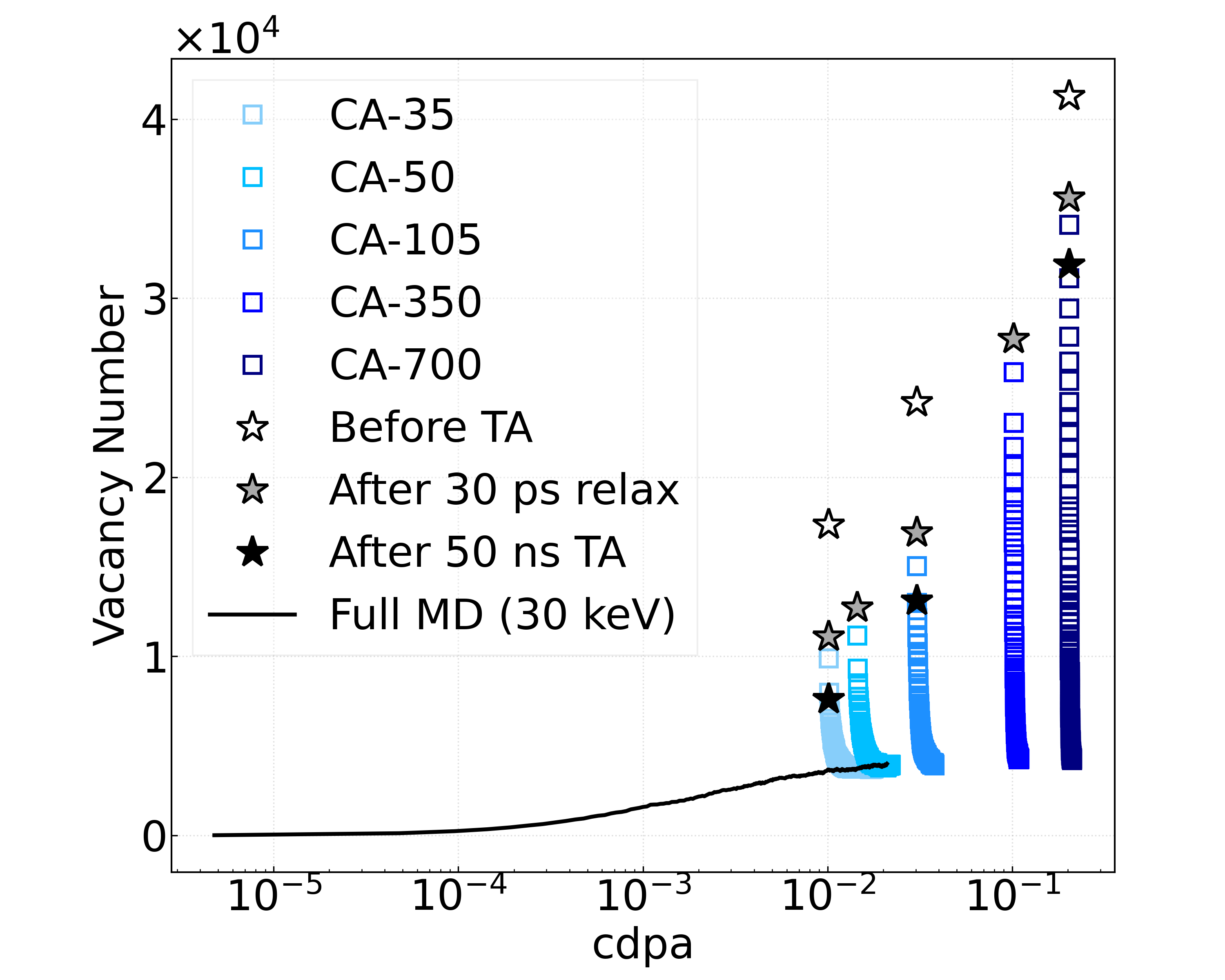}} 
\hspace{10mm}
\subfloat[size = 2]{\includegraphics[width=.35\linewidth]{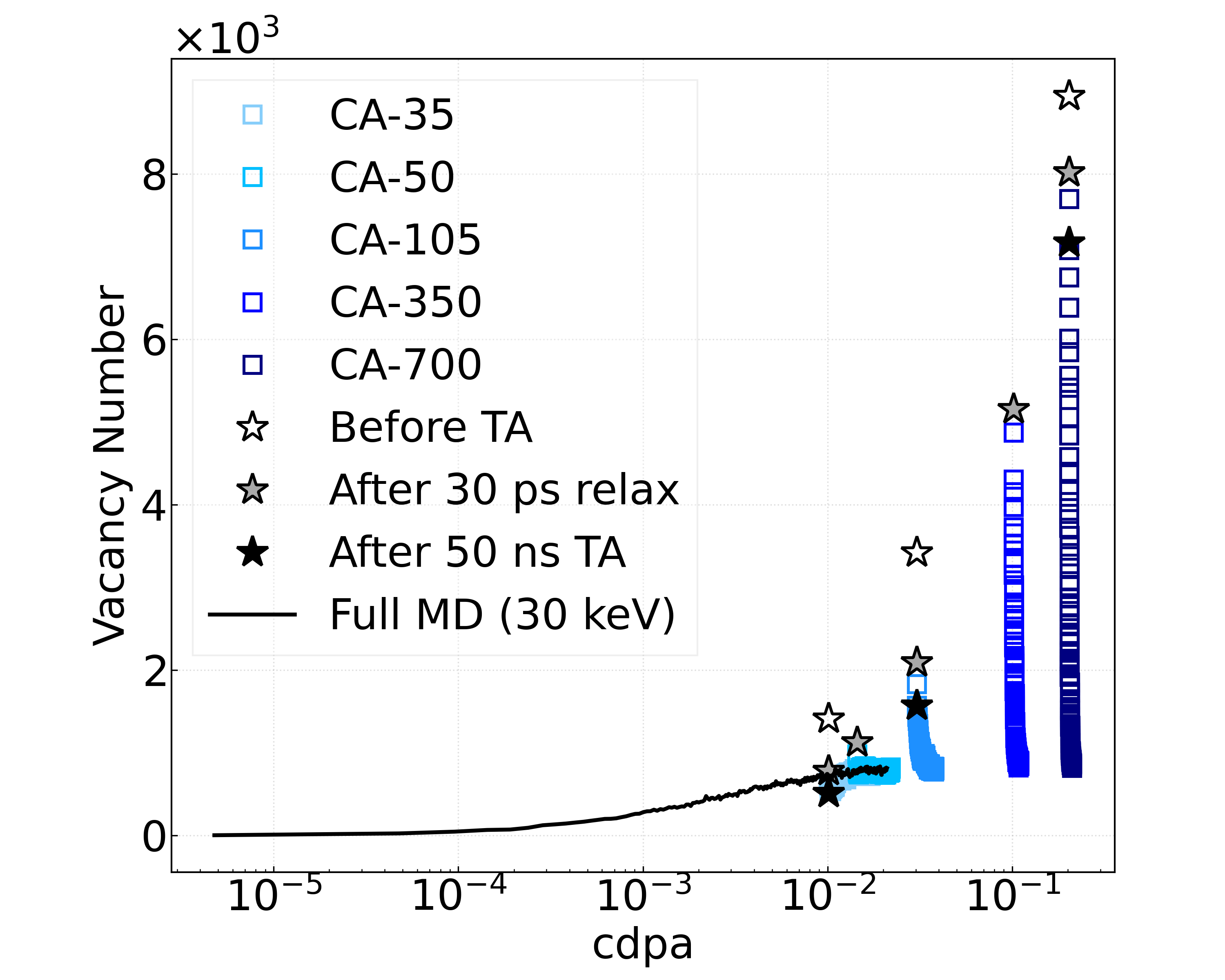}} \\
\subfloat[size = 3]{\includegraphics[width=.35\linewidth]{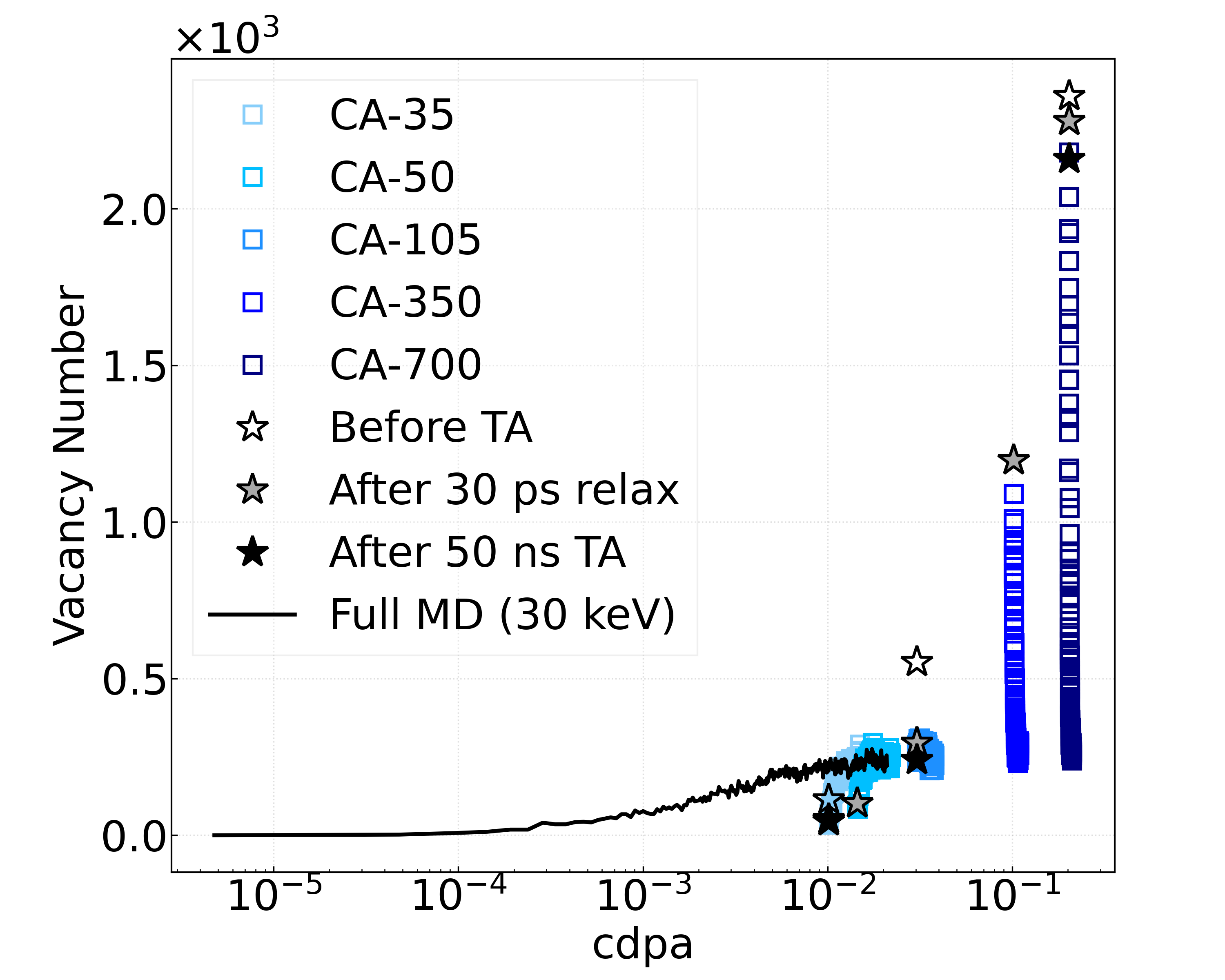}}
\hspace{10mm}
\subfloat[size = 4-5]{\includegraphics[width=.35\linewidth]{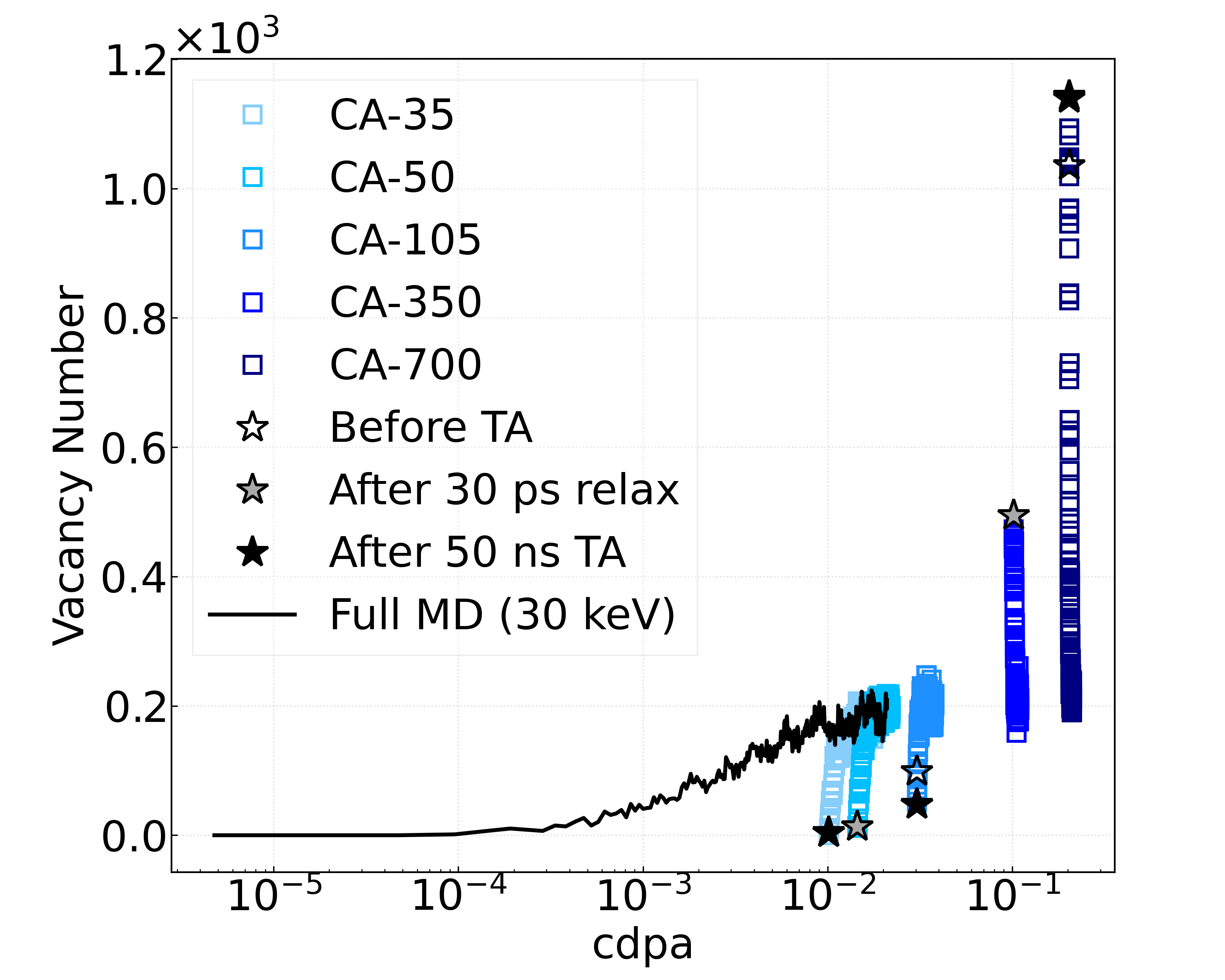}} \\
\subfloat[size = 6-10]{\includegraphics[width=.35\linewidth]{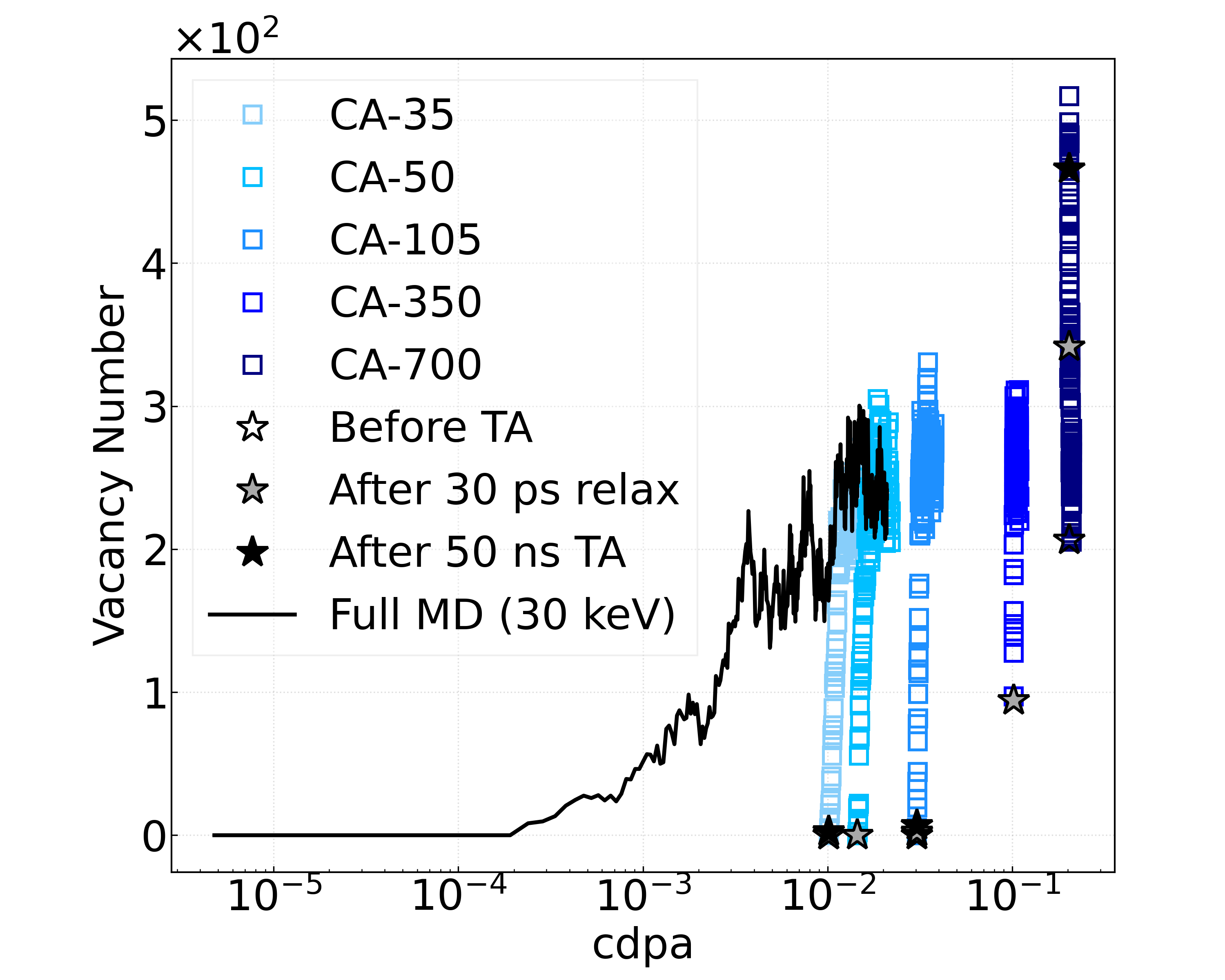}}
\hspace{10mm}
\subfloat[size = 10+]{\includegraphics[width=.35\linewidth]{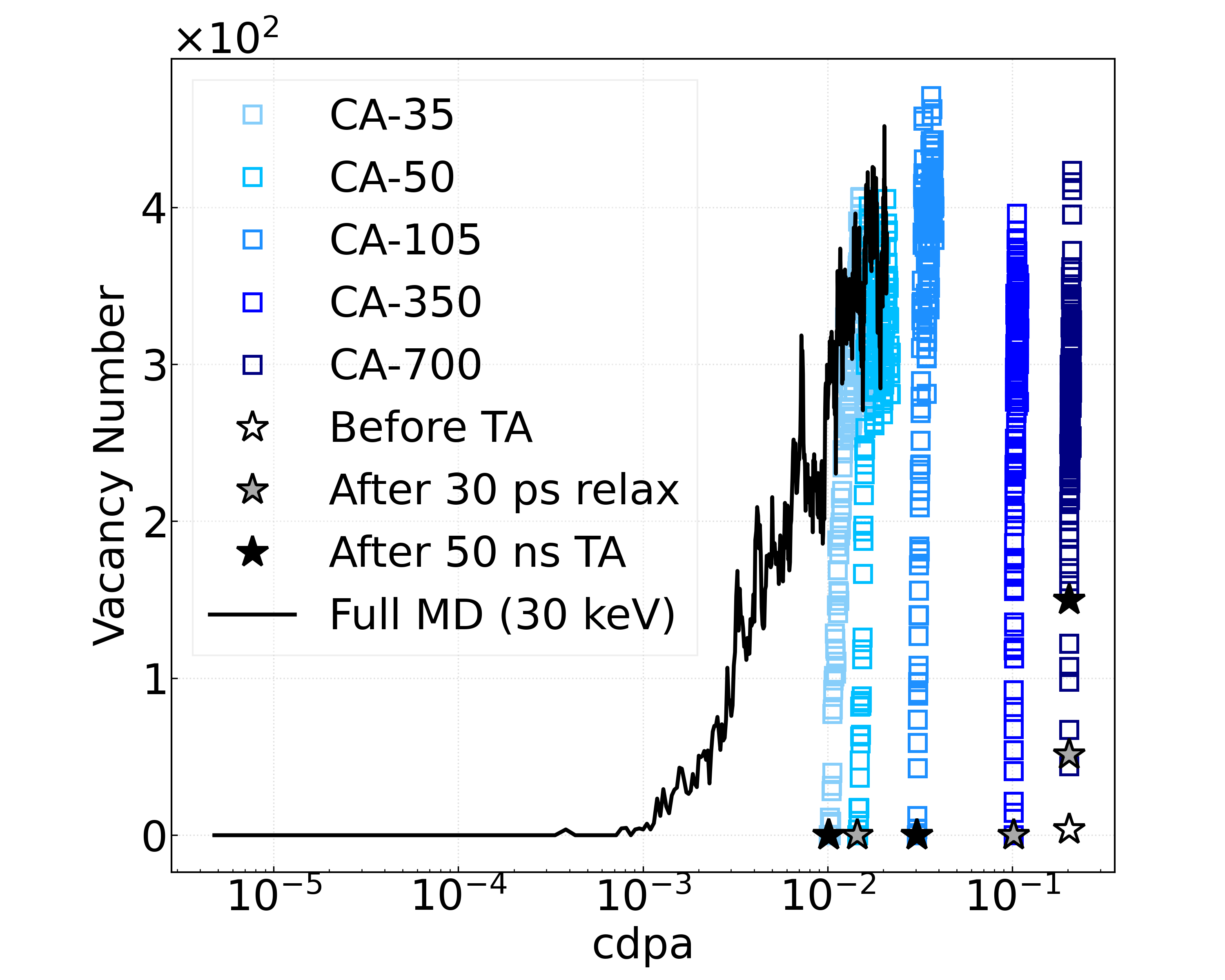}} \\
\end{center}
\caption{Number of vacancies found in different-sized vacancy clusters by the VID method. Solid lines: full MD simulations. Different shades of blue squares: the vacancy evolution during CA. Open star markers: CRA simulations only. Gray filled star markers: 30 ps relaxation at room temperature after the CRA. Black filled star markers: after 50 ns TA at room temperature.}
\label{VID_clusterSize}
\end{figure*}

\begin{figure*}
\begin{center} 
\subfloat[size = 1]{\includegraphics[width=.35\linewidth]{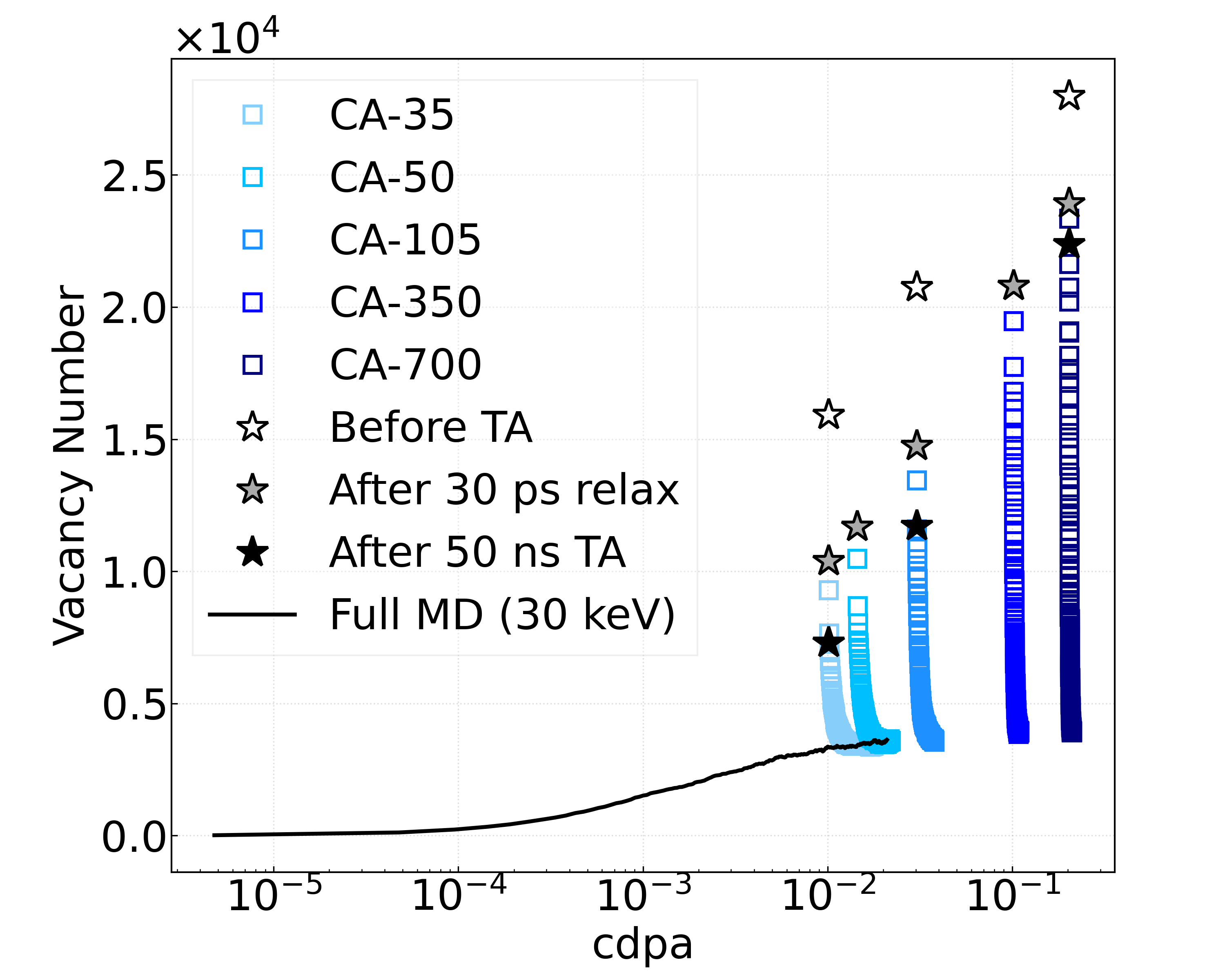}} 
\hspace{10mm}
\subfloat[size = 2]{\includegraphics[width=.35\linewidth]{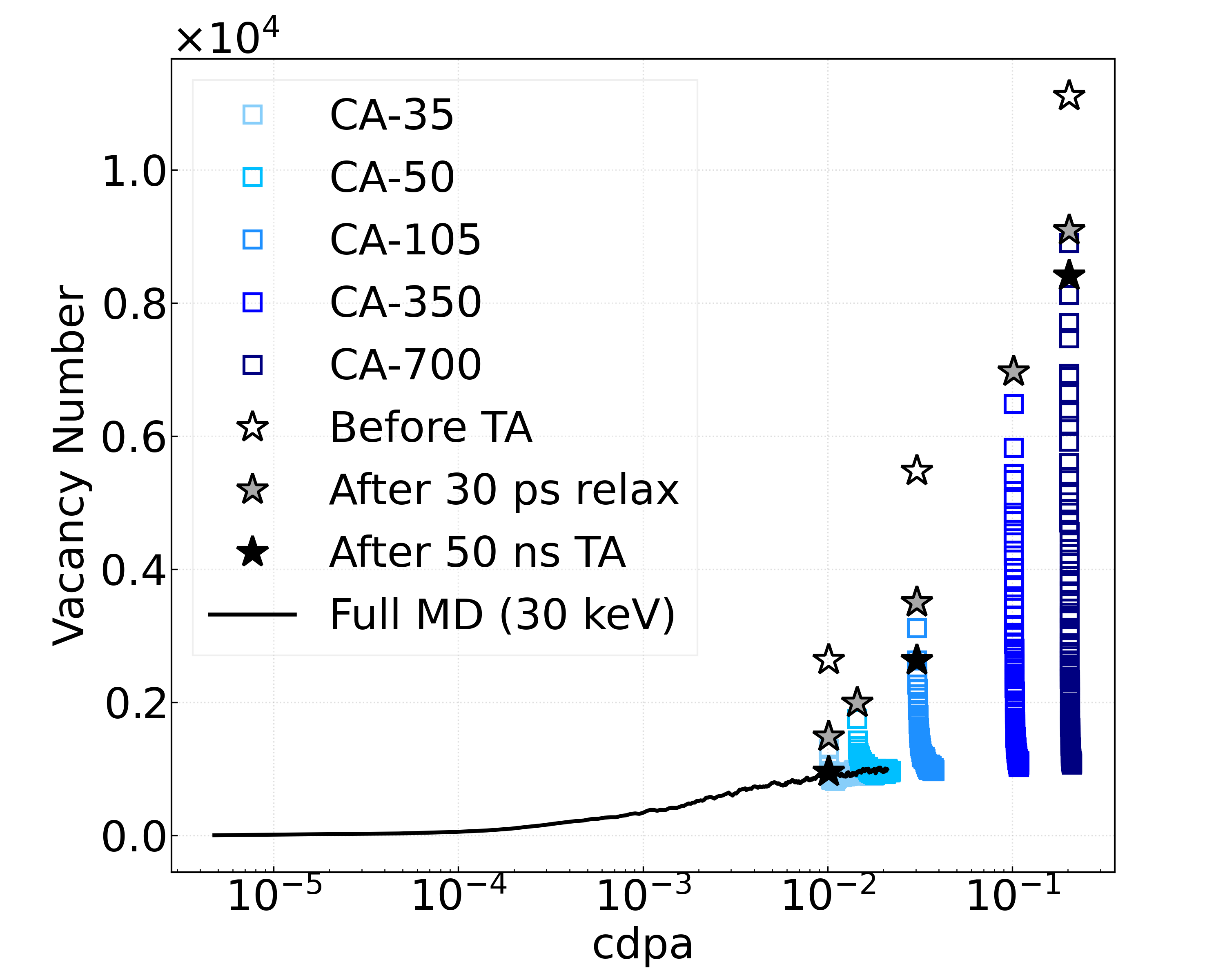}} \\
\subfloat[size = 3]{\includegraphics[width=.35\linewidth]{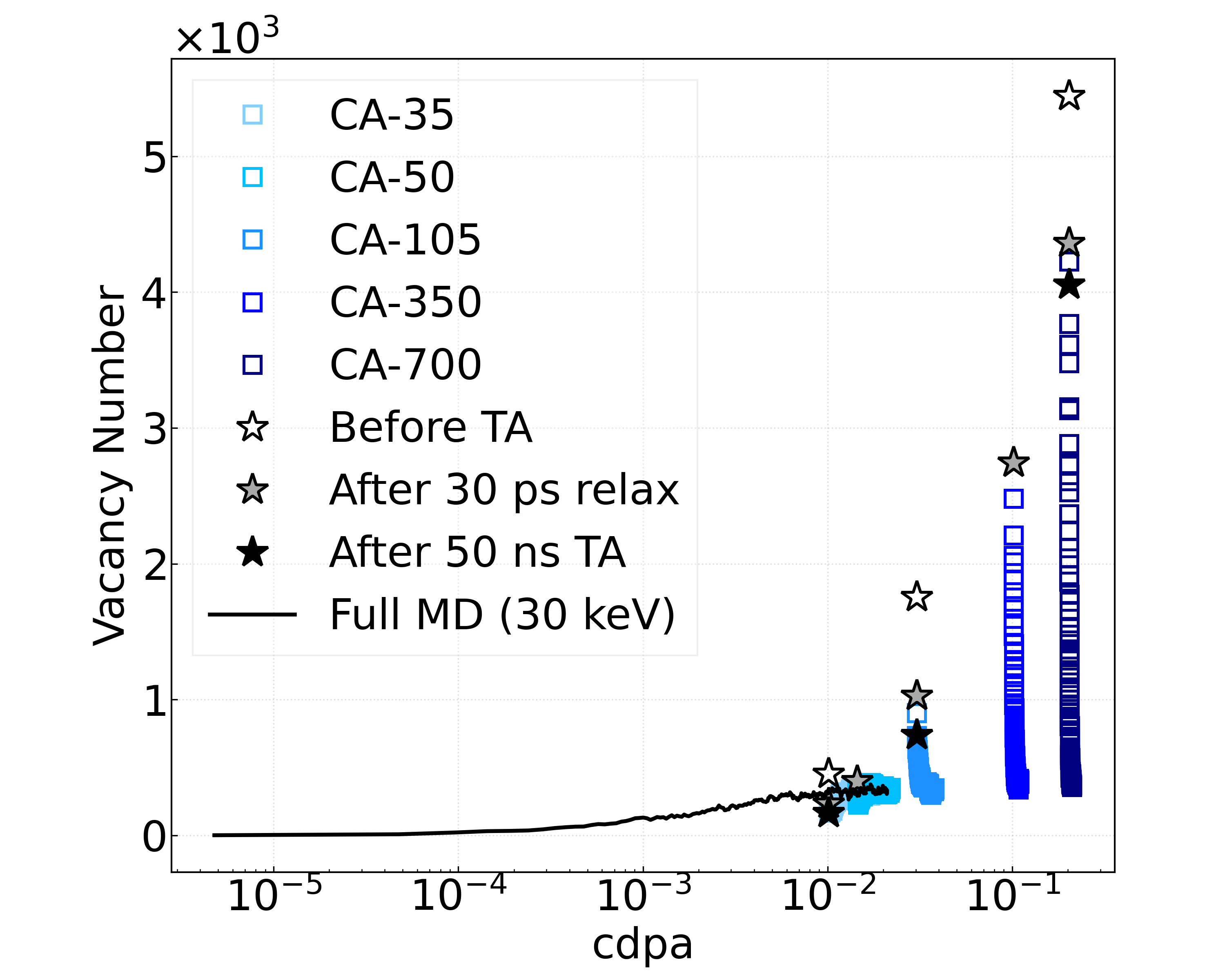}}
\hspace{10mm}
\subfloat[size = 4-5]{\includegraphics[width=.35\linewidth]{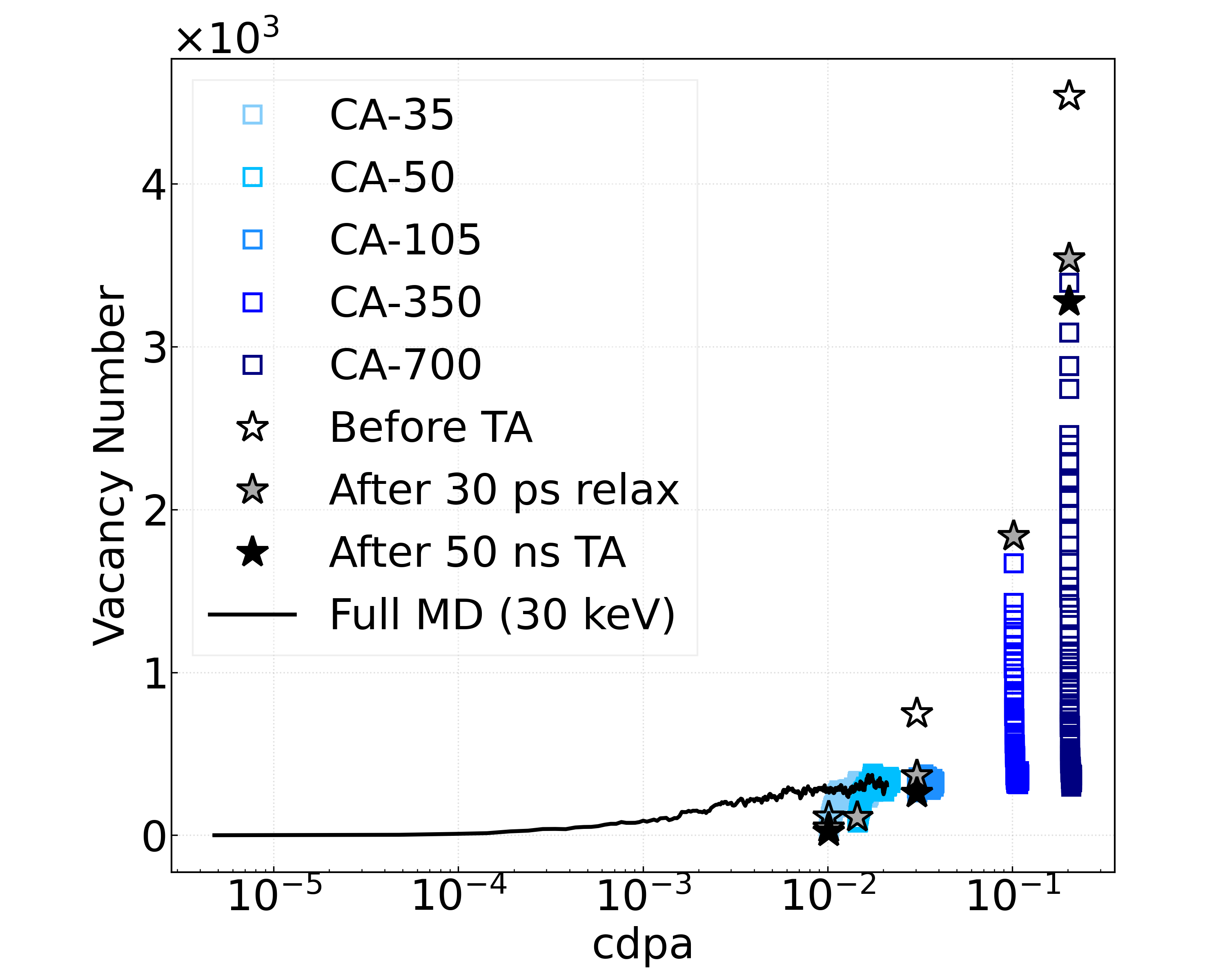}} \\
\subfloat[size = 6-10]{\includegraphics[width=.35\linewidth]{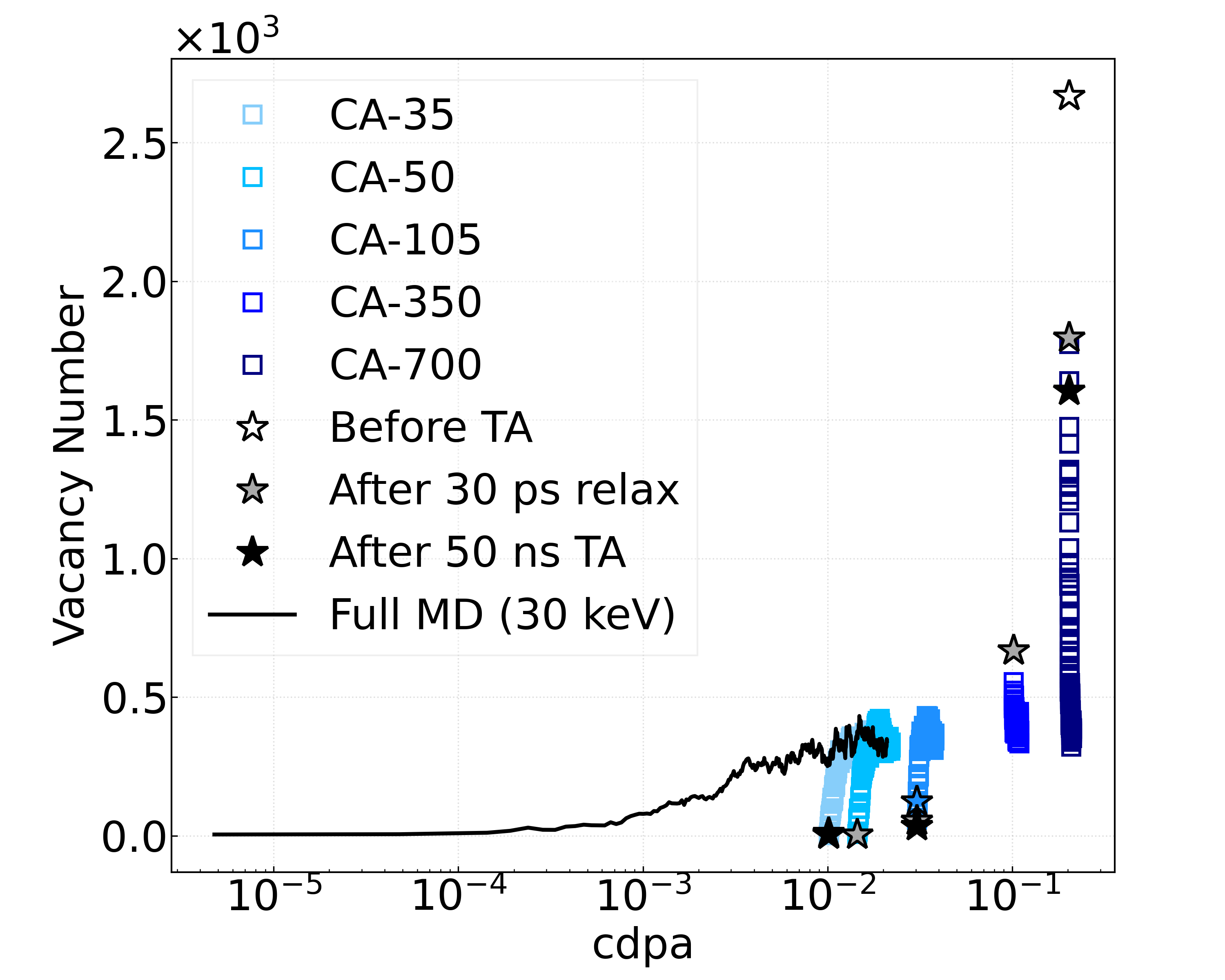}}
\hspace{10mm}
\subfloat[size = 10+]{\includegraphics[width=.35\linewidth]{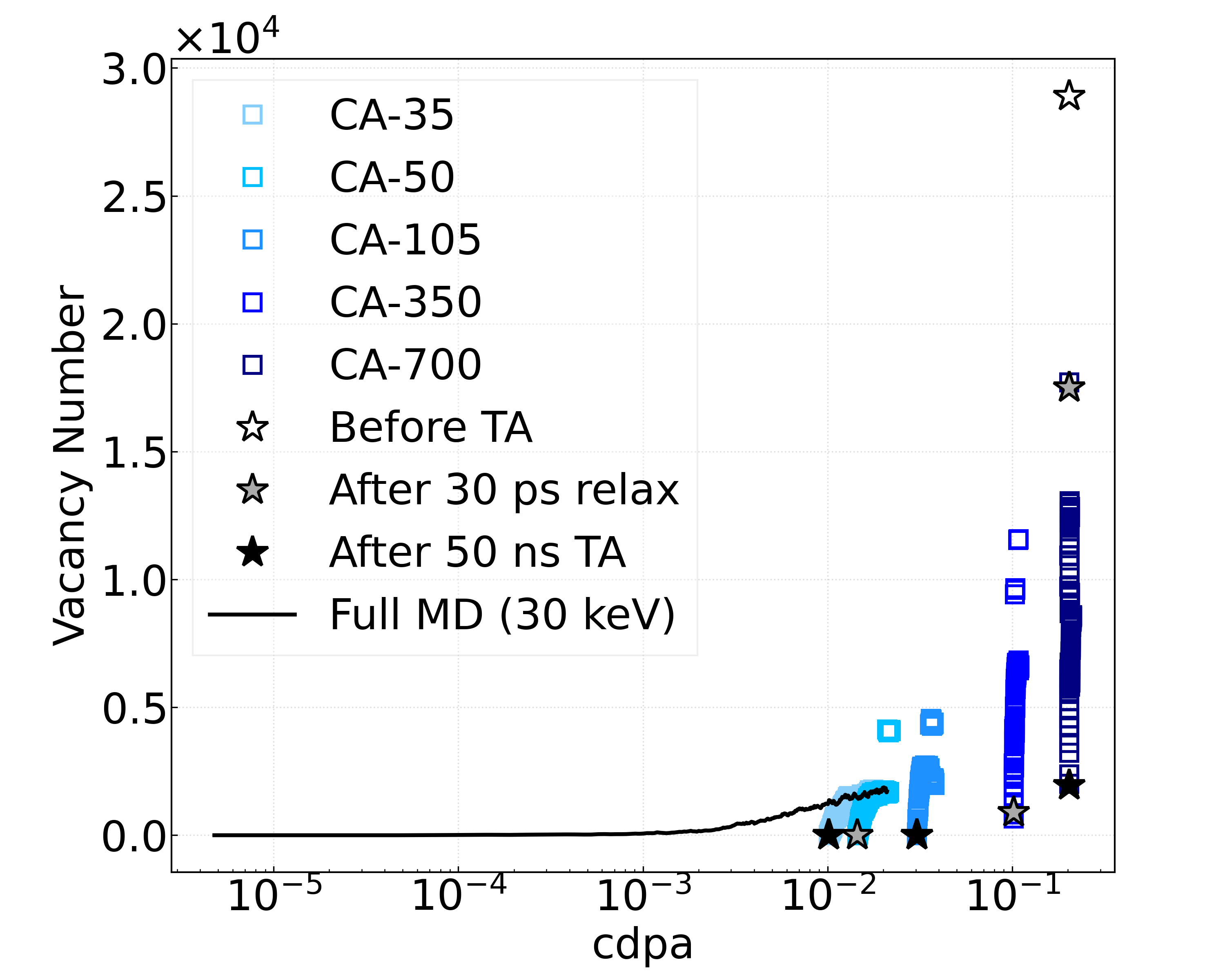}} \\
\end{center}
\caption{Number of vacancies found in different-sized vacancy clusters by the W-S method. Solid lines: full MD simulations. Different shades of blue squares: the vacancy evolution during CA. Open star markers: CRA simulations only. Gray filled star markers: 30 ps relaxation at room temperature after the CRA. Black filled star markers: after 50 ns TA at room temperature.}
\label{Va_WS_clusterSize}
\end{figure*}

Fig.~\ref{VID_clusterSize} shows the number of vacancies found in different-sized clusters by the VID method. Compared to Fig.~\ref{Va_WS_clusterSize}, the vacancies obtained by W-S method, except for the discontinuous fluctuation of the largest size clusters (Fig.~\ref{Va_WS_clusterSize}(f)), the evolution of vacancy clusters with different sizes detected by both methods show good agreements between the full MD and CRA+CA cases. As mentioned in the main text, the defects are often spatially separated from each other utilizing only CRA. This can be well observed in Supplementary Figs.~\ref{VID_clusterSize}(a) and~\ref{Va_WS_clusterSize}(a): The number of mono-vacancies is observed to decrease with the subsequent CA, and finally stabilize at the saturation level. Di-vacancies are also initially decreasing during CA for the CA-35 set, and after around 30 times of CA, the vacancy concentration then climbs to the saturation level of full MD. In Fig.~\ref{VID_clusterSize}(b), di-vacancy clusters in the CA-50 set initially drops quickly to the full MD level and remained almost unchanged in the subsequent 1600 times irradiation. The other higher initial dose sets showed a similar trend as the CA-50 set, but required more CA steps before converging to the full MD results. For the tri-vacancy case, the two sets with the lowest initial doses show a climbing trend during CA, while the value of CA-105 set fluctuates at the saturation level throughout the whole process, according to the VID method as shown in Fig.~\ref{VID_clusterSize}(c). Notably, the W-S method with a cluster size of 4-5 also exhibits the same trend as observed in the VID method, where the two lowest initial dose sets display an upward trend while the CA-105 case remains at a fluctuation level. By analogy, until the 10+ size set, the emergence of large clusters shows an upward trend under all five sets of different initial doses. It needs to be emphasized again that the CA is of great importance to regress the evolution of vacancy clusters to the correct saturation level, including the formation and evolution of interstitial clusters as well. For the interstitial clustering discussion, please see the main text. 

\section{Strain and corresponding defect evolution}
\begin{figure*}
\begin{center} 
\subfloat[Atomic strain]{\includegraphics[width=.35\linewidth]{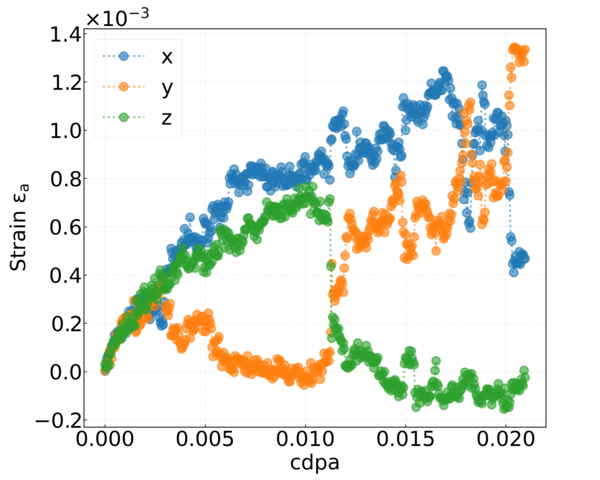}} 
\hspace{10mm}
\subfloat[Defects]{\includegraphics[width=.35\linewidth]{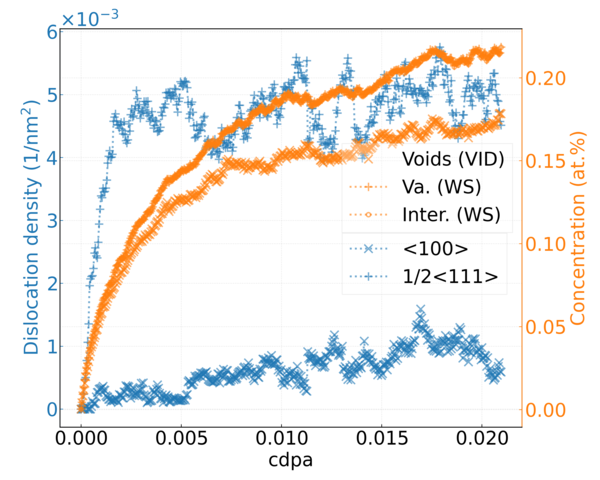}}
\end{center}
\caption{Atomic strain (a) and defect (b) evolution (point defects detected by the W-S and VID method, and dislocations by DXA) for the 1st full MD case.}
\label{eps30_1}
\end{figure*}

\begin{figure*}
\begin{center} 
\subfloat[Atomic strain]{\includegraphics[width=.35\linewidth]{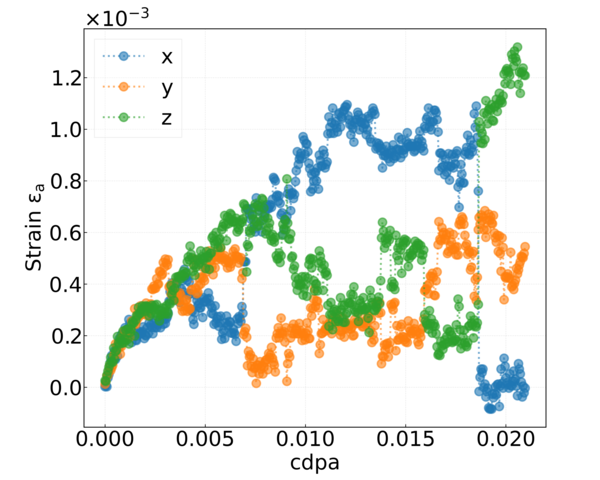}} 
\hspace{10mm}
\subfloat[Defects]{\includegraphics[width=.35\linewidth]{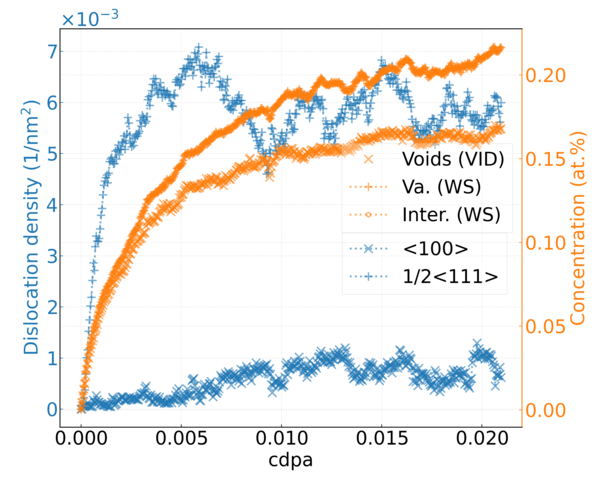}}
\end{center}
\caption{Atomic strain (a) and defect (b) evolution (point defects detected by the W-S and VID method, and dislocations by DXA) for the 2nd full MD case.}
\label{eps30_2}
\end{figure*}

\begin{figure*}
\begin{center} 
\subfloat[Atomic strain]{\includegraphics[width=.35\linewidth]{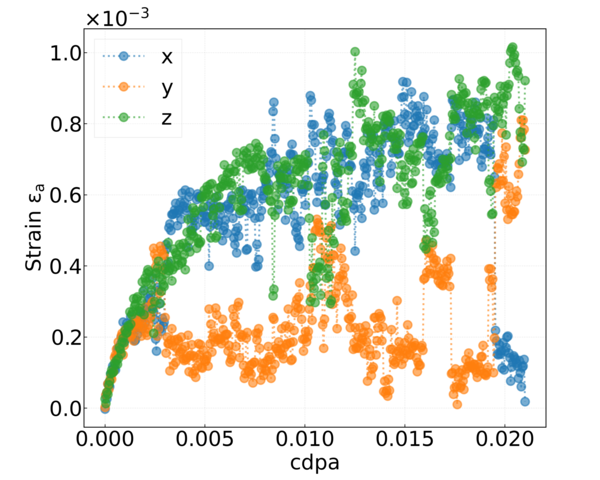}} 
\hspace{10mm}
\subfloat[Defects]{\includegraphics[width=.35\linewidth]{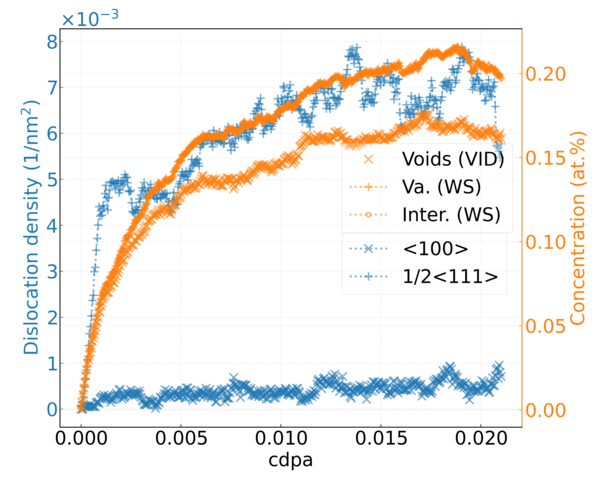}}
\end{center}
\caption{Atomic strain (a) and defect (b) evolution (point defects detected by the W-S and VID method, and dislocations by DXA) for the 3rd full MD case.}
\label{eps30_3}
\end{figure*}

\begin{figure*}
\begin{center} 
\subfloat[Atomic strain]{\includegraphics[width=.35\linewidth]{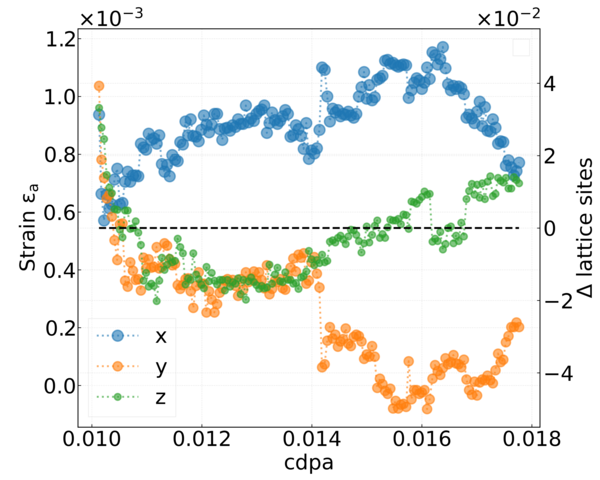}} 
\hspace{10mm}
\subfloat[Defects]{\includegraphics[width=.35\linewidth]{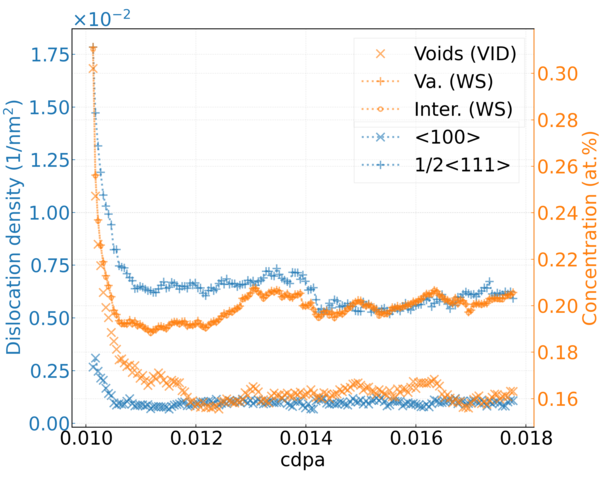}}
\end{center}
\caption{Atomic strain (a) and defect (b) evolution (point defects detected by the W-S and VID method, and dislocations by DXA) for the 1st CRA+CA case at a starting dose of 0.01 cdpa.}
\label{eps35_1}
\end{figure*}

\begin{figure*}
\begin{center} 
\subfloat[Atomic strain]{\includegraphics[width=.35\linewidth]{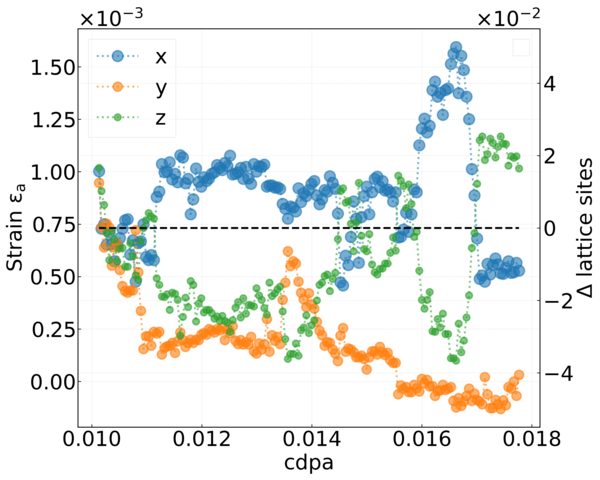}} 
\hspace{10mm}
\subfloat[Defects]{\includegraphics[width=.35\linewidth]{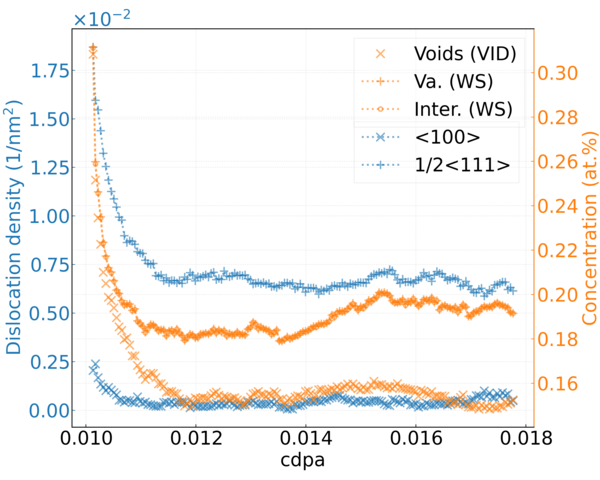}}
\end{center}
\caption{Atomic strain (a) and defect (b) evolution (point defects detected by the W-S and VID method, and dislocations by DXA) for the 2nd CRA+CA case at a starting dose of 0.01 cdpa.}
\label{eps35_2}
\end{figure*}

\begin{figure*}
\begin{center} 
\subfloat[Atomic strain]{\includegraphics[width=.35\linewidth]{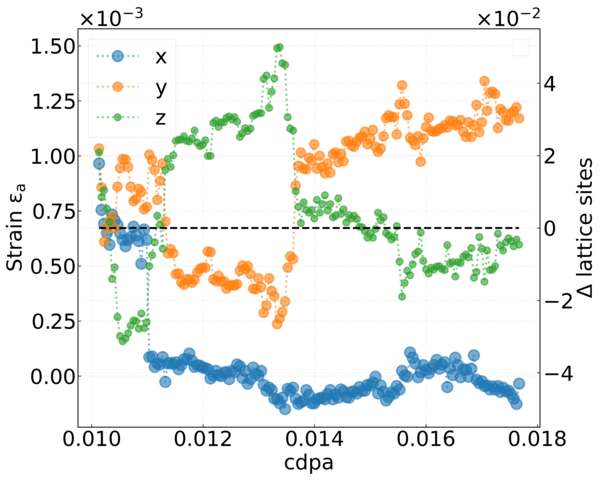}} 
\hspace{10mm}
\subfloat[Defects]{\includegraphics[width=.35\linewidth]{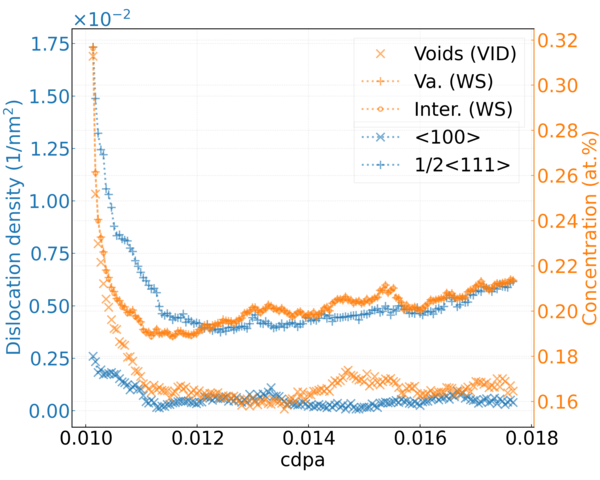}}
\end{center}
\caption{Atomic strain (a) and defect (b) evolution (point defects detected by the W-S and VID method, and dislocations by DXA) for the 3rd CRA+CA case at a starting dose of 0.01 cdpa.}
\label{eps35_3}
\end{figure*}

\begin{figure*}
\begin{center} 
\subfloat[Atomic strain]{\includegraphics[width=.35\linewidth]{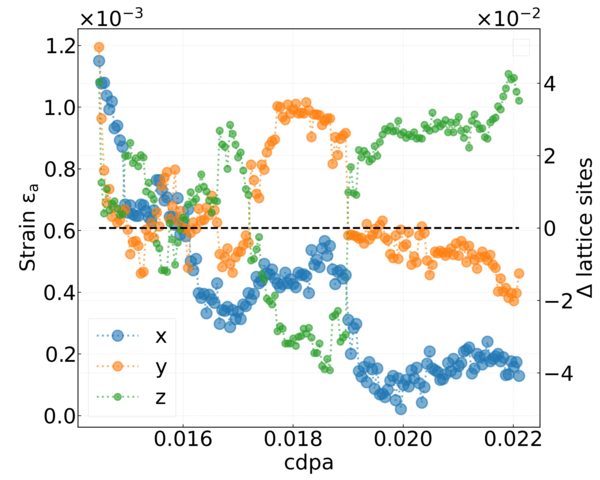}} 
\hspace{10mm}
\subfloat[Defects]{\includegraphics[width=.35\linewidth]{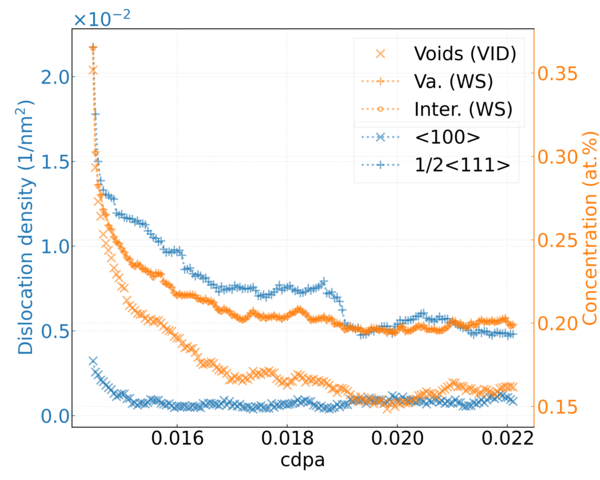}}
\end{center}
\caption{Atomic strain (a) and defect (b) evolution (point defects detected by the W-S and VID method, and dislocations by DXA) for the 1st CRA+CA case at a starting dose of 0.0143 cdpa.}
\label{eps50_1}
\end{figure*}

\begin{figure*}
\begin{center} 
\subfloat[Plastic strain]{\includegraphics[width=.35\linewidth]{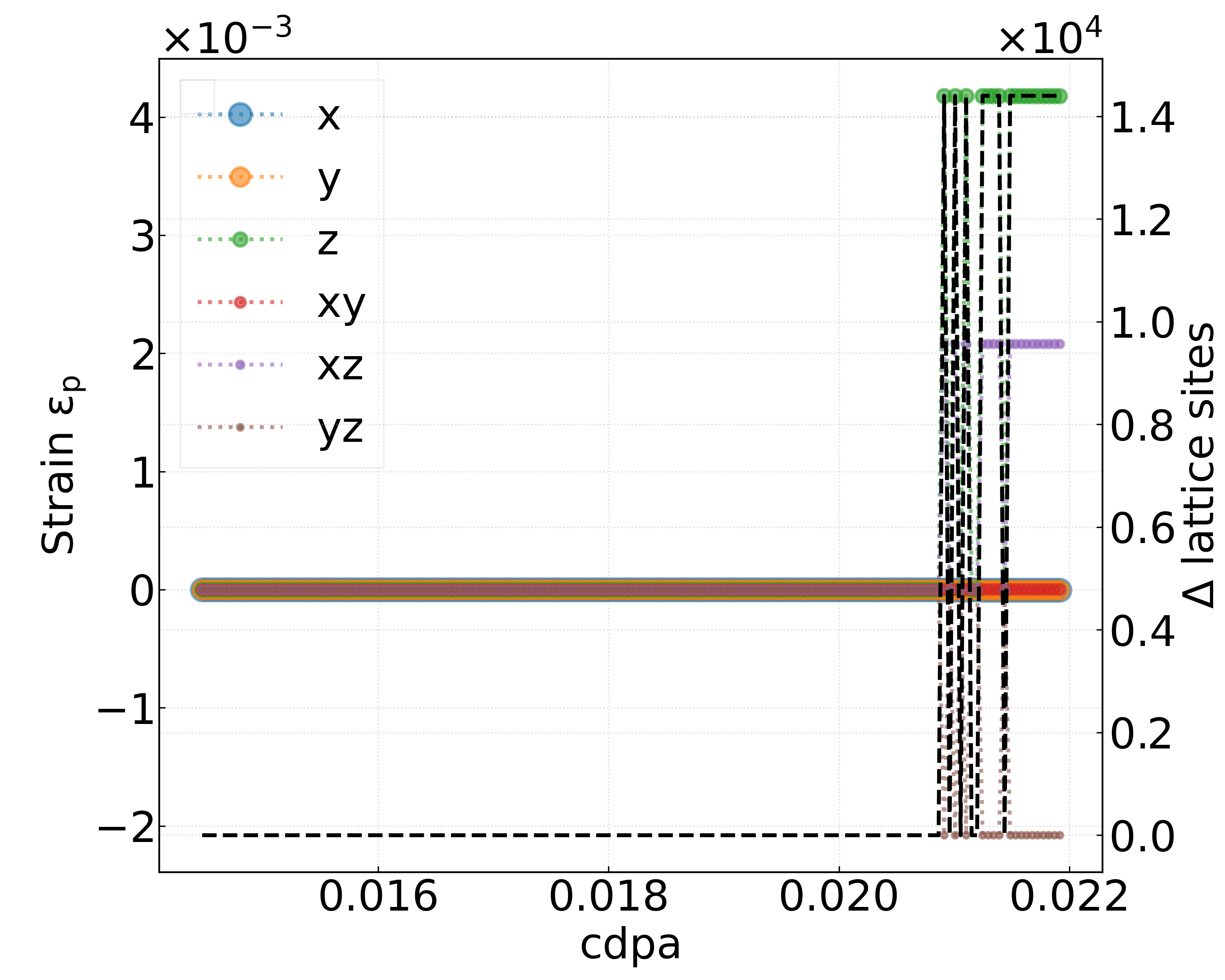}} 
\hspace{10mm}
\subfloat[Atomic strain]{\includegraphics[width=.35\linewidth]{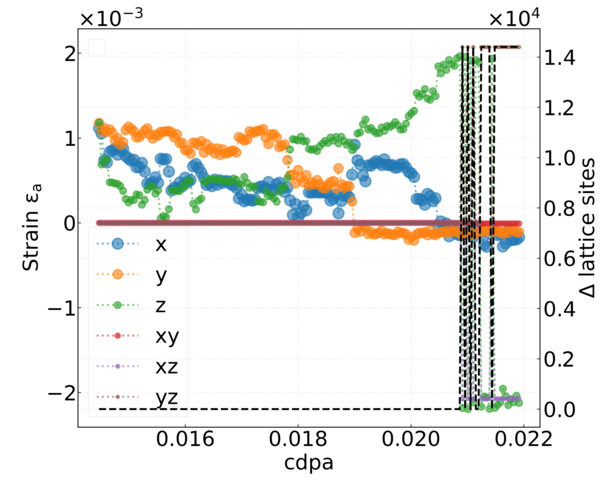}} \\
\subfloat[Rotation]{\includegraphics[width=.35\linewidth]{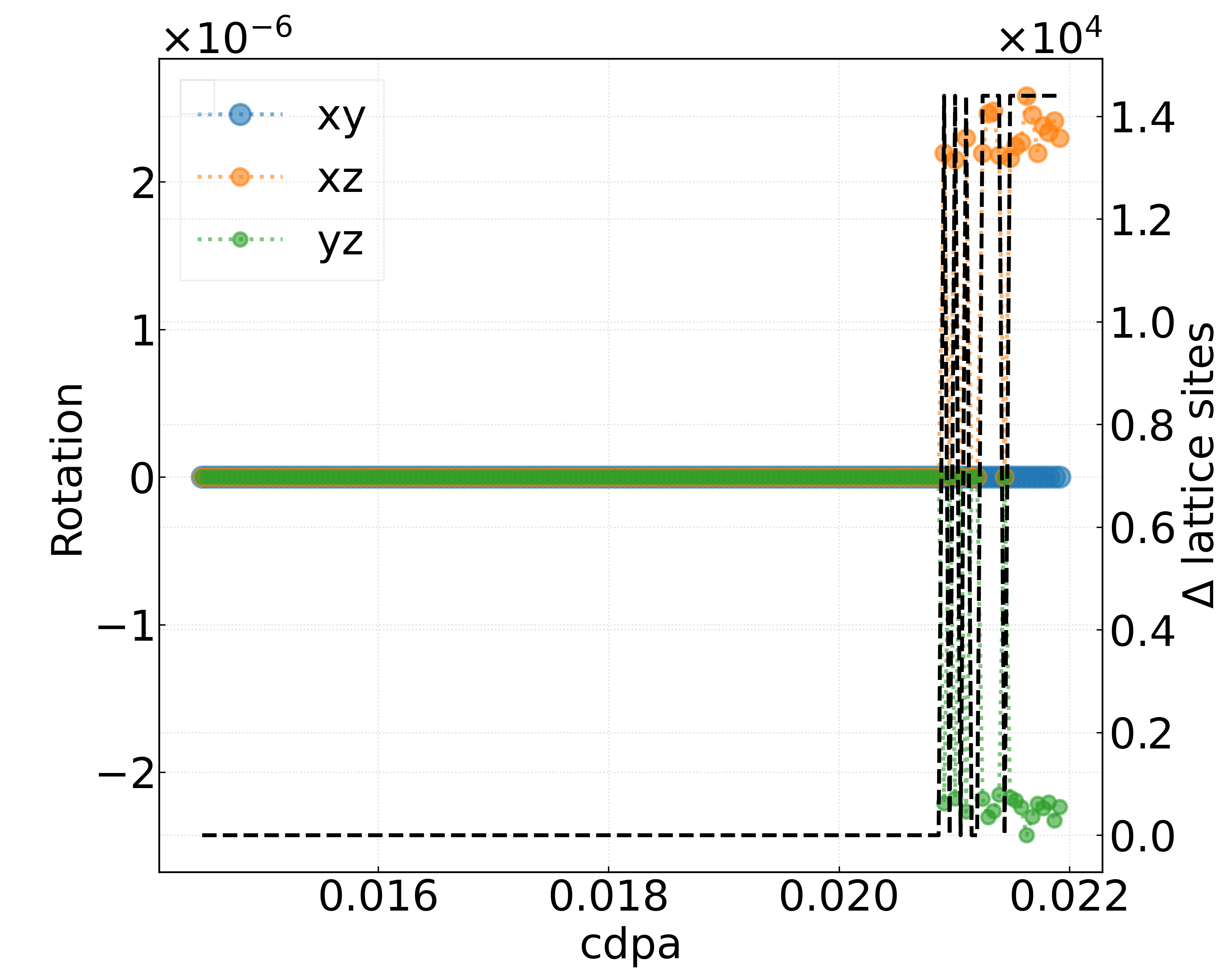}} 
\hspace{10mm}
\subfloat[Defects]{\includegraphics[width=.35\linewidth]{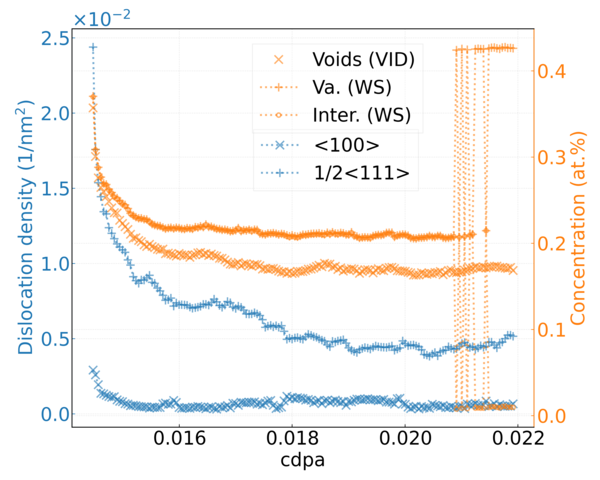}}
\end{center}
\caption{Plastic strain (a), atomic strain (b), rigid rotation (c) and defect (d) evolution (point defects detected by the W-S and VID method, and dislocations by DXA) for the 2nd CRA+CA case at a starting dose of 0.0143 cdpa.}
\label{eps50_2}
\end{figure*}

\begin{figure*}
\begin{center} 
\subfloat[Atomic strain]{\includegraphics[width=.35\linewidth]{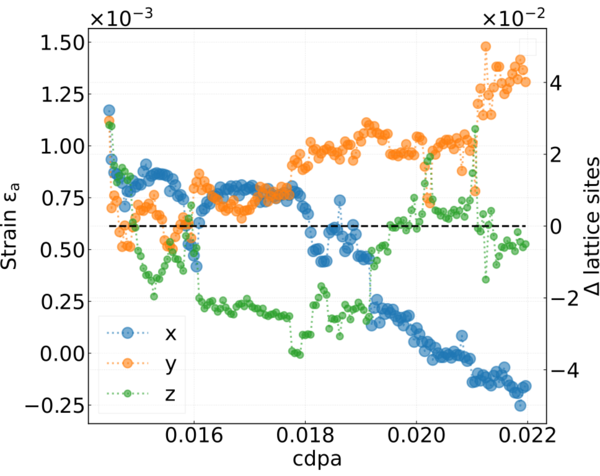}} 
\hspace{10mm}
\subfloat[Defects]{\includegraphics[width=.35\linewidth]{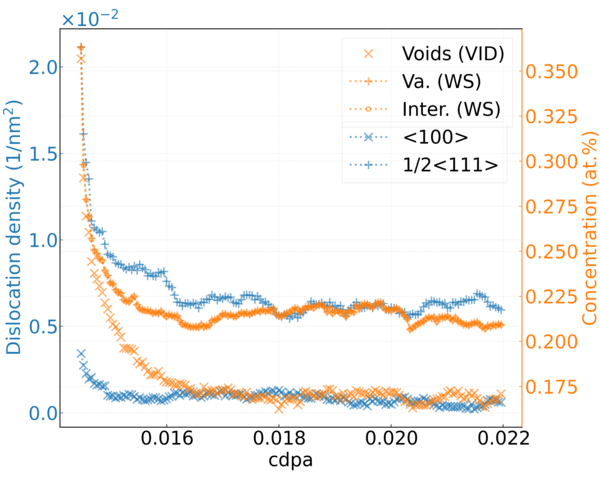}}
\end{center}
\caption{Atomic strain (a) and defect (b) evolution (point defects detected by the W-S and VID method, and dislocations by DXA) for the 3rd CRA+CA case at a starting dose of 0.0143 cdpa.}
\label{eps50_3}
\end{figure*}

\begin{figure*}
\begin{center} 
\subfloat[Atomic strain]{\includegraphics[width=.35\linewidth]{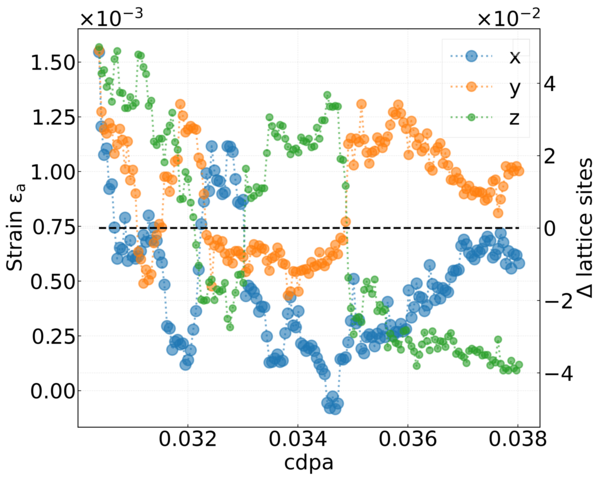}} 
\hspace{10mm}
\subfloat[Defects]{\includegraphics[width=.35\linewidth]{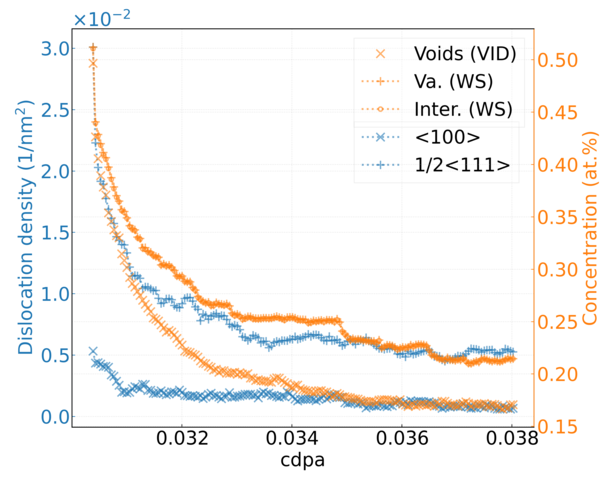}}
\end{center}
\caption{Atomic strain (a) and defect (b) evolution (point defects detected by the W-S and VID method, and dislocations by DXA) for the 1st CRA+CA case at a starting dose of 0.03 cdpa.}
\label{eps105_1}
\end{figure*}

\begin{figure*}
\begin{center} 
\subfloat[Atomic strain]{\includegraphics[width=.35\linewidth]{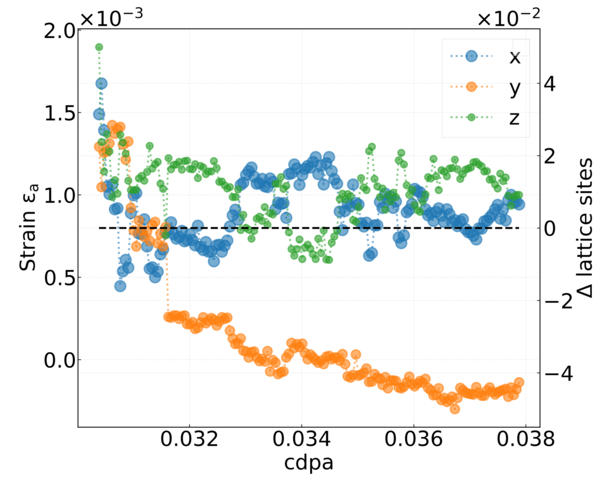}} 
\hspace{10mm}
\subfloat[Defects]{\includegraphics[width=.35\linewidth]{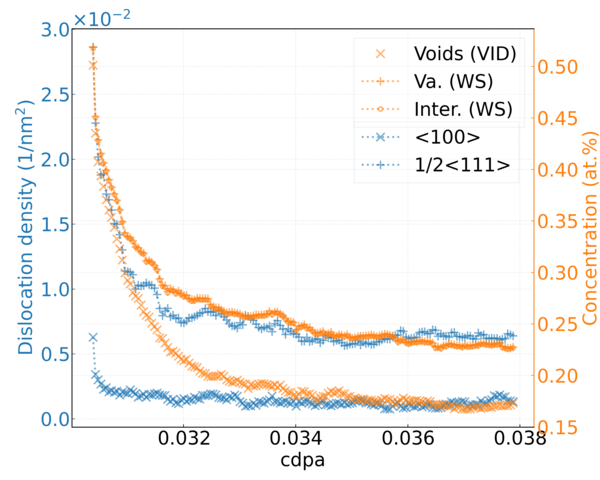}}
\end{center}
\caption{Atomic strain (a) and defect (b) evolution (point defects detected by the W-S and VID method, and dislocations by DXA) for the 2nd CRA+CA case at a starting dose of 0.03 cdpa.}
\label{eps105_2}
\end{figure*}

\begin{figure*}
\begin{center} 
\subfloat[Plastic strain]{\includegraphics[width=.35\linewidth]{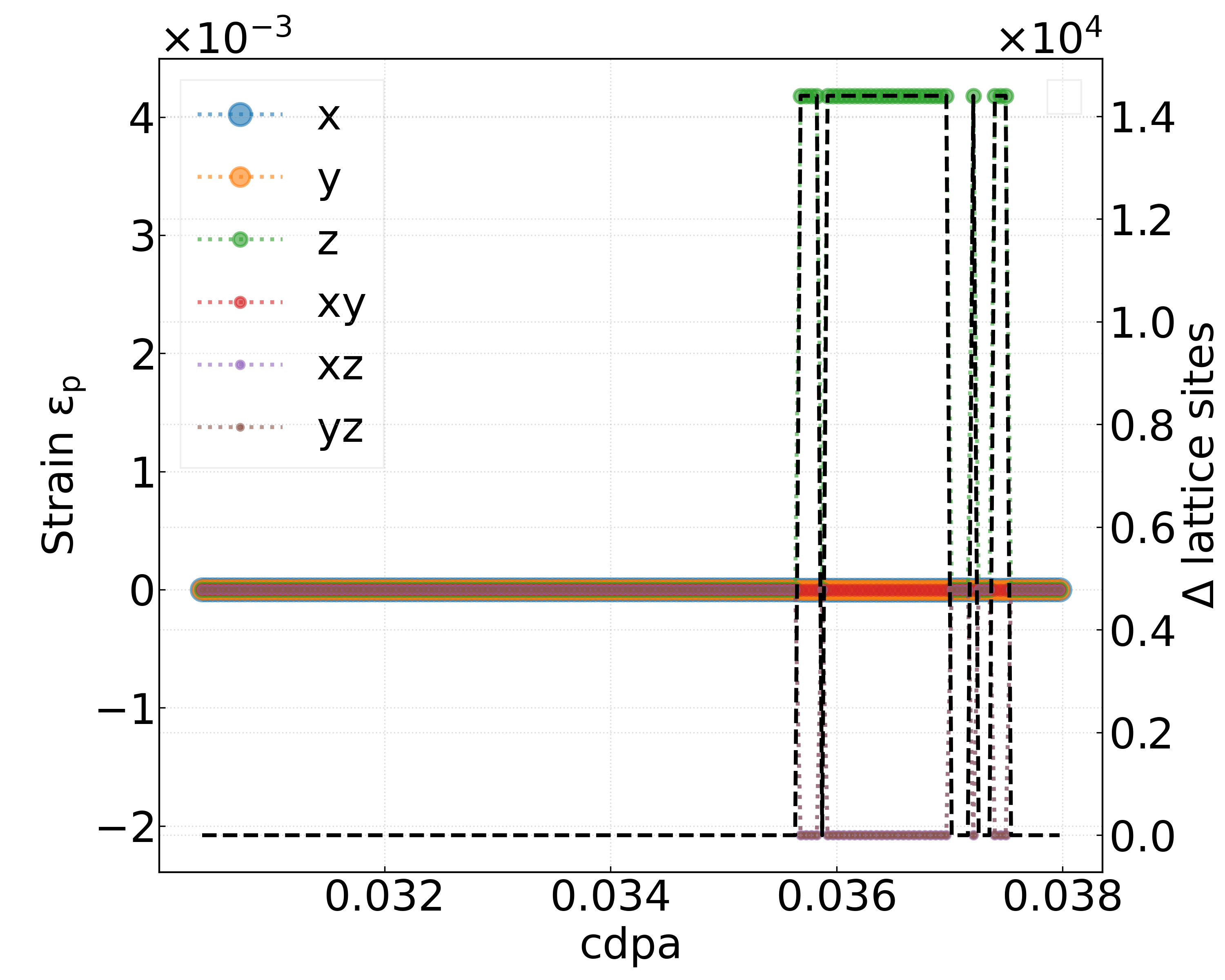}} 
\hspace{10mm}
\subfloat[Atomic strain]{\includegraphics[width=.35\linewidth]{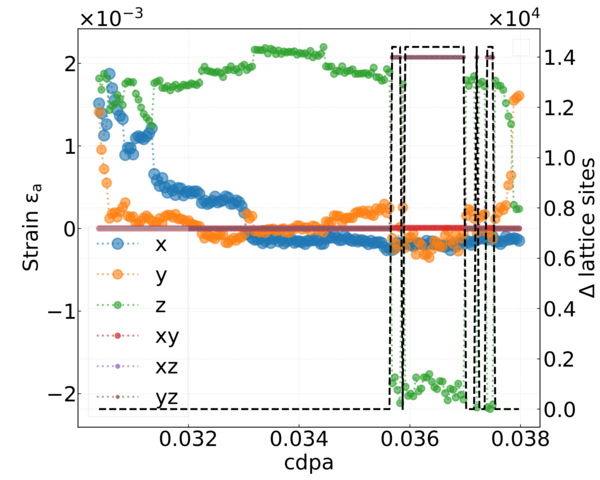}} \\
\subfloat[Rotation]{\includegraphics[width=.35\linewidth]{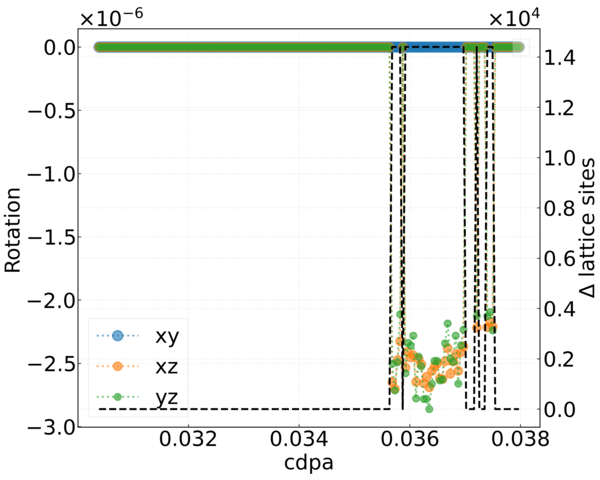}} 
\hspace{10mm}
\subfloat[Defects]{\includegraphics[width=.35\linewidth]{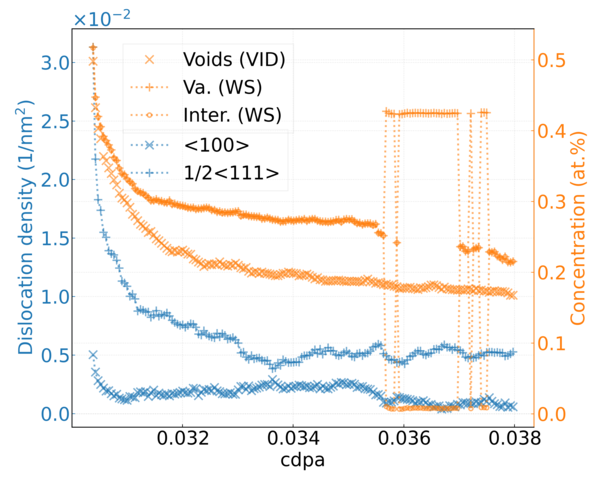}}
\end{center}
\caption{Plastic strain (a), atomic strain (b), rigid rotation (c) and defect (d) evolution (point defects detected by the W-S and VID method, and dislocations by DXA) for the 3rd CRA+CA case at a starting dose of 0.03 cdpa.}
\label{eps105_3}
\end{figure*}

\begin{figure*}
\begin{center} 
\subfloat[Plastic strain]{\includegraphics[width=.35\linewidth]{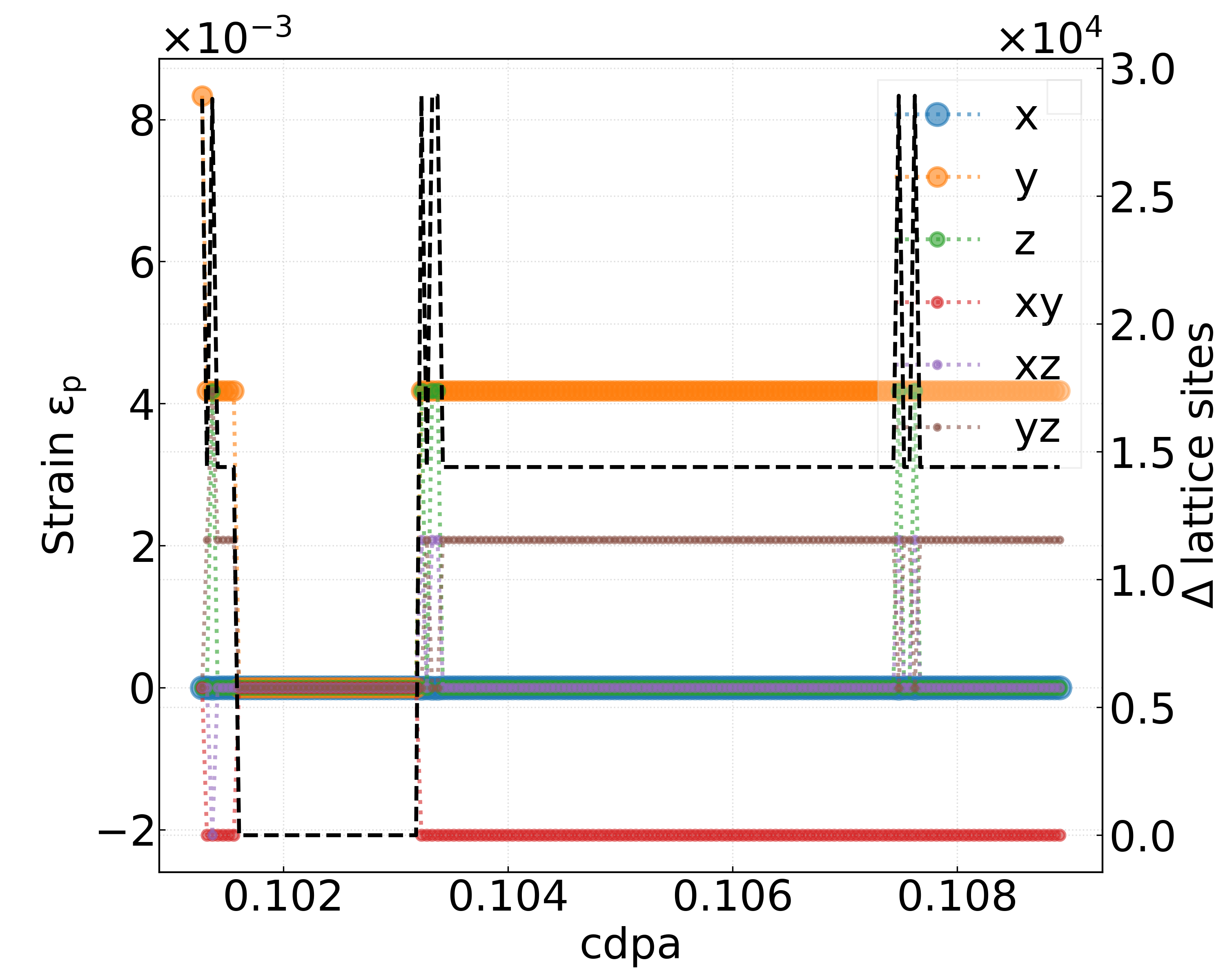}} 
\hspace{10mm}
\subfloat[Atomic strain]{\includegraphics[width=.35\linewidth]{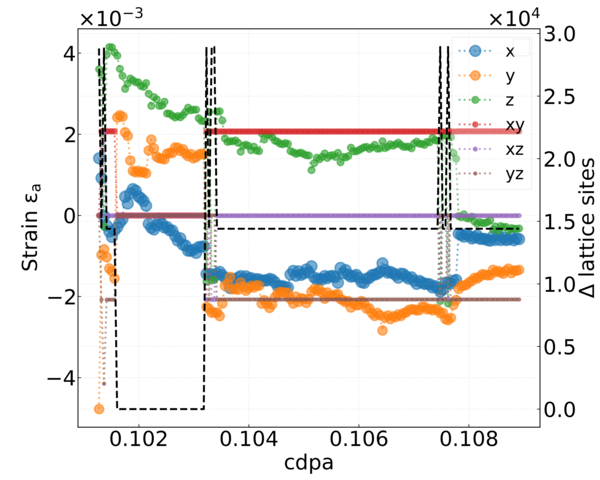}} \\
\subfloat[Rotation]{\includegraphics[width=.35\linewidth]{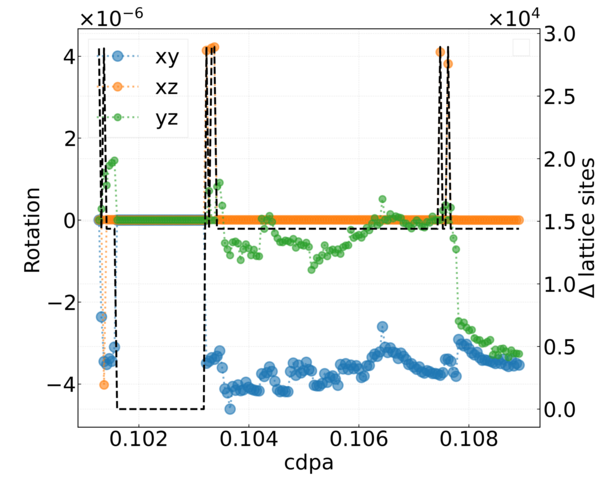}} 
\hspace{10mm}
\subfloat[Defects]{\includegraphics[width=.35\linewidth]{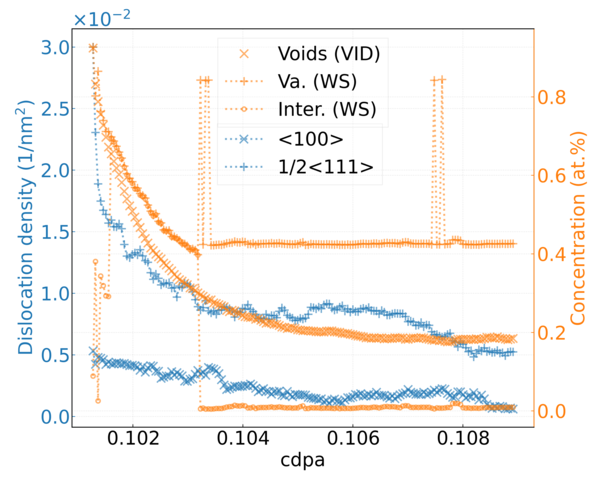}}
\end{center}
\caption{Plastic strain (a), atomic strain (b), rigid rotation (c) and defect (d) evolution (point defects detected by the W-S and VID method, and dislocations by DXA) for the 1st CRA+CA case at a starting dose of 0.1 cdpa.}
\label{eps350_1}
\end{figure*}

\begin{figure*}
\begin{center} 
\subfloat[Plastic strain]{\includegraphics[width=.35\linewidth]{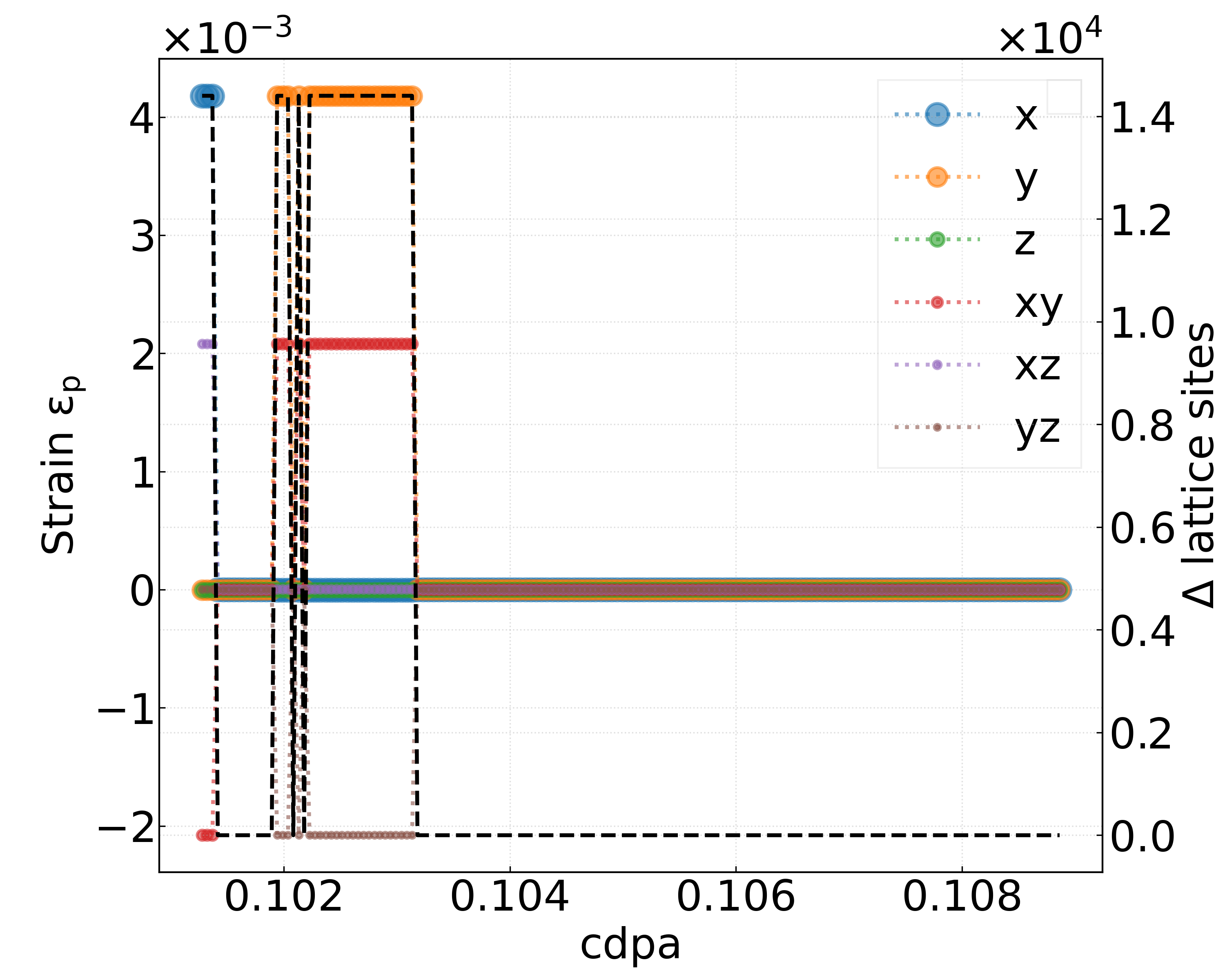}} 
\hspace{10mm}
\subfloat[Atomic strain]{\includegraphics[width=.35\linewidth]{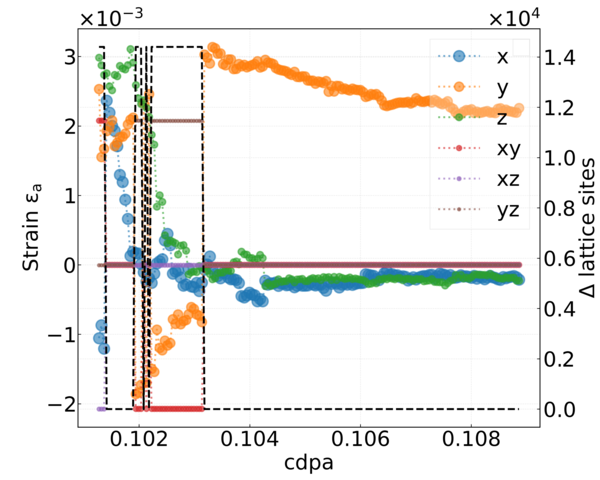}} \\
\subfloat[Rotation]{\includegraphics[width=.35\linewidth]{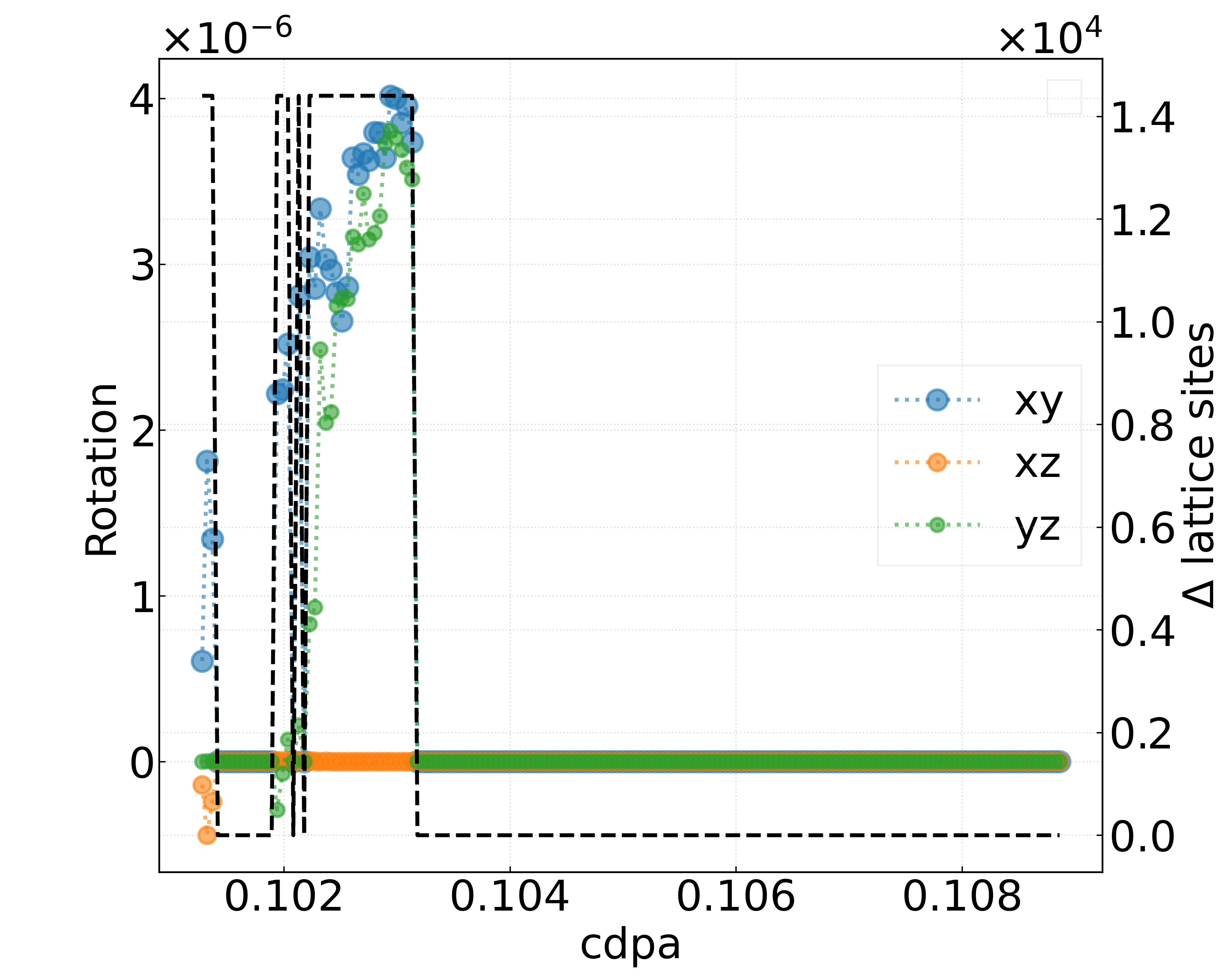}} 
\hspace{10mm}
\subfloat[Defects]{\includegraphics[width=.35\linewidth]{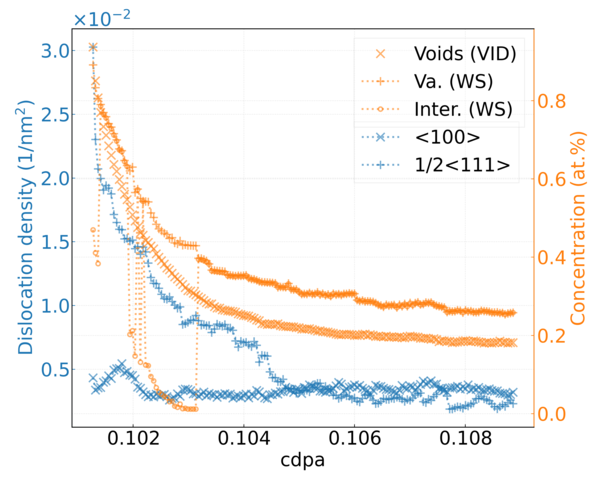}}
\end{center}
\caption{Plastic strain (a), atomic strain (b), rigid rotation (c) and defect (d) evolution (point defects detected by the W-S and VID method, and dislocations by DXA) for the 2nd CRA+CA case at a starting dose of 0.1 cdpa.}
\label{eps350_2}
\end{figure*}

\begin{figure*}
\begin{center} 
\subfloat[Plastic strain]{\includegraphics[width=.35\linewidth]{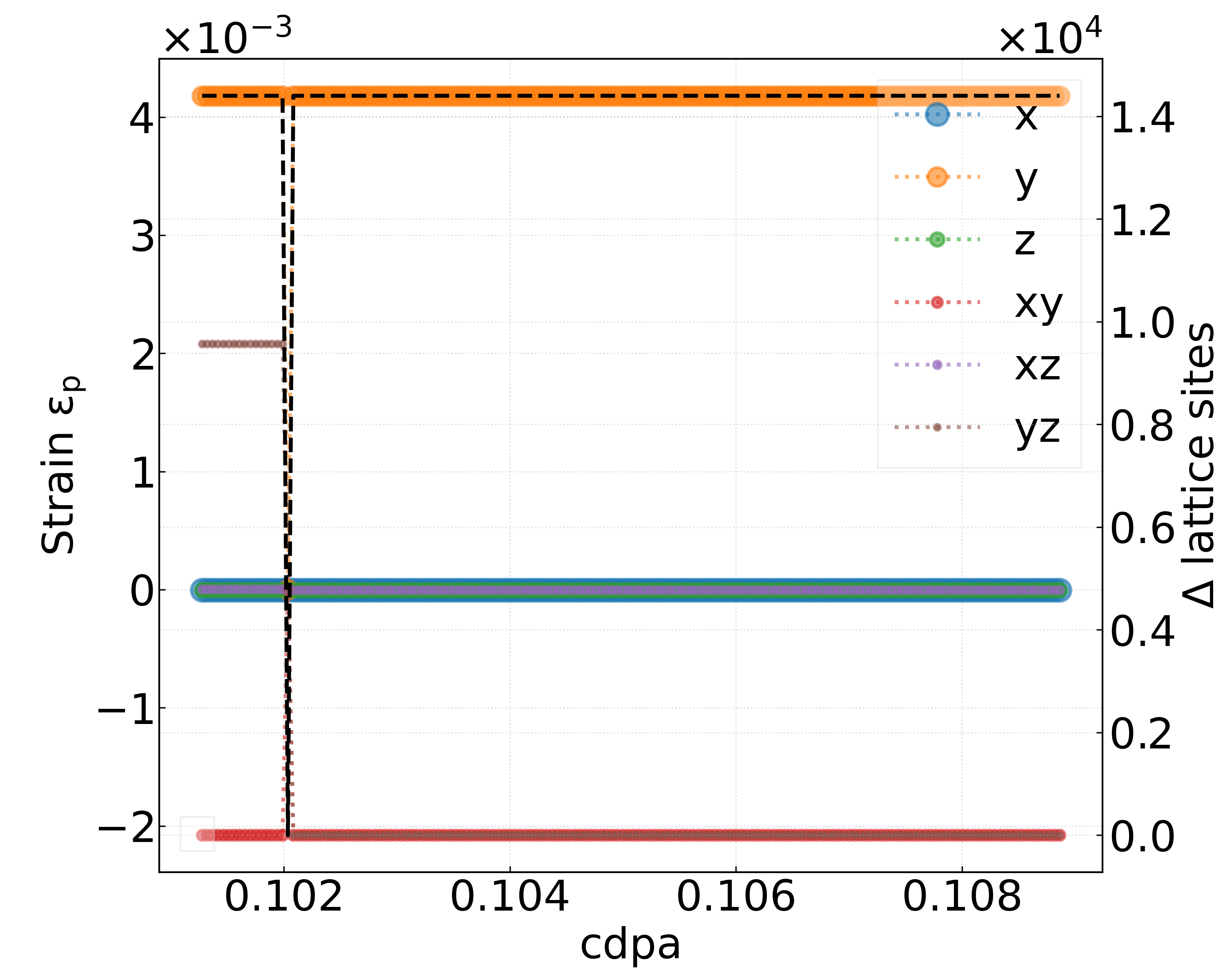}} 
\hspace{10mm}
\subfloat[Atomic strain]{\includegraphics[width=.35\linewidth]{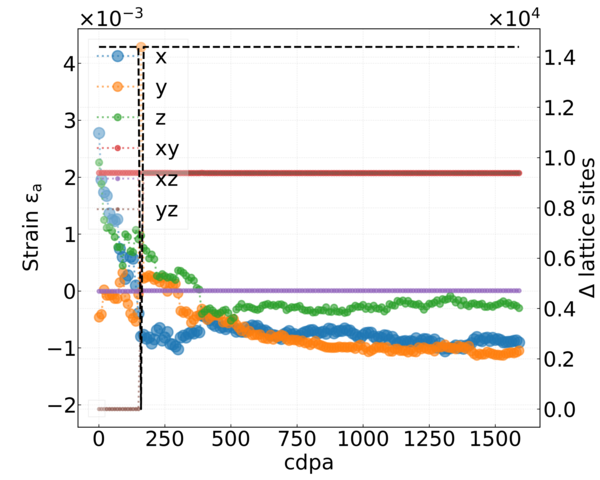}} \\
\subfloat[]{\includegraphics[width=.35\linewidth]{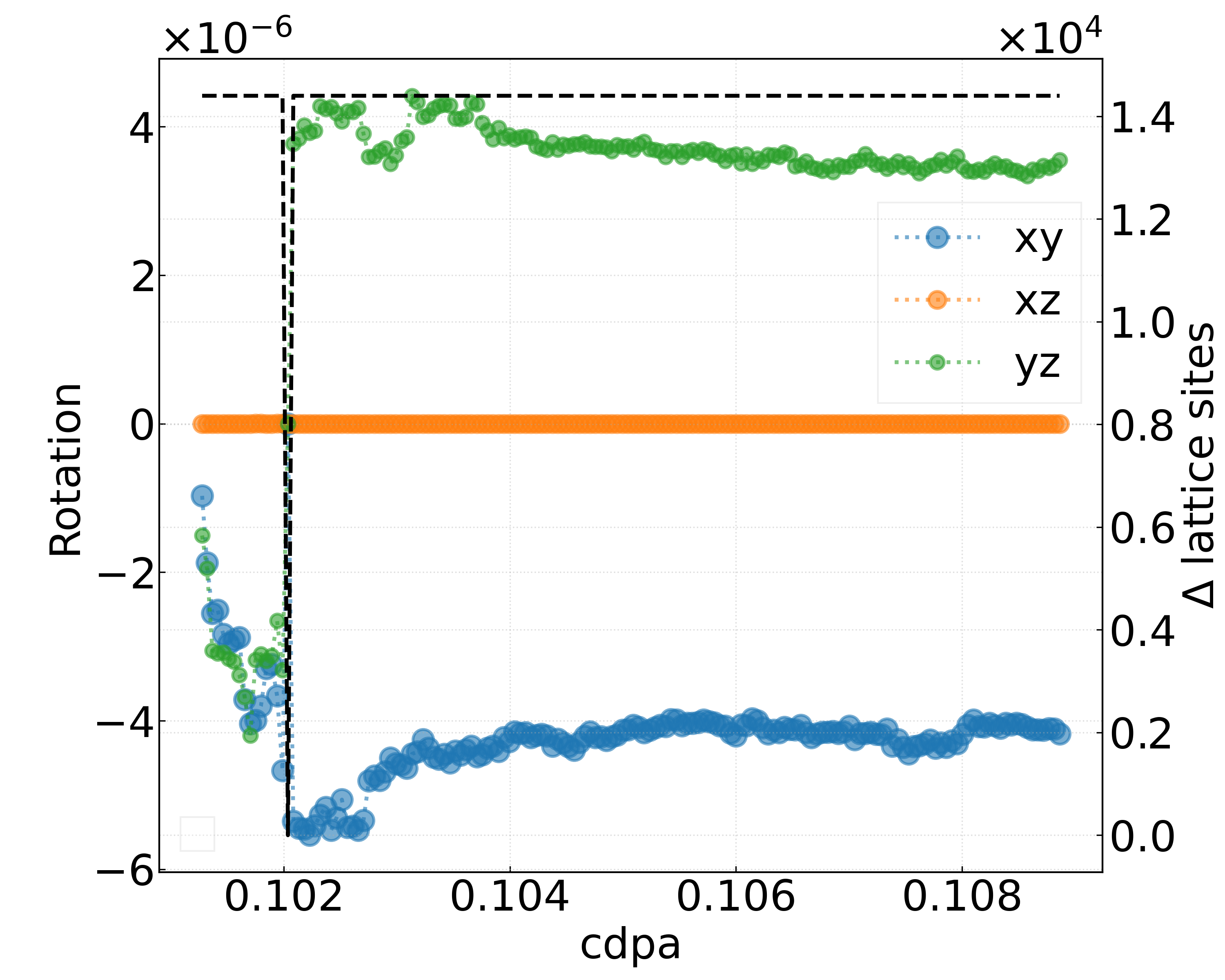}} 
\hspace{10mm}
\subfloat[Defects]{\includegraphics[width=.35\linewidth]{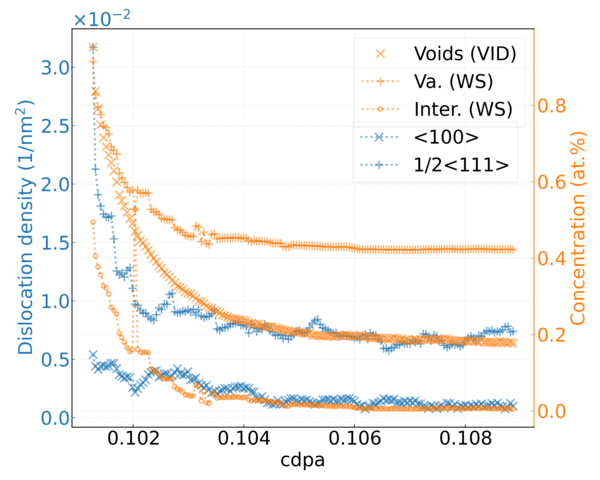}}
\end{center}
\caption{Plastic strain (a), atomic strain (b), rigid rotation (c) and defect (d) evolution (point defects detected by the W-S and VID method, and dislocations by DXA) for the 3rd CRA+CA case at a starting dose of 0.1 cdpa.}
\label{eps350_3}
\end{figure*}

\begin{figure*}
\begin{center} 
\subfloat[Plastic strain]{\includegraphics[width=.35\linewidth]{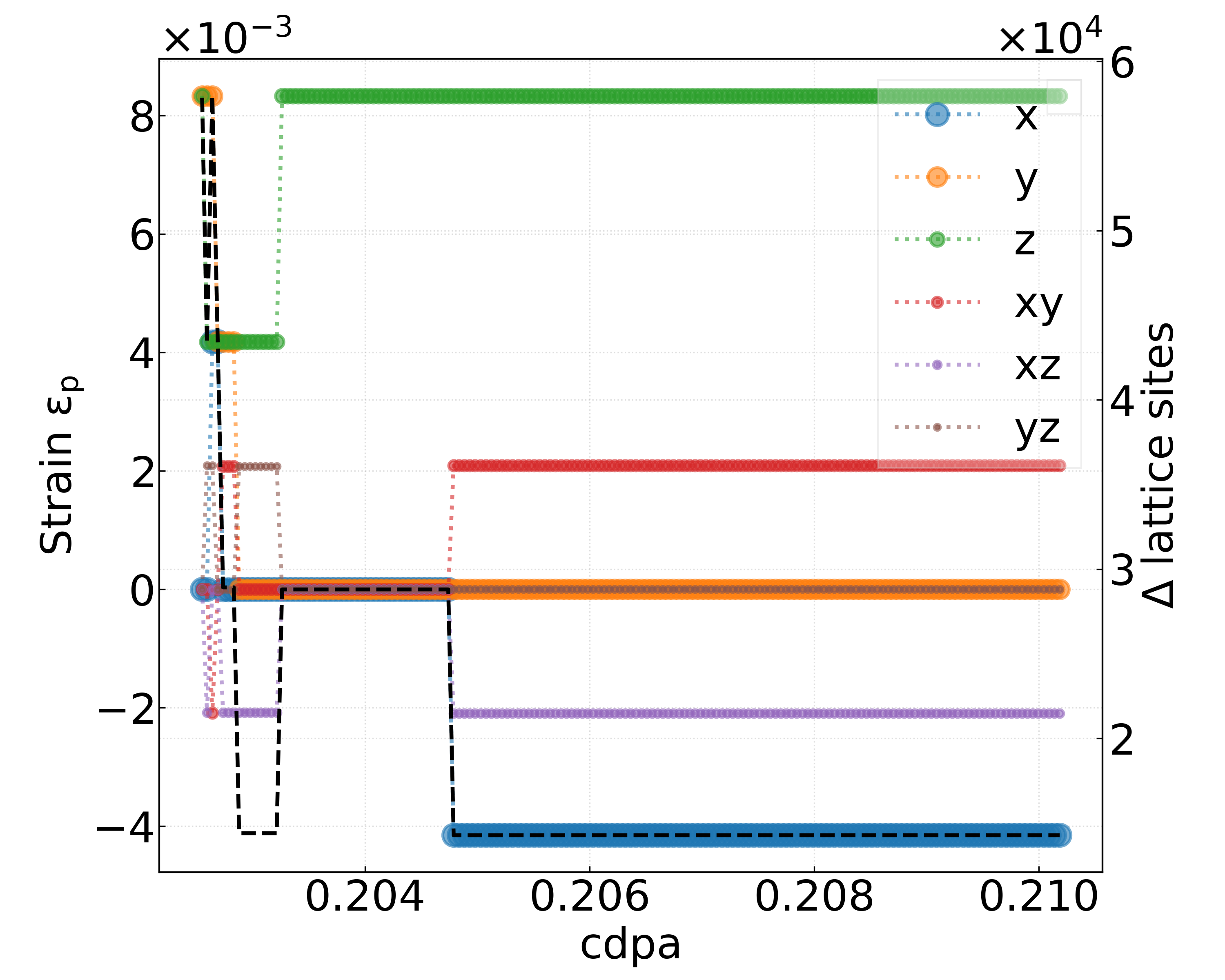}} 
\hspace{10mm}
\subfloat[Atomic strain]{\includegraphics[width=.35\linewidth]{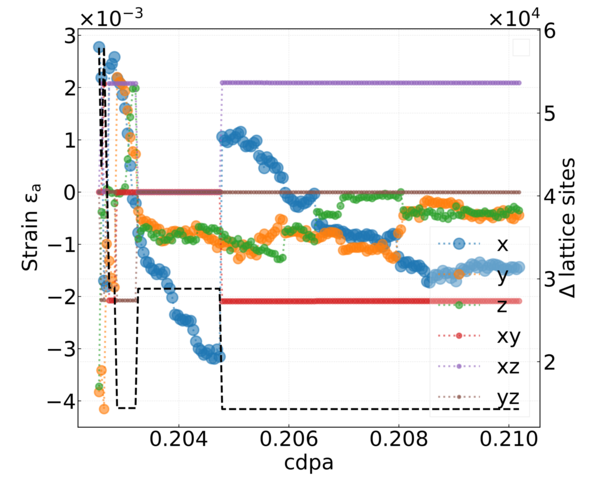}} \\
\subfloat[Rotation]{\includegraphics[width=.35\linewidth]{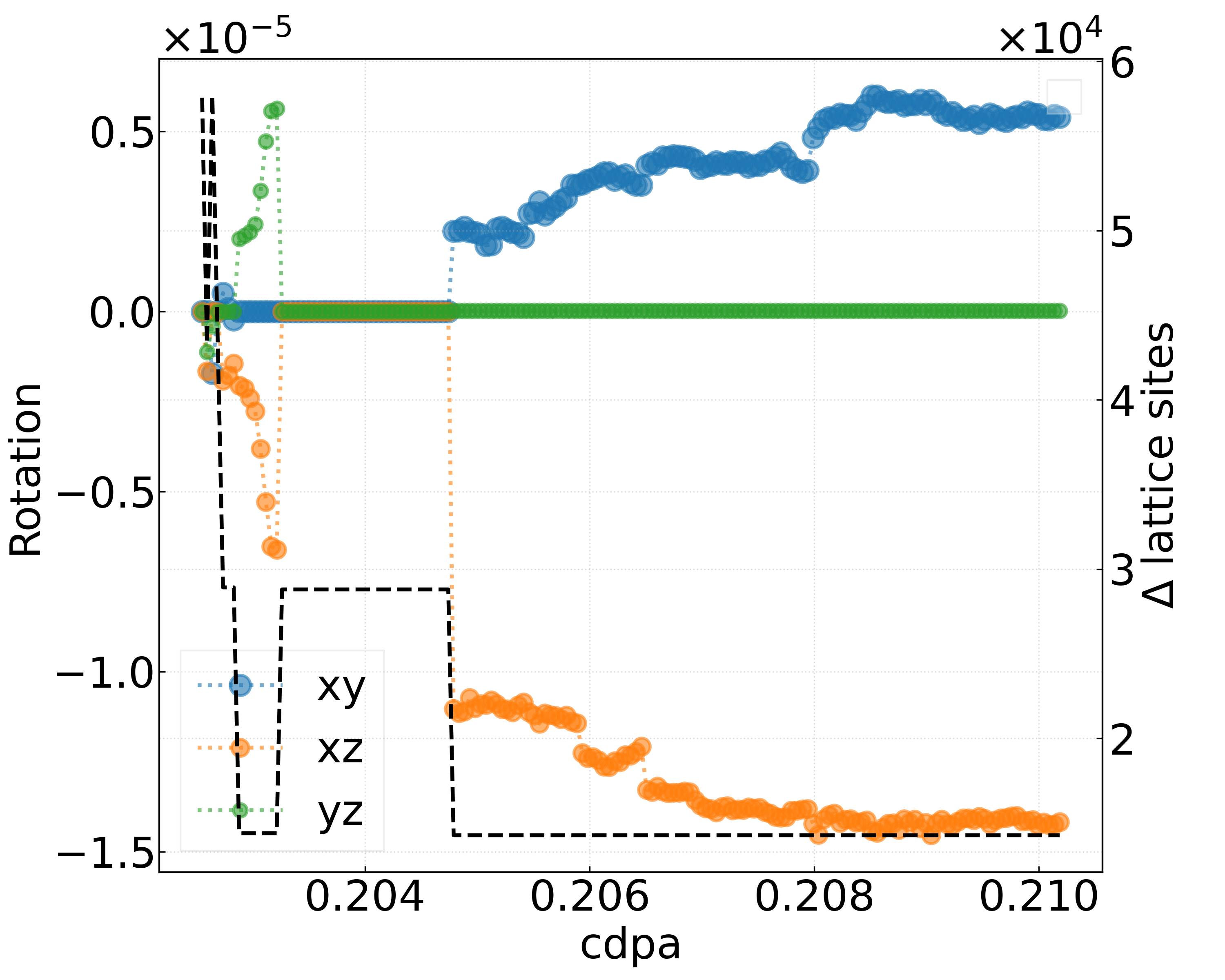}} 
\hspace{10mm}
\subfloat[Defects]{\includegraphics[width=.35\linewidth]{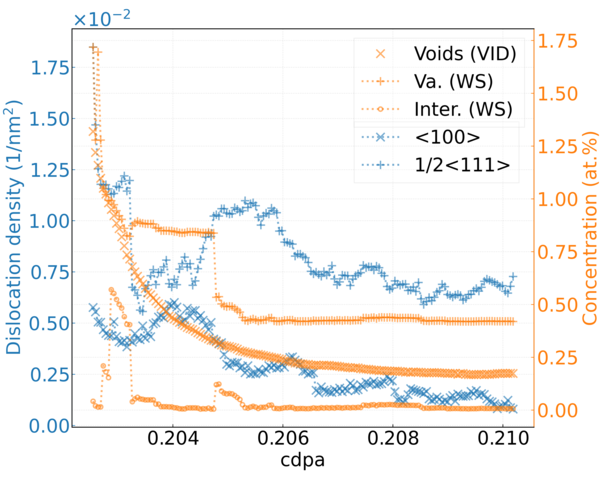}}
\end{center}
\caption{Plastic strain (a), atomic strain (b), rigid rotation (c) and defect (d) evolution (point defects detected by the W-S and VID method, and dislocations by DXA) for the 1st CRA+CA case at a starting dose of 0.2 cdpa.}
\label{eps700_1}
\end{figure*}

\begin{figure*}
\begin{center} 
\subfloat[Plastic strain]{\includegraphics[width=.35\linewidth]{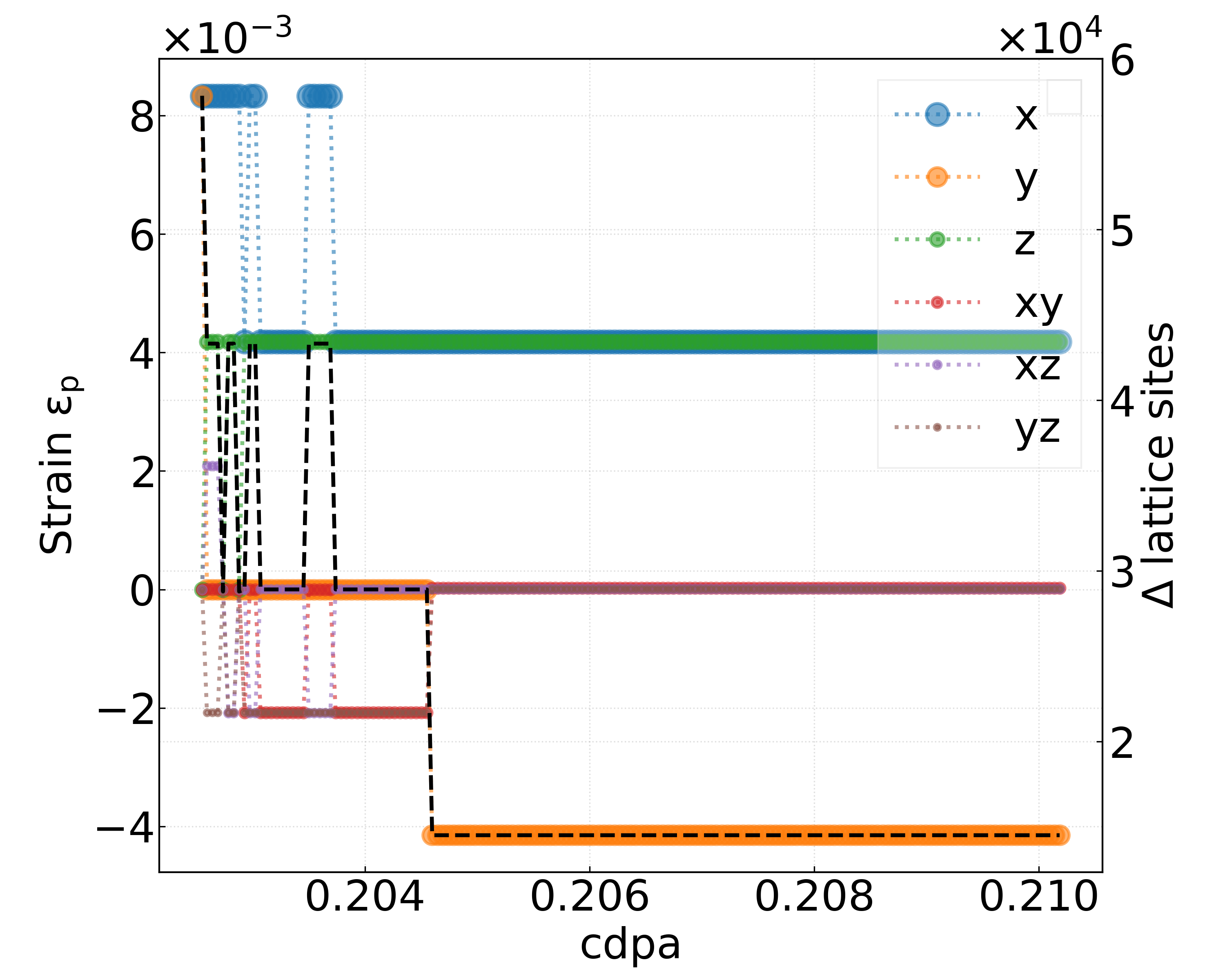}} 
\hspace{10mm}
\subfloat[Atomic strain]{\includegraphics[width=.35\linewidth]{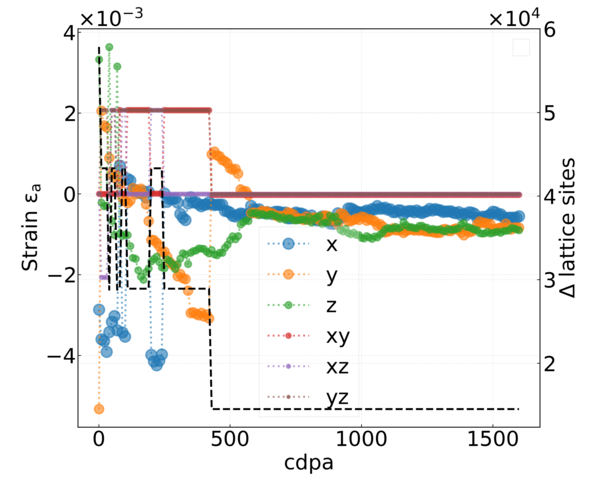}} \\
\subfloat[Rotation]{\includegraphics[width=.35\linewidth]{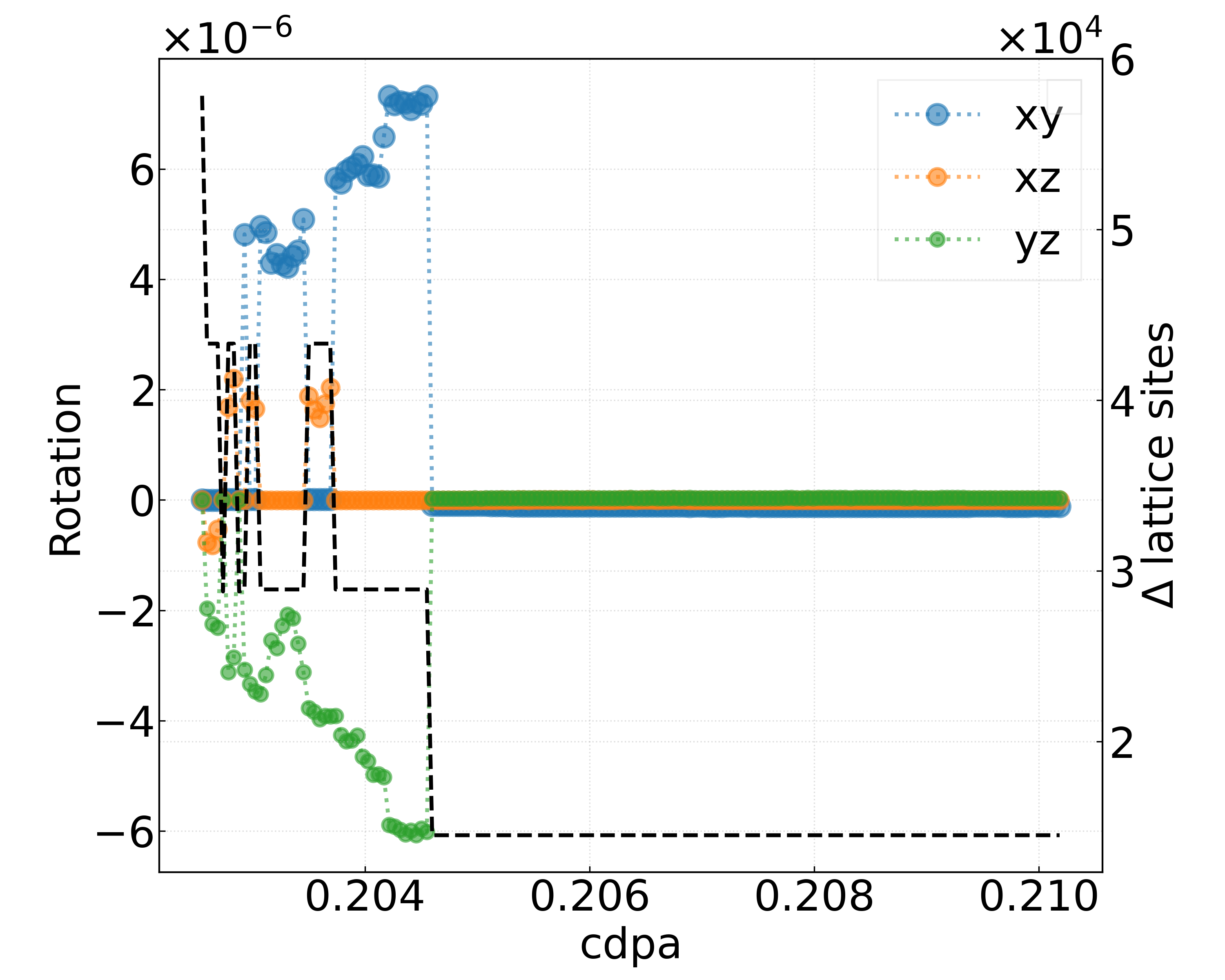}} 
\hspace{10mm}
\subfloat[Defects]{\includegraphics[width=.35\linewidth]{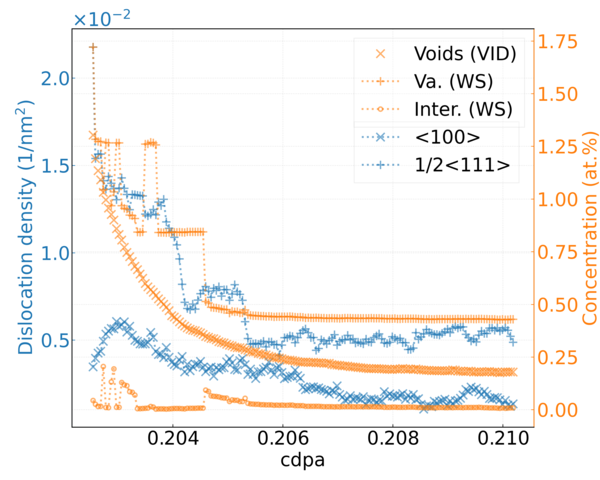}}
\end{center}
\caption{Plastic strain (a), atomic strain (b), rigid rotation (c) and defect (d) evolution (point defects detected by the W-S and VID method, and dislocations by DXA) for the 2nd CRA+CA case at a starting dose of 0.2 cdpa.}
\label{eps700_2}
\end{figure*}

\begin{figure*}
\begin{center} 
\subfloat[Plastic strain]{\includegraphics[width=.35\linewidth]{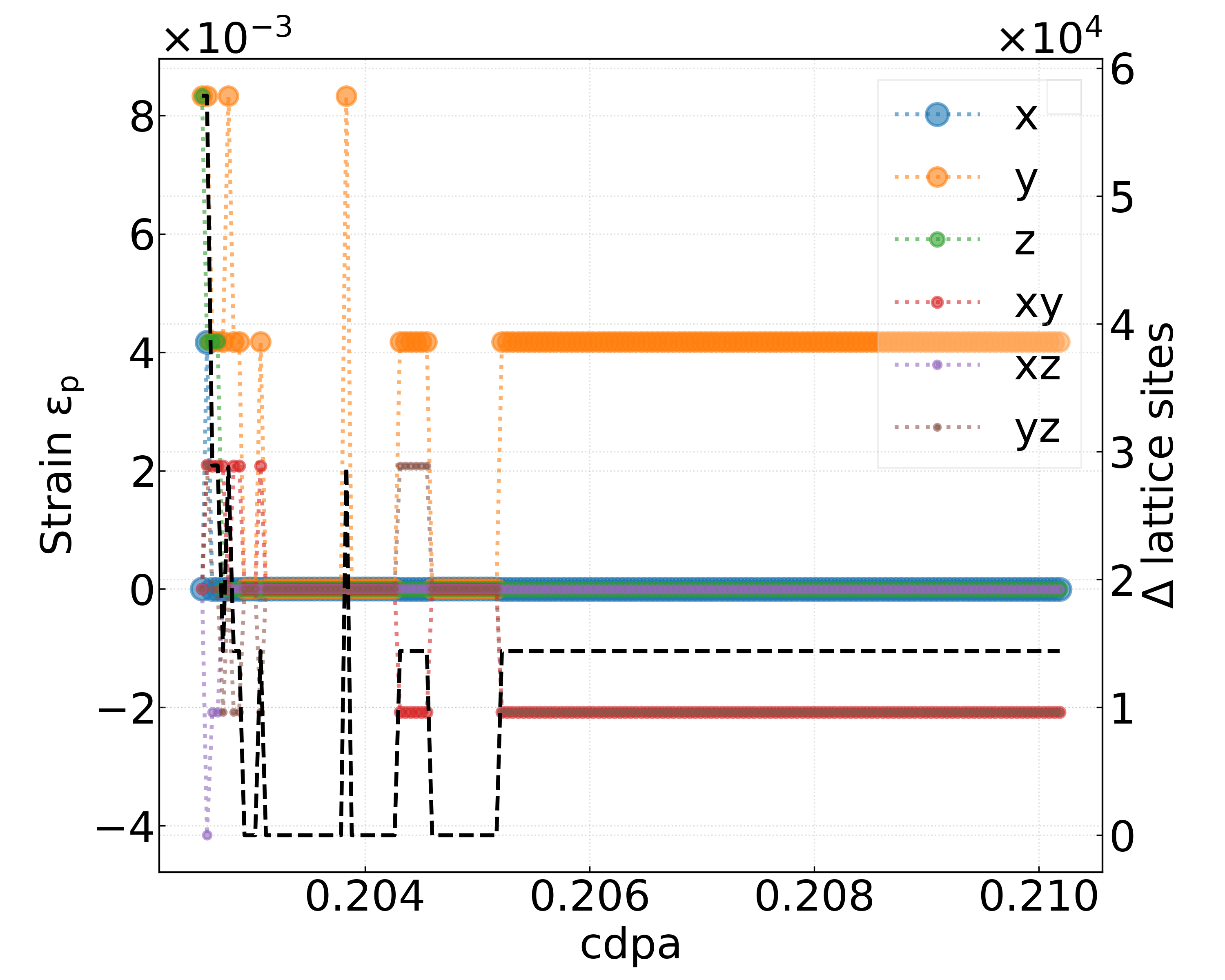}} 
\hspace{10mm}
\subfloat[Atomic strain]{\includegraphics[width=.35\linewidth]{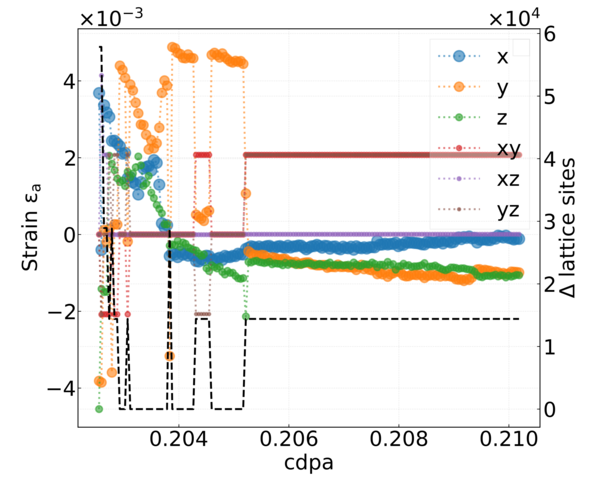}} \\
\subfloat[Rotation]{\includegraphics[width=.35\linewidth]{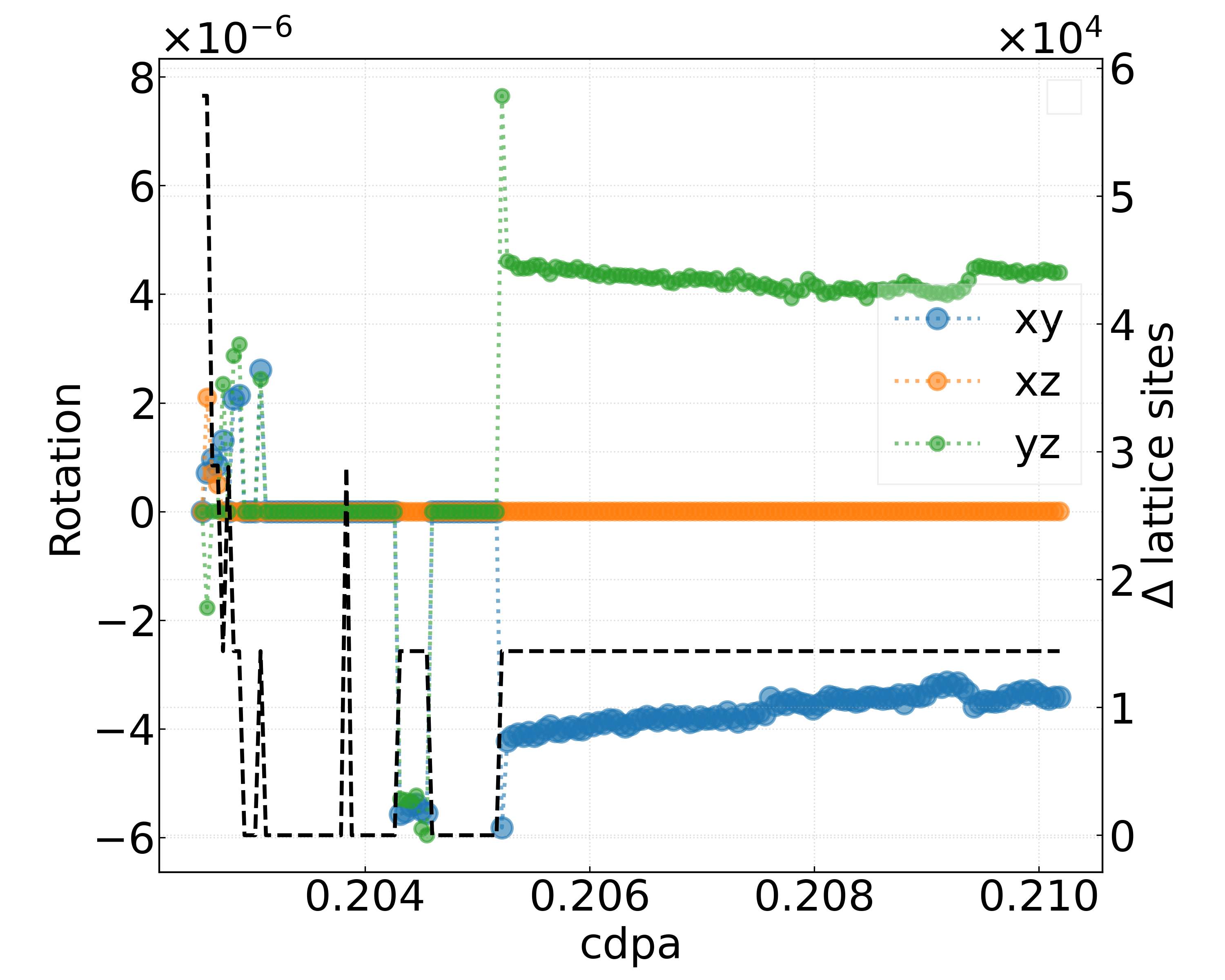}} 
\hspace{10mm}
\subfloat[Defects]{\includegraphics[width=.35\linewidth]{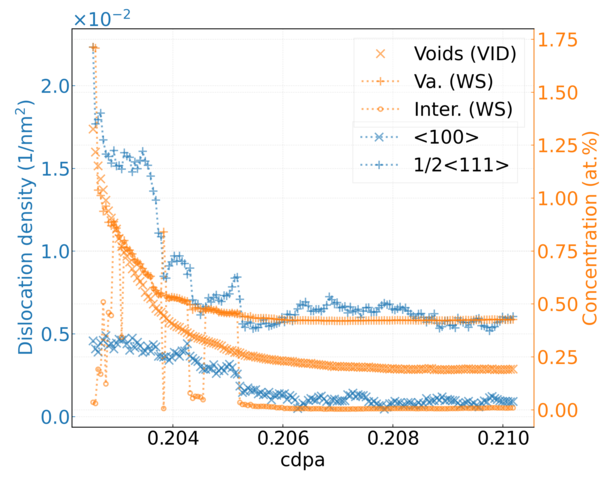}}
\end{center}
\caption{Plastic strain (a), atomic strain (b), rigid rotation (c) and defect (d) evolution (point defects detected by the W-S and VID method, and dislocations by DXA) for the 3rd CRA+CA case at a starting dose of 0.2 cdpa.}
\label{eps700_3}
\end{figure*}

Fig.~\ref{eps30_1} to Fig.~\ref{eps30_3} illustrate strain responses of full MD simulations, which also show dislocation habit plane changes (which introduce the sudden fluctuate of the atomic strain). The corresponding defect evolution was analyzed using the VID and W-S methods to identify vacancies and interstitials. It should be noted that the concentration of vacancies and interstitials remains constant throughout the simulation, while the concentration of voids is slightly lower than the defects detected by the W-S method. The results of the CA simulations are shown in Fig.~\ref{eps35_1} to~\ref{eps700_3}, for all cases investiagted. It can be observed that in the simulations with a starting dose of 0.01, 0.0143, and 0.03 cdpa, the probability of excess plane formation is quite low, due to the lower amount of defects. Out of the three independent simulations conducted for each of these cases, only two cases showed excess plane formation, specifically the second case with a starting dose of 0.0143 cdpa and the third case with a starting dose of 0.03 cdpa. No plastic deformation occurred in the other simulations at this low dose regime. However, when the starting dose exceeded 0.1 cdpa, elasto-plastic deformation and rigid rotation were observed in all simulations, most of them accompanying with swelling event (formation of extra lattice sites).

\bibliographystyle{apsrev4-2}
\bibliography{article2}

\newpage

\begin{table}[t]
\centering
\begin{tabular}{|c|c|c|} 
\hline
$E_{\mathrm{PKA}}$ (keV)    & Potential                                    & $E_{\mathrm{dep}}$ (eV/atom)  \\ 
\hline
\multirow{2}{*}{10}         & MNB (Ref.~\cite{MasonPRM_VoidDetection})  & 7.17 - 8.13                   \\ 
\cline{2-2}\cline{3-3}
                            & AT-ZN (Ref.~\cite{Granberg2022})          & 7.90 - 11.07                  \\ 
\hline
30                          & AT-ZN                                        & 8.27 - 9.55                   \\
\hline
\end{tabular}
\caption{Energy deposition per atom in the cell when the defect concentration reached saturation after CA, according to different interatomic potentials and irradiation energies. Some of the values are taken or derived from the literature~\cite{MasonPRM_VoidDetection,Granberg2022}}
\label{tab:Edeposit}
\end{table}

\begin{figure}[]
\begin{center} 
\subfloat[1/2\hkl<111>]{\includegraphics[width=.45\columnwidth]{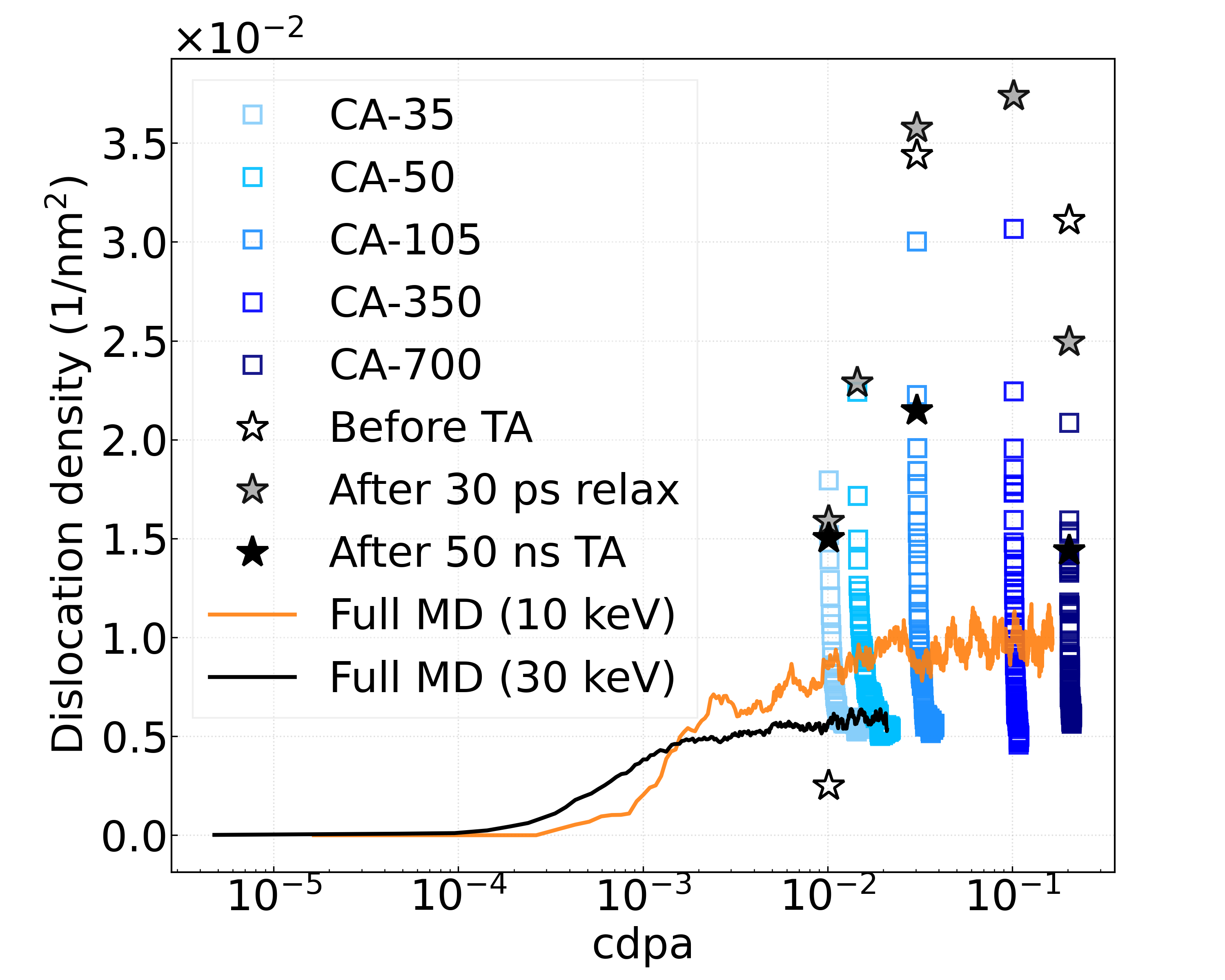}} ~~~~~~~
\subfloat[\hkl<100>]{\includegraphics[width=.45\columnwidth]{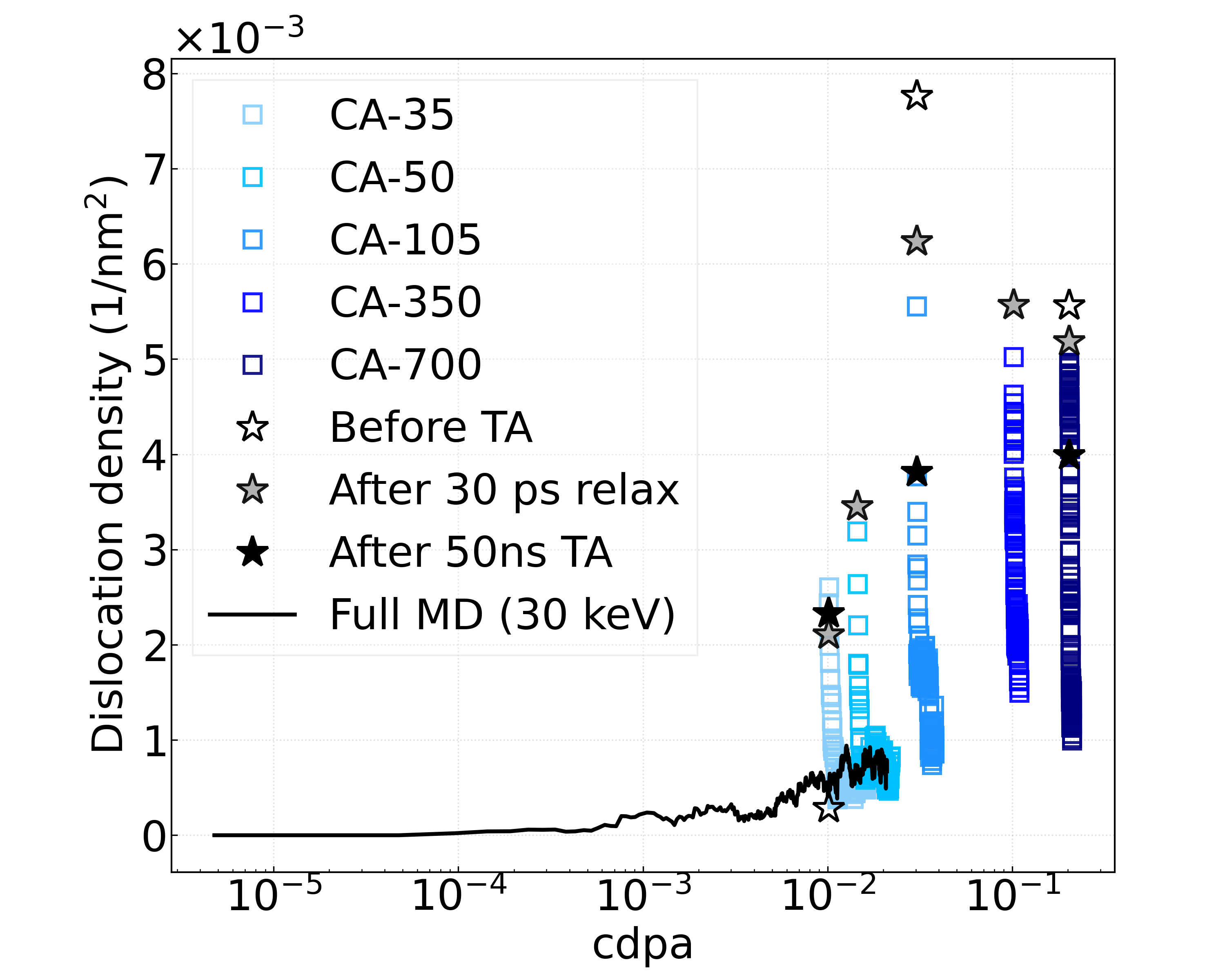}}  \\
\subfloat[Snapshot of interstitials and vacancies]{\includegraphics[width=.4\columnwidth]{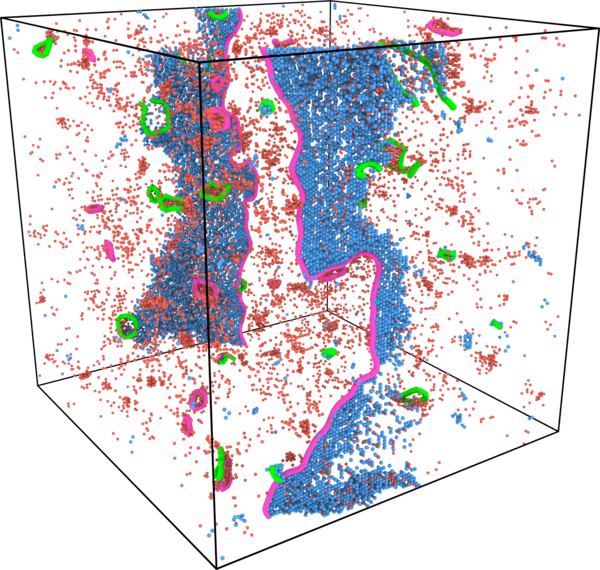}}
\end{center}
\caption{Evolution of the total (a) 1/2\hkl<111> and (b) \hkl<100> type dislocation density as a function of dose. Solid lines: full MD simulations. Different shades of blue squares: the vacancy evolution during CA. Open star markers: CRA simulations only. Gray filled star markers: 30 ps relaxation at room temperature after the CRA. Black filled star markers: after 50 ns TA at room temperature. The orange line data from Ref.~\cite{Granberg2022}. (c) One of the three simulations shows the distribution of interstitial clusters wrapped by \hkl<100> type dislocations, shown as pink lines.}
\label{Dislocation}
\end{figure}

\begin{figure*}
\begin{center} 
\subfloat[size = 1]{\includegraphics[width=.3\linewidth]{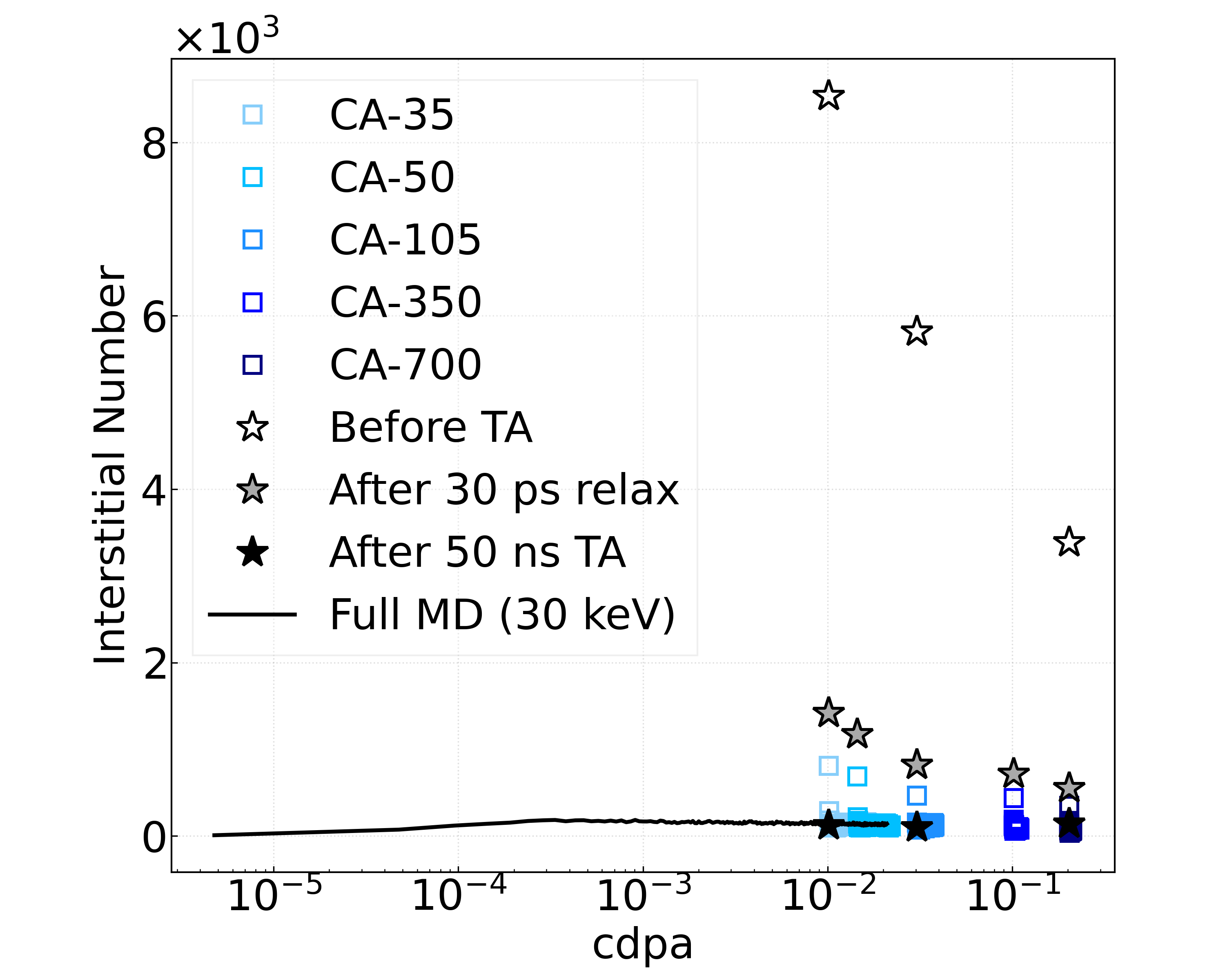}} 
\hspace{10mm}
\subfloat[size = 2]{\includegraphics[width=.3\linewidth]{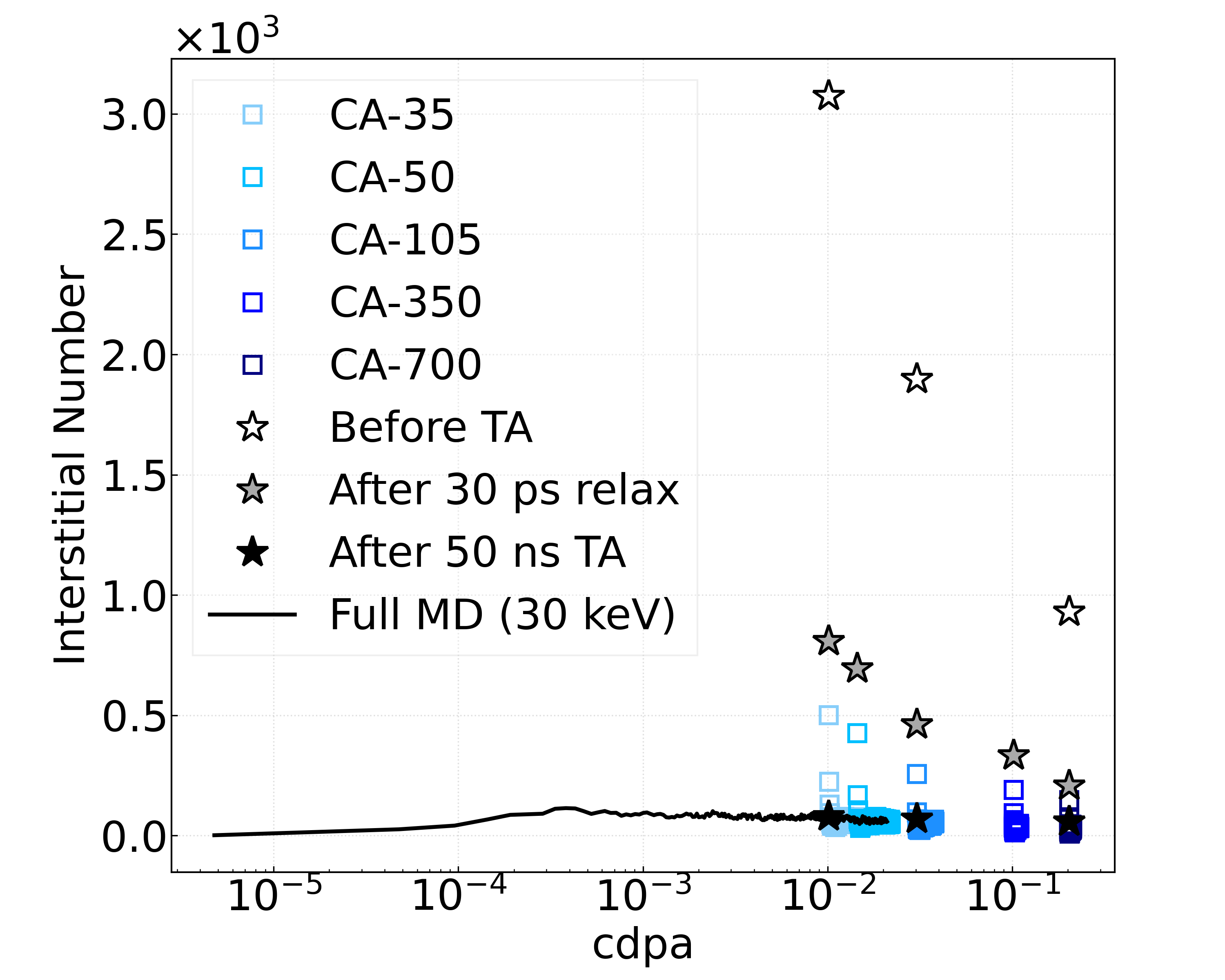}} \\
\subfloat[size = 3-4]{\includegraphics[width=.3\linewidth]{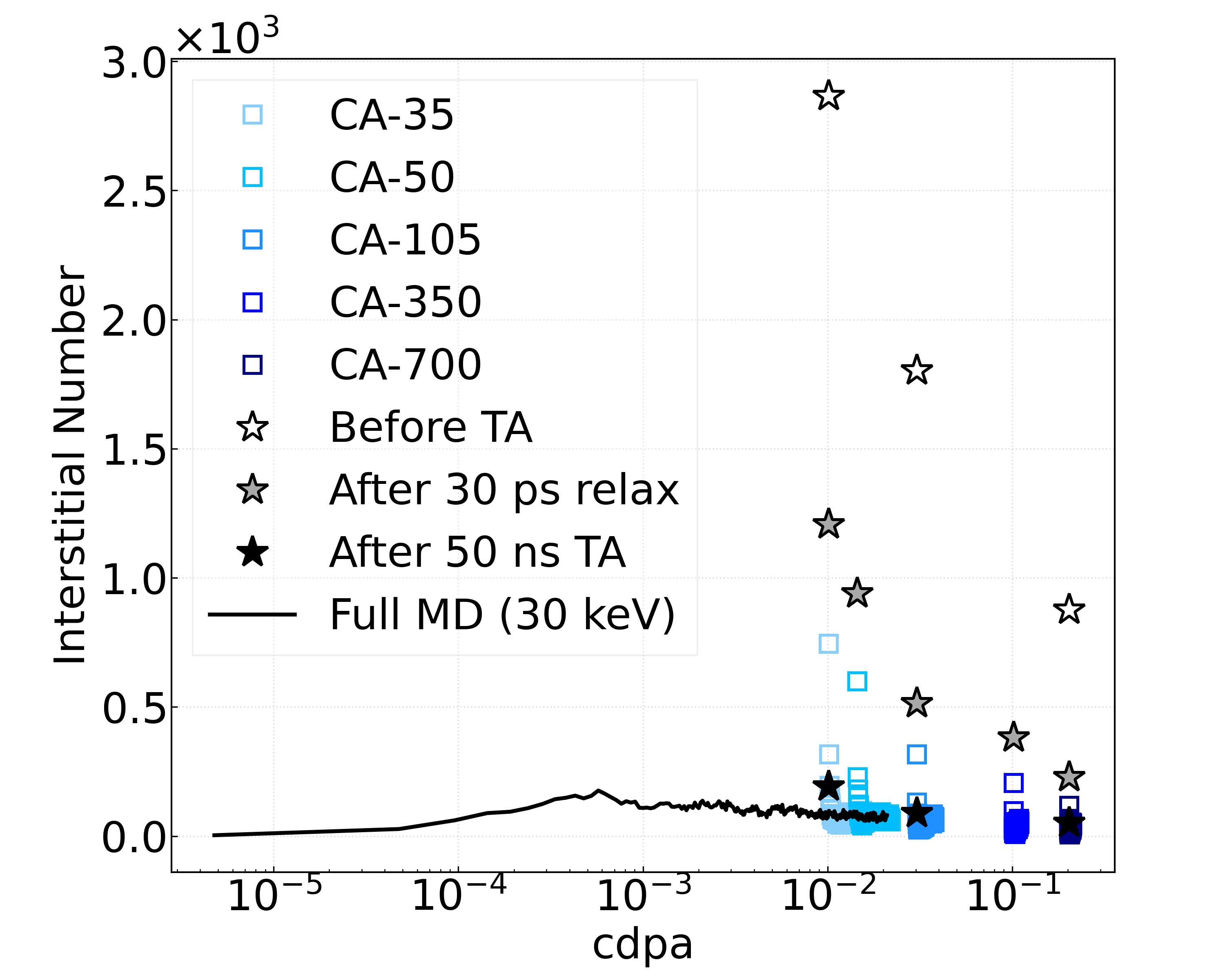}}
\hspace{10mm}
\subfloat[size = 5-10]{\includegraphics[width=.3\linewidth]{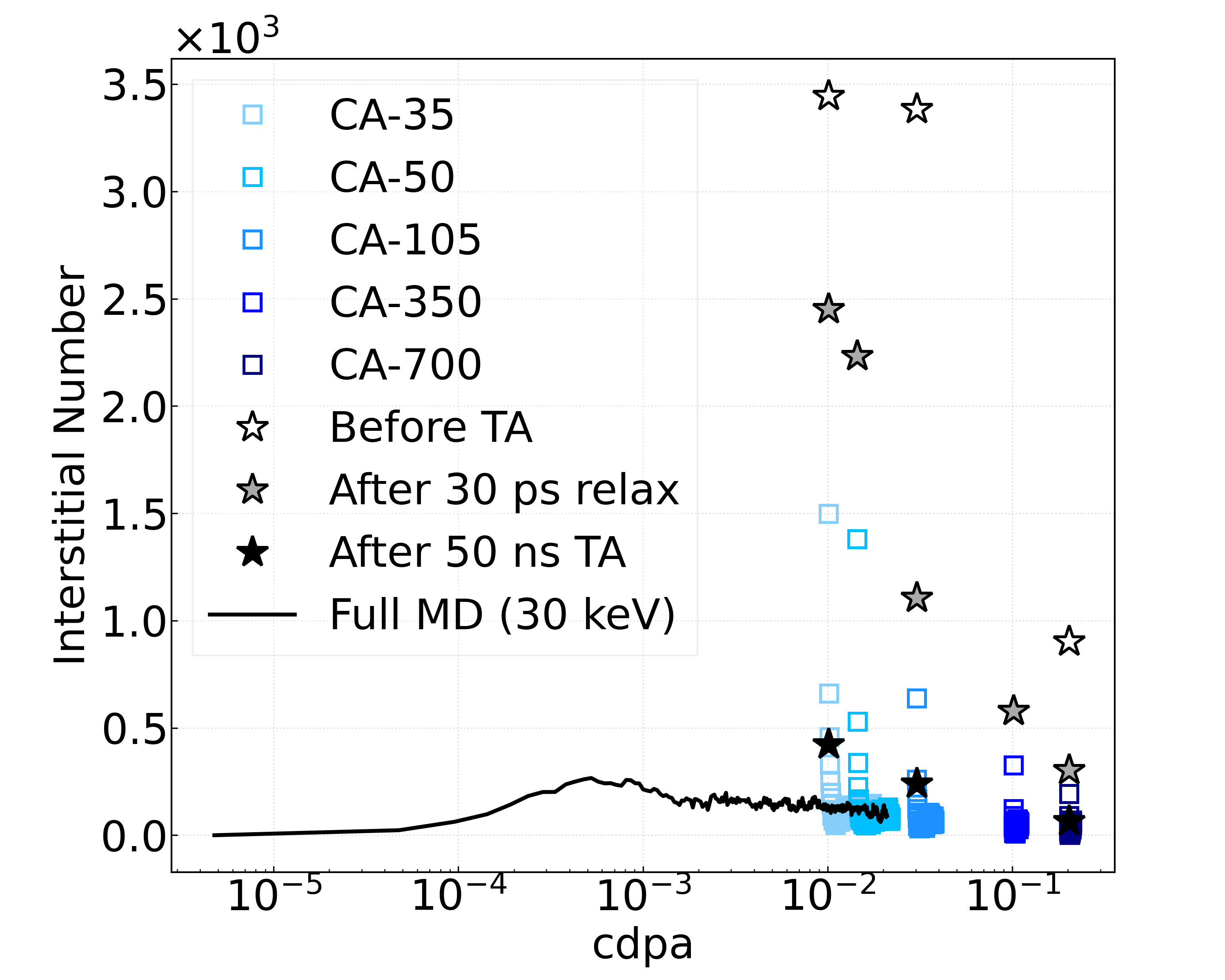}} \\
\subfloat[size = 11-20]{\includegraphics[width=.3\linewidth]{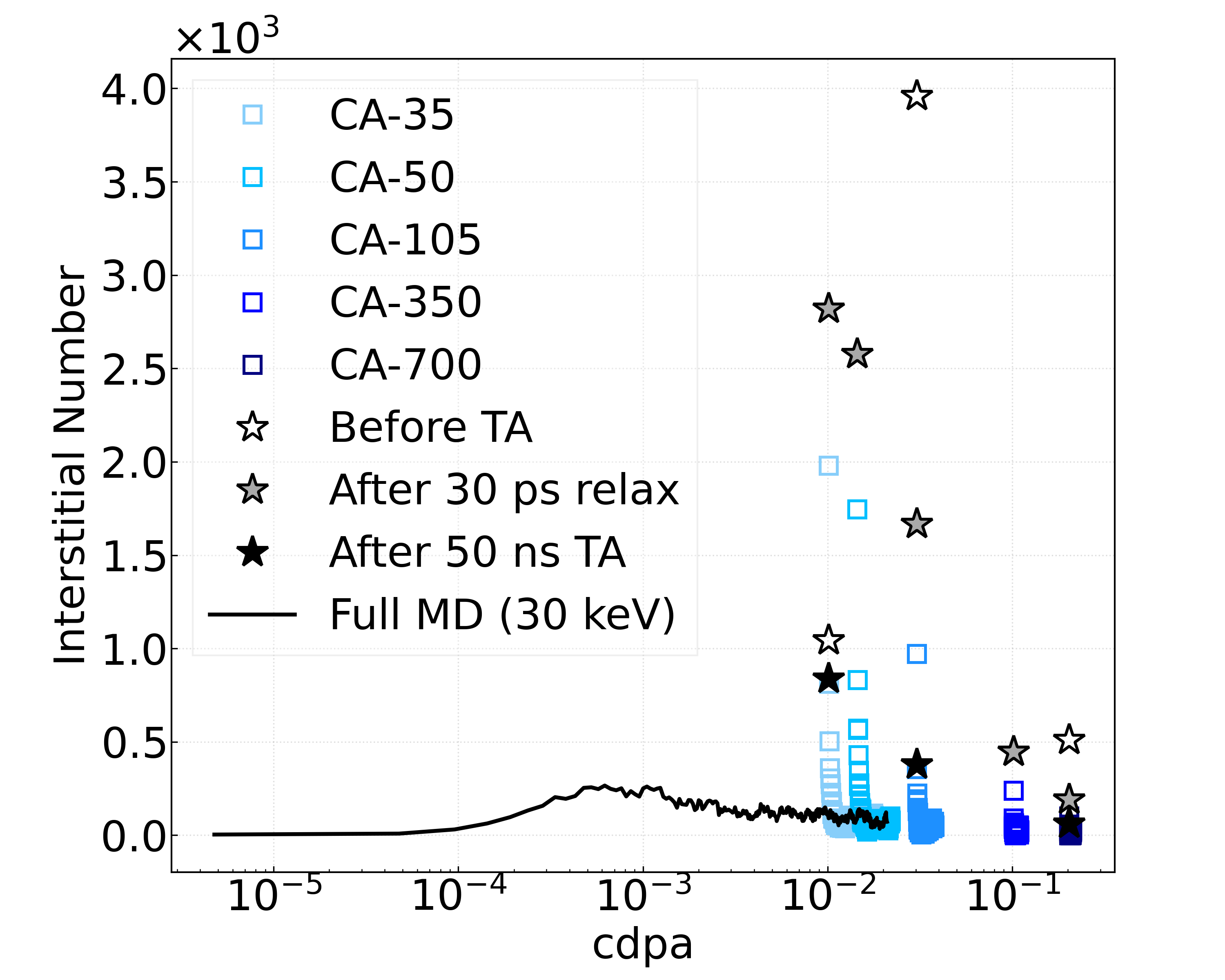}}
\hspace{10mm}
\subfloat[size = 21-50]{\includegraphics[width=.3\linewidth]{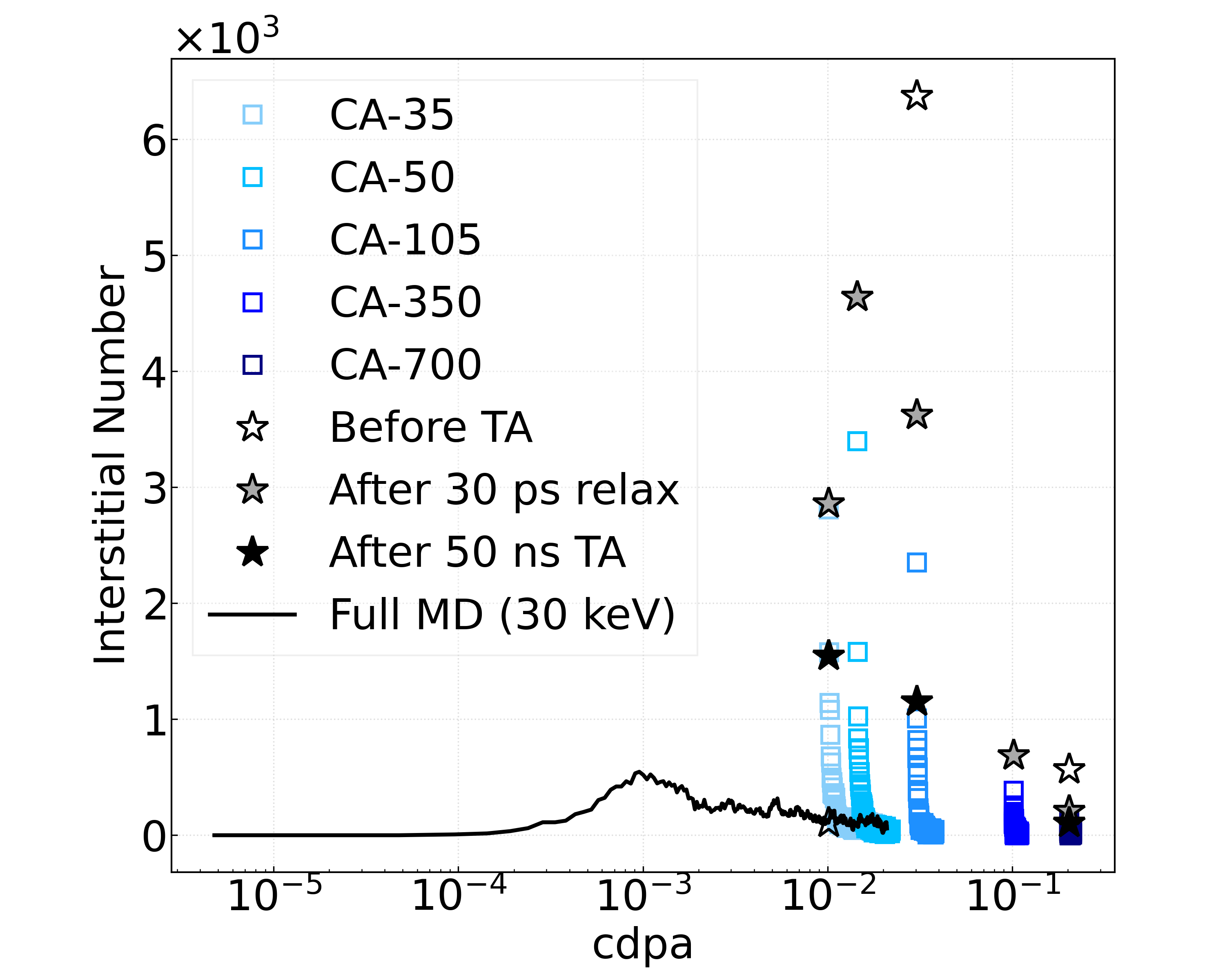}} \\
\subfloat[size = 51+]{\includegraphics[width=.3\linewidth]{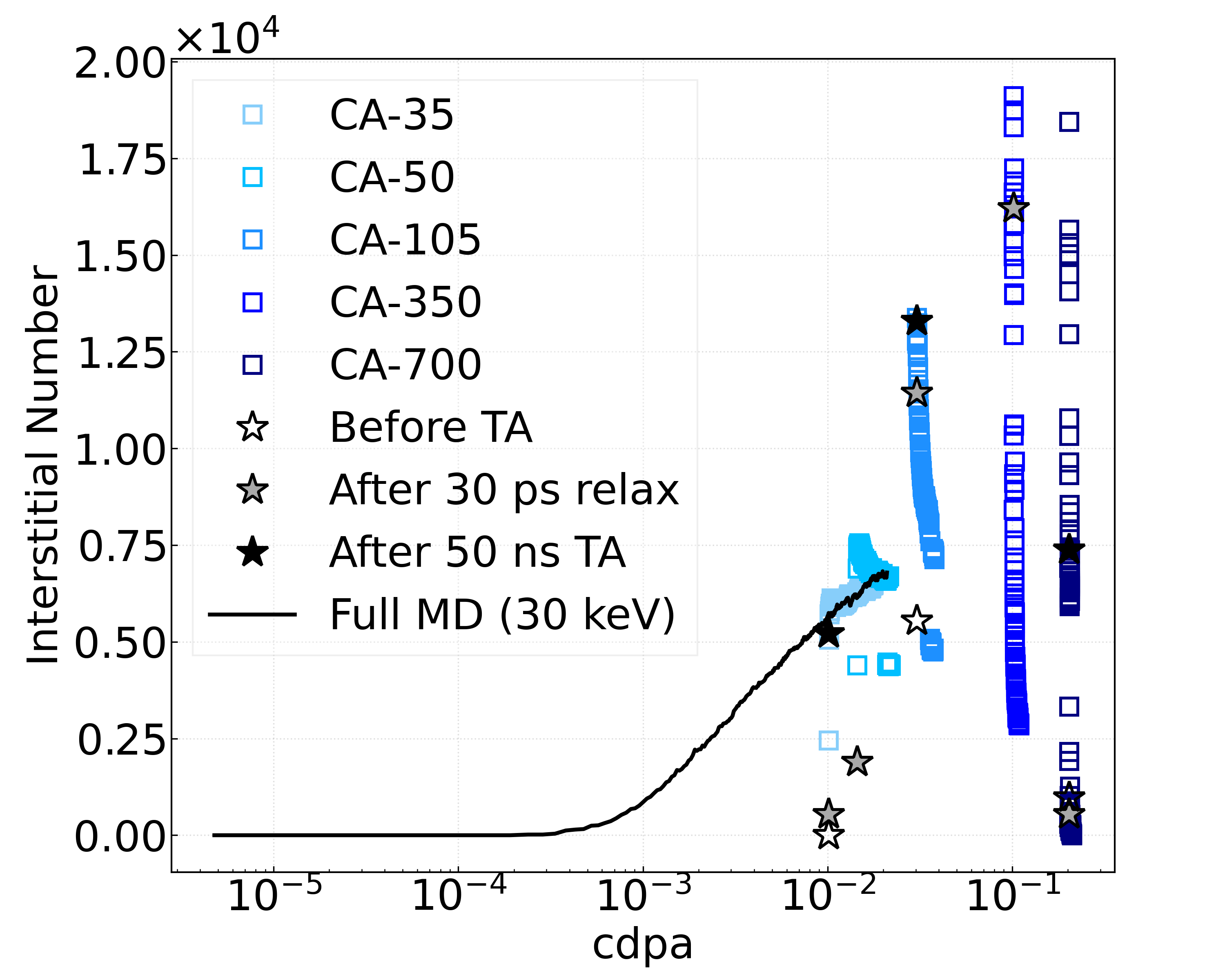}} \\
\end{center}
\caption{Number of interstitials found in different-sized vacancy clusters by the W-S method. Solid lines: full MD simulations. Different shades of blue squares: the vacancy evolution during CA. Open star markers: CRA simulations only. Gray filled star markers: 30 ps relaxation at room temperature after the CRA. Black filled star markers: after 50 ns TA at room temperature.}
\label{I_WS_clusterSize}
\end{figure*}

\newpage
\clearpage

\newpage
\clearpage

\newpage
\clearpage

\newpage
\clearpage

